\documentclass[twoside,a4paper]{article}
\usepackage{amsmath,graphicx,amssymb,fancyhdr,amsthm,amsfonts,cite,verbatim,epsfig}
\usepackage{dsfont}
\newtheorem{thm}{Theorem}[section]

\newtheorem{prop}[thm]{Proposition}

\theoremstyle{definition}
\newtheorem{defn}[thm]{Definition}
\theoremstyle{remark}
\newtheorem{rem}[thm]{Remark}
\def\beq{\begin{eqnarray}}
\def\eeq{\end{eqnarray}}
\def\bsp{\begin{split}}
\def\esp{\end{split}}

\def\k{\kappa}
\def\Om{\Omega}

\def\d{\mathrm{d}}
\def\diag{\mathrm{diag}}
\def\Dth{D^{6<}}
\def\s{\sigma}
\def\Dtw{D^{45<}}

\def\e{\varepsilon}
\def\eb{{\bf e}}

\def\vc{\check{v}}
\def\nc{\check{n}}
\def\Lc{\check{L}}
\def\Kc{\check{K}}
\def\Hc{\check{H}}
\def\vh{\hat{v}}
\def\nh{\hat{n}}
\def\Hh{\hat{H}}
\def\Lh{\hat{L}}
\def\Kh{\hat{K}}
\def\Hb{\bar{H}}
\def\Rb{\bar{R}}
\def\Sb{\bar{S}}
\def\wb{\bar{w}}
\def\Ch{\hat{C}}
\def\Cc{\check{C}}
\def\Cck{{\sf \Cc_{-K}}}
\def\Cckt{{\sf \Cc_{K}}{}^t}
\def\a{\alpha}
\def\b{\beta}
\def\F{{\bf F}}
\def\U{{\bf U}}
\def\V{{\bf V}}
\def\W{{\bf W}}
\def\X{{\bf X}}
\def\Y{{\bf Y}}

\def\Zc{\bar{Z}}
\def\Wc{\bar{W}}
\def\diff{\mbox{diff}}
\def\diag{\mbox{diag}}
\def\Ker{\mbox{Ker}}
\def\Im{\mbox{Im}}

\def\rank{\rho}
\def\id{\mbox{id}}
\def\gg{{\bf g}}
\def\g{{\mbox g}}
\def\p{\bot}
\def\para{\parallel}
\def\l{\lambda}

\def\mm{{\bf m}}
\def\mmc{\bar{\mm}}
\def\bl {\mbox{\boldmath{$\ell$}}}
\def\bn {\mbox{\boldmath{$n$}}}

\newcommand{\mf}[1]{{\mathfrak #1}}

\newcommand{\mb}[1]{{\mathbb #1}}

\newcommand{\mbold}[1]{\mbox{\boldmath{\ensuremath{#1}}}}

\newcommand{\vrk}{\hfill $\Box$}
\newcommand{\be}{\begin{equation}}
\newcommand{\ee}{\end{equation}}
\newcommand{\beqn}{\begin{eqnarray}}
\newcommand{\eeqn}{\end{eqnarray}}

\begin{document}

\title{\Large\textbf{Refinements of the Weyl tensor classification in five dimensions}}

\author{{\large Alan Coley$^\diamond$, Sigbj\o rn Hervik$^\diamondsuit$, Marcello Ortaggio$^\star$ and Lode Wylleman$^{\diamondsuit,\dagger,\natural}$}\\
\vspace{0.05cm} \\
{\small $^\diamond$ Department of Mathematics and Statistics, Dalhousie University}, {\small Halifax, Nova Scotia, Canada B3H 3J5}  \\
{\small $^\diamondsuit$ Faculty of Science and Technology, University of Stavanger}, {\small  N-4036 Stavanger, Norway}  \\
{\small $^\star$ Institute of Mathematics, Academy of Sciences of the Czech Republic}, {\small \v Zitn\' a 25, 115 67 Prague 1, Czech Republic} \\
{\small $^\dagger$ Faculty of Applied Sciences TW16, Ghent
University},  {\small  9000 Gent, Belgium}\\
{\small $^\natural$ Department of Mathematics, Utrecht University, 3584 CD Utrecht, The Netherlands}\\
{\small E-mail: \texttt{aac@mathstat.dal.ca, sigbjorn.hervik@uis.no,
ortaggio@math.cas.cz, lode.wylleman@ugent.be}} }

\date{\today}
\maketitle
\pagestyle{fancy}
\fancyhead{} 
\fancyhead[EC]{A. Coley, S. Hervik, M. Ortaggio and L. Wylleman}
\fancyhead[EL,OR]{\thepage}
\fancyhead[OC]{The Weyl operator}
\fancyfoot{} 

\begin{abstract}

We refine the null alignment classification of the Weyl tensor of a
five-dimensional spacetime.  The paper focusses on the algebraically
special alignment types {\bf {N}}, {\bf {III}}, {\bf {II}} and {\bf {D}}, while types {\bf {I}} and {\bf {G}} are briefly
discussed.  A first refinement is provided by the notion of spin type of
the components of highest boost weight.  Second, we analyze the Segre
types of the Weyl operator acting on bivector space and examine the
intersection with the spin type classification.  We present a full
treatment for types {\bf {N}} and {\bf {III}}, and illustrate the classification from
different viewpoints (Segre type, rank, spin type) for types {\bf {II}} and {\bf {D}},
paying particular attention to possible nilpotence,  which is a new feature of higher dimensions.  We also
point out other essential differences with the four-dimensional case. In passing, we exemplify the refined classification by mentioning the special subtypes associated to certain important spacetimes, such as Myers-Perry black holes, black strings, Robinson-Trautman spacetimes, and purely electric/magnetic type {\bf {D}} spacetimes.

\end{abstract}


\newpage\tableofcontents\newpage

\section{Introduction}\label{section: Introduction}

Lorentzian spacetimes with more than four dimensions are of current
interest in mathematical physics.  It is consequently useful to have
higher dimensional
generalizations of the classification schemes which
have been successfully employed in four dimensions.  In particular, the
introduction of the alignment theory
\cite{Coleyetal04,Milsonetal05,Coley08}, based on the concept of {\em boost
weight} (abbreviated {\em b.w.} throughout this
paper), has made it possible to algebraically classify any tensor in a
Lorentzian spacetime of arbitrary dimensions by its {\em (null) alignment
type}, including the classification of the Weyl tensor and the Ricci
tensor.  To complement this, a higher dimensional generalization of the
Newman-Penrose formalism has been presented, which consists of the
Bianchi \cite{Pravdaetal04} and Ricci identities \cite{OrtPraPra07} and
of the commutator relations \cite{Coleyetal04vsi} written out for a null
frame.  More recently, the corresponding GHP formalism has also been
developed  \cite{Durkeeetal10}.

However, other mathematical tools for the study of higher dimensional
Lorentzian spacetimes can also be developed including, e.g., the classification of
tensors utilizing {\em bivectors}.  From this viewpoint, the algebraic
(Segre type) classification of the Weyl tensor, considered as a linear
operator on bivector space, turns out to be equivalent to the algebraic
classification by alignment type in the special case of four dimensions
(i.e., the Petrov classification \cite{Stephanibook}); however, these two classification schemes are non-equivalent in  higher dimensions.  In particular, the alignment
classification is rather course, and developing the algebraic
classification of the Weyl bivector operator may lead to a more refined
scheme.

For this purpose the bivector formalism in higher dimensional
Lorentzian spacetimes was developed in \cite{ColHer09}. The Weyl
bivector operator was defined in a manner consistent with its b.w.
decomposition. Components of fixed b.w. were then
characterized in terms of basic constituents which transform under
irreducible representations of the spins.
This leads to another refinement of the alignment classification,
based on geometric relations between the highest b.w. constituents.
The types arising will be referred to as \emph{spin types}.

In this paper we study the general scheme of \cite{ColHer09} (and thus the two classification refinements mentioned above, and their interplay)
 in the case of
five-dimensional (5D) Lorentzian spacetimes.  These are of
particular interest for a number of reasons.  First, they provide
the simplest arena in which properties of gravity qualitatively
differ from the well-known 4D case.  Certain important new solutions
such as black rings, which are intrinsically higher-dimensional,
seem to admit a closed exact form only in five dimensions (see,
e.g., \cite{EmpRea08} and references therein).  An alternative
spinor classification of the Weyl tensor has also been developed in
5D, and its connection with the b.w. approach has been
discussed~\cite{Desmet02,GarMar09,God10}. Since the two
classifications are not equivalent, the refinements we propose may
also be useful in the spinor classification.  Finally, as a peculiar
feature of five dimensions (see also \cite{ColHer09}), the highest
b.w. constituents are represented by square matrices, vectors and a
single scalar.  In this way both refinements of the 5D Weyl tensor
classification can be carried out in a fully explicit manner and its
main properties can be easily displayed.

The structure of the paper is as follows.  In section \ref{section:
preliminaries} we review the algebraic properties of the 5D Weyl
tensor (i.e., the constituents, the null alignment types and the
Weyl bivector operator). The definitions and the main ideas
regarding the spin type and the Weyl operator refinements are
presented in section \ref{section:  refinements}.  In sections
\ref{section:  type N}--\ref{section:  type II} we elaborate on both
types of refinement and their intersections separately for each of
the primary alignment types {\bf {N}}, {\bf {III}} and {\bf {II}},
where special attention is given to the type {\bf {D}} subcase of
{\bf {II}}.  In section \ref{section:  type IG} we discuss the split
of the Weyl operator in electric and magnetic parts, which is most
useful for types {\bf {I}} and {\bf {G}}.  We conclude in
section~\ref{section: Conclusion} with a discussion and we make some
brief remarks regarding future work. Finally, there are three
appendices. Appendix A summarizes some useful basic facts about the
Jordan normal form of square matrices. Appendix B provides further
details of the Weyl operator classification for type {\bf {III}}.
The intersection of the spin type and eigenvalue structure
classifications in the type {\bf {II}} case is exemplified in
appendix C.

\section{Preliminaries}\label{section: preliminaries}

In this section we recapitulate the necessary definitions and results
from earlier work, meanwhile introducing the notation and
conventions to be used.

\subsection{Boost weight and Weyl tensor constituents}\label{subsec: bw and constituents}

Consider a point $p$ of a 5D spacetime $(M,\gg)$ with Lorentzian metric signature $3$,
and assume that the Weyl tensor at $p$ is non-zero.
By definition this tensor inherits the basic Riemann tensor symmetries and is moreover traceless:
\be
 C_{abcd}=C_{cdab},\quad C_{abcd}=-C_{bacd},\quad C_{a[bcd]}=0,\quad C^a{}_{bad}=0 , \label{symm1}
\ee
where round (square) brackets denote complete (anti-)symmetrization as usual.
Let $(\bl,\bn,{\bf m}_i,\,i=3\,..\,5)$ be a null frame of $T_p M$,
consisting of two null vectors $\bl$ and $\bn$, normalized by $l^an_a=1$,
and three spacelike orthonormal vectors $\mm_i$, orthogonal to the null vectors
($\mm_i{}^a\mm_j{}_a=\delta_{ij},\,\mm_i{}^a\bl_a=\mm_i{}^a\bn_a=0$, with $\delta_{ij}$
the Kronecker-delta). We take 0 and 1 to be the frame indices corresponding to $\bl$ and $\bn$,
respectively (e.g., $T_a l^a=T_0$), whereas $i,\,j,\,k,\ldots$ denote
spacelike indices, running from 3 to 5.~\footnote{We omit the
index 2. This is in accordance with \cite{ColHer09}, but in contrast
with \cite{Coleyetal04,Pravdaetal04} and \cite{GarMar09}, where (01,234), (12,345) are
used, respectively.} For a joint notation of the null frame indices we use capital Roman letters $A_i$ in the sequel.

Under a boost in the $(\bl,\bn)$-plane  the frame vectors transform according to
\begin{eqnarray}
\bl\mapsto\l\bl,\quad \bn\mapsto\l^{-1}\bn,\quad {\bf m}_i\mapsto {\bf m}_i,\qquad \l\in\mathbb{R}\setminus \{0\},\label{boost}
\end{eqnarray}
such that the components of a rank $p$ tensor $T_{a_1\ldots a_p}$ change as follows:
\begin{equation}\label{boost tensor transform }
T_{A_1\ldots A_p}\mapsto \l^{b_{A_1\ldots A_p}}T_{A_1\ldots A_p},\qquad b_{A_1\ldots A_p}\equiv \sum_{i=1}^p (\delta_{A_i0}-\delta_{A_i1}),
\end{equation}
where $\delta_{AB}$ is the Kronecker delta symbol.
Thus the integer $b_{A_1\ldots A_p}$ is the difference between the numbers of 0- and 1-indices, and is called the
{\em boost weight} (b.w.) of the frame component $T_{A_1\ldots A_p}$ (or, rather, of the $p$-tuple $(A_1,\ldots,A_p)$).
For the Weyl tensor, the conditions (\ref{symm1}) imply that all
components of b.w.\ $\pm 4$ or $\pm 3$ are zero, as well as
algebraic relations between the Weyl components of fixed b.w.
($-2\leq \mbox{b.w.}\leq 2$)~\cite{Coleyetal04,Pravdaetal04}:
\begin{eqnarray}
&&\textrm{b.w.}\ 2:\,\,C_{0}{}^i{}_{0i}=0;\qquad \qquad \textrm{b.w.}\ -2:\,\,C_{1}{}^i{}_{1i}=0;\label{bw 2 rel}\\
&&\textrm{b.w.}\ 1:\,\,C_{010i}=C_{0}{}^j{}_{ij};\qquad \qquad \textrm{b.w.}\ -1:\,\,C_{101i}=C_{1}{}^j{}_{ij};\label{bw 1 rel}\\
&&\textrm{b.w.}\ 0:\,\,2C_{0(ij)1}=C_{i}{}^k{}_{jk},\qquad 2C_{0[ij]1}=-C_{01ij},\qquad 2C_{0101}=-C^{ij}{}_{ij}=2C_{0}{}^i{}_{1i}.\label{bw 0 rel}
\end{eqnarray}

Consider now the spin group, which is isomorphic to $O(3)$ and acts on the null frame according to
\begin{eqnarray}
\bl\mapsto\bl,\quad \bn\mapsto\bn,\quad {\bf m}_j\mapsto {\bf m}_iG^i{}_j,\qquad
G^i{}_jG_k{}^j=\delta^i_k.\label{spin}
\end{eqnarray}
The independent Weyl tensor components of a fixed b.w.\ $q$ define objects which transform under irreducible representations of the spin group.
These objects were presented in \cite{ColHer09}, for general spacetime dimension $n+2$, and are here referred to as the b.w.\ $q$ {\em (Weyl) constituents}.
~\footnote{We will use the notation of the Weyl tensor components and constituents used in \cite{ColHer09}. It may be useful to compare it with that employed in other works. For
negative b.w. components Ref.~\cite{Pravdaetal04} defined
$2\Psi_{ij}=C_{1i1j}$, $2\Psi_{jki}=C_{1ijk}$, and
$\Psi_{i}=-C_{011i}$, while for zero b.w. components, Ref.~\cite{PraPraOrt07} introduced $\Phi_{ij}=C_{0i1j}$,
with $\Phi_{ij}^S$ and $\Phi_{ij}^A$ for its symmetric, respectively, antisymmetric part
and $\Phi$ for its trace. These quantities have then appeared in several subsequent papers. A different set of symbols for the full set of Weyl components has been defined in \cite{Durkeeetal10}.
}
In particular, the components $C_{ijkl}$ (b.w.\ 0) and $C_{1ijk}$ (b.w.\ $-1$) are decomposed as follows ($n\geq 3$):
\begin{eqnarray}
&&C^{ij}{}_{kl}\equiv \Hb^{[ij]}{}_{[kl]}=\bar{C}^{ij}{}_{kl}+\frac{4}{n-2}\delta^{[i}_{[k}\Sb^{j]}{}_{l]}+\frac{2}{n(n-1)}\Rb\delta^{[i}_{[k}\delta^{j]}_{l]},\label{Cijkl}\\
&&C_{1ijk}\equiv \Lc_{i[jk]}=2\delta_{i[j}\check{v}_{k]}+\check{T}_{ijk},\qquad \check{T}^i{}_{ik}=0=\check{T}_{i(jk)}. \label{C1ijk}
\end{eqnarray}
Here $\Hb_{ijkl}$ symbolizes a $n$-dimensional Riemann-like tensor (i.e., a tensor exhibiting all the properties (\ref{symm1}), except for the last one), while $\bar{C}_{ijkl}$, $\Rb\equiv \Hb^{ij}{}_{ij}$ and $\Sb_{ij}\equiv \Hb^k{}_{ikj}-\tfrac 1n \Rb\delta_{ij}$ stand for the associated Weyl tensor, Ricci scalar and tracefree Ricci tensor, respectively. For the case of five dimensions ($n=3$) to be treated here, we have that the b.w.~0 Weyl constituent $\bar{C}_{ijkl}$ vanishes identically,
\begin{equation}\label{Cijkl=0}
\bar{C}_{ijkl}= 0,
\end{equation}
while the b.w.~-1 constituent $\check{T}$ is equivalent to a traceless symmetric matrix $\nc$:
\begin{equation}
\nc_{ij}\equiv \tfrac 12 \e^{kl}{}_{(i}\check{T}_{j)kl}\quad\Leftrightarrow\quad \check{T}^i{}_{jk}=\e_{jkl}\check{n}^{il},\qquad \nc_{ij}=\nc_{(ij)},\quad \nc^i{}_i=0,\label{redef Tcheck}
\end{equation}
where $\e_{ijk}$ denotes the sign of the permutation $(ijk)$ of (345). Analogously for $C_{0ijk}$, giving rise to b.w.~1 constituents $\hat v$ and $\nh$. Regarding the b.w.~0 constituent $A$, defined for general dimensions by $A_{ij}\equiv C_{01ij}$, we will use one more simplification specific to $n=3$ (not made explicit in \cite{ColHer09}): as $A_{ij}$ is antisymmetric in $ij$, we will use its dual vector $\wb$ as the equivalent Weyl constituent:
\begin{equation}\label{redef A}
\wb_i=\tfrac 12 \e_{ijk}A^{jk}\quad\Leftrightarrow\quad A_{ij}= \e_{ijk}\wb^k.
\end{equation}
The symbols of the Weyl constituents and their relation with the Weyl components  are summarized in table \ref{dim5}, where the relations (\ref{bw 2 rel})-(\ref{bw 0 rel}) have been implicitly included. The two-index constituents $\Hh$, $\nh$, $\Sb$, $\nc$ and $\Hc$ are traceless and symmetric $3\times 3$ matrices;
those with one index define $3\times 1$ column vectors $\vh$, $\wb$ and $\vc$. 
Together with $\Rb$ they add up to the $25+9+1=35$ independent components of the 5D Weyl tensor.

\begin{table}[ht]
\begin{tabular}{|r|l|l|}
\hline
b.w. & Constituents & \;\;\;\;\;\;\;\;Weyl tensor components \\
\hline
$+ 2$ & $\hat{H}_{ij}$  & \;\;$ C_{0i0j}\equiv \hat{H}_{ij}$ \\
\hline
$+1$  & $\hat{n}_{ij}$, $\hat{v}_i$ &
\begin{tabular}{l}$C_{0ijk}\equiv \Lh_{i[jk]}=2\delta_{i[j}\hat{v}_{k]}+\nh_i{}^l\varepsilon_{ljk}$\\
$C_{010i}\equiv \Kh_{i}=-2\hat{v}_i $\end{tabular} \\
\hline
$0$  & $\Sb_{ij}$, $\wb_i$, $\Rb$ & \begin{tabular}{l}
$C^{ij}{}_{kl}\equiv \Hb^{[ij]}{}_{[kl]}=4\delta^{[i}_{[k}\Sb^{j]}{}_{l]}+\frac{1}{3}\Rb\delta^{[i}_{[k}\delta^{j]}_{l]}$ \\
$C_{1i0j}\equiv M_{ij}=-\tfrac
12\Sb_{ij}-\tfrac 16 \Rb\delta_{ij}-\tfrac 12\e_{i
jk}\wb^k$\\ $C_{01ij}\equiv A_{[ij]}=\e_{ijk}\wb^k$\\
$C_{0101}\equiv \Phi=-\tfrac 12 \Rb$
\end{tabular}\\
\hline
$-1$ &  $\check{v}_i$, $\check{n}_{ij}$ &
\begin{tabular}{l}$C_{1ijk}\equiv \Lc_{i[jk]}=2\delta_{i[j}\check{v}_{k]}+\check{n}_i{}^{l}\varepsilon_{ljk}$\\
$C_{101i}\equiv \Kc_i=-2\check{v}_i$\end{tabular} \\
\hline
$-2$ &   $\check{H}_{ij}$  & \;\;$ C_{1i1j}\equiv\check{H}_{ij}$ \\
\hline
\end{tabular}
\caption{5D Weyl tensor components and constituents.}
\label{dim5}
\end{table}

\subsection{Null alignment type}\label{subsec: alignment type}

Given the null frame $(\bl,\bn,{\bf m}_i)$ of $T_p M$, the {\em boost order} of a rank $p$ tensor
$T_{a_1\ldots a_p}$ with respect to the frame is defined to be the maximal b.w.\ of its non-vanishing components
in the frame decomposition~\cite{Milsonetal05}. This integer is invariant under the subgroup $\mbox{Fix}([\bl])$ of Lorentz transformations fixing the null direction $[\bl]$.~\footnote{The subgroup of $\mbox{Fix}([\bl])$, consisting of special Lorentz transformations, is also known as Sim$(n)$ in $n+2$ dimensions.}
It follows that the boost order is a function of $[\bl]$ only, denoted by $b_T([\bl])$. For the Weyl tensor $C_{abcd}$ and for {\em generic} $\bl$ we have $b_C([\bl])=2$. If a null direction $[\bl]$ exists for which $b_C([\bl])\leq 1$, it is called a {\em Weyl aligned null direction} (WAND) of alignment order $1-b_C([\bl])$. A WAND is called single if its alignment order is 0, and multiple (double, triple, quadruple) if the alignment order is greater than zero (1, 2, 3). The integer
\begin{equation}\label{def zeta}
\zeta\equiv \min_{\bl}\, b_C([\bl])
\end{equation}
is a pointwise invariant of $(M,\gg)$, defining the {\em (Weyl) primary}
or {\em principal alignment type} $2-\zeta$ at $p$; if $\zeta=2,1,0,1$ or
$-2$ this type is still denoted by {\bf {G}}, {\bf {I}}, {\bf {II}}, {\bf {III}} or {\bf {N}},
respectively~\cite{Milsonetal05,Coleyetal04}.~\footnote{This generalizes
the Petrov types {\bf {I}}, {\bf {II}}, {\bf {D}}, {\bf {III}} and {\bf {N}} from four to
higher dimensions.
Petrov types {\bf {II}} and {\bf {D}} together correspond to primary alignment type {\bf {II}}.
Type {\bf {G}} does not occur in four dimensions, but is the generic situation in
higher dimensions~\cite{Milsonetal05}.  Type {\bf {O}} corresponds to the trivial
case $C_{abcd}=0$, which we have excluded here.}  For types {\bf {N}} and {\bf {III}} the
quadruple, respectively triple, WAND $[\bl]$ is in fact the unique multiple
WAND~\cite{Coleyetal04}.~\footnote{If there was another multiple WAND
$[\bl^*]$ we would get the contradiction $C_{abcd}=0$ by considering
components with respect to a null frame $(\bl,\bl^*,\mm_i)$.} If there is a unique double WAND
in the type {\bf {II}} case we will denote this by {\bf {II}}$_{0}$; if there are more
of them we denote this by {\bf {D}} $\equiv$  {\bf {II}}$_{ii}$, in accordance with
the secondary alignment type notation introduced in
\cite{Coleyetal04,Milsonetal05}.
In the present paper
we will focus on the {\em algebraically special types}
{\bf {N}}, {\bf {III}} and {\bf {II}} (or
{\bf {II}}$_0$ and {\bf {D}} separately) in the Weyl algebraic classification scheme (and
where the context is clear, we will refer
to these algebraically special types simply by
type {\bf {II}} or one of its specializations). Similarly, for type {\bf {I}} we will write {\bf {I}}$_0$ if there is a unique
single WAND, and {\bf {I}}$_i$ if there more than one.  In general, a spacetime admits
no WANDs, and we
denote the general case by type {\bf {G}}.

\subsection{The Weyl bivector operator}\label{subsec: Weyl operator}

Let $\wedge^2T_pM$ be the 10-dimensional real vector space of
contravariant bivectors (antisymmetric two-tensors
$F^{ab}=F^{[ab]}$) at $p$. By the first couple of equations in
(\ref{symm1}), the map
\begin{equation}\label{Cop}
{\sf C}:\quad F^{ab}\mapsto \tfrac 12 C^{ab}{}_{cd}F^{cd}=\tfrac 12
F^{cd} C_{cd}{}^{ab}
\end{equation}
determines a linear operator (or endomorphism) on $\wedge^2T_pM$,
which we shall refer to as the {\em Weyl operator}.

A null frame of $T_p M$
\begin{equation}\label{5D null frame}
(\bl,\bn,\mm_i) ,
\end{equation}
where $i=3,4,5$, induces a null basis of $\wedge^2T_pM$:
\begin{equation}\label{calB bivector basis}
(\U_3,\U_4,\U_5,\W,\W_{[45]},\W_{[53]},\W_{[34]},\V_3,\V_4,\V_5),
\end{equation}
consisting of the simple bivectors (in abstract index notation):
\begin{equation}\label{calB def}
\U_i^{ab}=\bn^{[a}\mm_i{}^{b]},\quad \W^{ab}=\bl^{[a}\bn^{b]},\quad \W_{[jk]}^{ab}=\mm_j{}^{[a}\mm_k{}^{b]},\quad \V_i^{ab}=\bl^{[a}\mm_i{}^{b]}.
\end{equation}
The only non-zero scalar products among these (as induced by the spacetime metric) are given by
\[
2\U_i^{ab}\V_{j\,ab}=\delta_{ij},\quad 2\W^{ab}\W_{ab}=-1,\quad 2\W_{[jk]}^{ab}\W_{[lm]\,ab}=\delta_{jl}\delta_{km}-\delta_{jm}\delta_{kl}.
\]
With the notation of table \ref{dim5}, the matrix representation of ${\sf C}$ with respect to (\ref{calB bivector basis})-(\ref{calB def}) 
can be written in a (3+4+3) block form~\cite{ColHer09}:
\begin{eqnarray}
\label{WeylOperator} 
&&{\cal C}\equiv\begin{bmatrix}
M^t & \Ch_K & \Hh \\
\Cc_K{}^t & \Omega & \Ch_{-K}{}^t \\
\Hc & \Cc_{-K}& M
\end{bmatrix},
\end{eqnarray}
where
\begin{eqnarray}
\Ch_{\pm K}\equiv [\pm\Kh\;\Lh],\quad\Omega\equiv\begin{bmatrix}-\Phi & -A^t \\
A & \bar{H} \end{bmatrix},\quad \Cc_{\pm K}\equiv [\pm\Kc\;\Lc].
\label{Omega_etc.}
\end{eqnarray}
Notice that, with respect to the given basis 
(\ref{calB def}),
the subspaces
\begin{eqnarray*}
{\cal U}\equiv
\langle \U_3,\U_4,\U_5\rangle,\qquad {\cal W}\equiv
\langle\W,\W_{[45]},\W_{[53]},\W_{[34]}\rangle,\qquad {\cal V}\equiv \langle\V_3,\V_4,\V_5\rangle
\end{eqnarray*}
precisely contain the bivectors of b.w.\ $-1$, 0 and $+1$, respectively.
If we denote the projection operator onto ${\cal X}$ by $p_{\cal X}$ and the restriction to ${\cal X}$ by $\cdot|_{{\cal X}}$, then the block entries of $C$ are the respective matrix representations of the maps
\begin{eqnarray}
&&{\sf M}^t\equiv  p_{\cal U}\circ {\sf C}|_{\cal U},\quad {\sf \Ch_{K}}\equiv  p_{\cal U}\circ {\sf C}|_{\cal W} ,\quad {\sf \Hh}\equiv  p_{\cal U}\circ {\sf C}|_{\cal V}, \label{p_U} \\
&&{\sf \Cc_{K}}{}^t\equiv p_{\cal W}\circ {\sf C}|_{\cal U},\quad  {\sf \Omega}\equiv p_{\cal W}\circ {\sf C}|_{\cal W},\quad {\sf \Ch_{-K}}{}^t\equiv p_{\cal W}\circ {\sf C}|_{\cal V},\\
&&{\sf \Hc}\equiv p_{\cal V}\circ {\sf C}|_{\cal U},\quad {\sf \Cc_{-K}}\equiv p_{\cal V}\circ {\sf C}|_{\cal W},\quad {\sf M}\equiv p_{\cal V}\circ {\sf C}|_{\cal V}. \label{p_V}
\end{eqnarray}

\section{Refinements of the null alignment classification}\label{section: refinements}

\subsection{Spin type refinement}\label{subsec: spin type}

\subsubsection{Weyl spin type of a null direction}

The boost order of the Weyl tensor with respect to a null frame
$(\bl,\bn,\mm_i)$ is defined to be the maximal b.w.\ of its
non-vanishing components and is a function of $[\bl]$ only, denoted
by $b_C([\bl])$ (cf.\ section \ref{subsec:  alignment type}).  The
last statement follows by considering the induced action of an
element $g\in$ Fix$([\bl])$ on the leading terms.  A more detailed
consideration leads to other invariants for this action, and thus to
other functions of (that is, properties associated with) $[\bl]$. To
this end, we (uniquely) decompose $g$ into a product
$S[G]\,A[\l]\,N[z]$, where $S[G]$ is a spin (\ref{spin}), $A[\l]$ a
boost (\ref{boost}), and $N(z)$ a null rotation about $[\bl]$,
acting on the null frame according to ($|z^2|\equiv z^iz_i$)
\begin{eqnarray}
\bl\mapsto\bl,\quad \bn\mapsto\bn+z^j{\bf m}_j-\frac{1}{2}|z|^2\bl,
\quad {\bf m}_i\mapsto{\bf m}_i-z_i\bl.\label{null rot l}
\end{eqnarray}
In conjunction with the vanishing of all b.w. $>b_C([\bl])$ terms,
(\ref{null rot l}) readily implies that $N(z)$ leaves the leading terms
(or, equivalently,  the b.w.\ $b_C([\bl])$  constituents) invariant~\cite{Milsonetal05}.
Moreover, $A[\l]$ has the effect of simply multiplying the $b_C([\bl])$ constituent
components by a common factor $\l^{b_C([\bl])}$. It follows that spin-invariant quantities
defined by the b.w. $b_C([\bl])$ constituents are properties associated with $[\bl]$. More specifically for 5D Weyl tensors, the spin-irreducible constituents of b.w.\ $b_C([\bl])$ consist of a traceless, symmetric matrix $X$ (for all values of $b_C([\bl])$), a column vector $x$ (for $-1\leq b_C([\bl])\leq 1$) and a scalar $\Rb$ (for $b_C([\bl])=0$ only).  This leads to the notion of {\em spin type of $[\bl]$}, as follows.\\

\underline{The matrix  $X$} is either zero or (symmetric and thus) diagonalizable by applying spins. By tracelessness this leads to four possible {\em primary spin types of $[\bl]$}, symbolized in a Segre-like notation by
\begin{equation}
\{(000)\}, \qquad \{(11)1\}, \qquad\{110\}, \qquad
\{111\}.\label{Segre types}
\end{equation}
Here a zero (one) indicates a zero (non-zero) eigenvalue, and round brackets indicate equal eigenvalues. Hence, $\{(000)\}$ corresponds to the trivial case $X=0$. For primary spin type $\{(11)1\}$ there is one non-degenerate eigendirection and an eigenplane orthogonal to it. The last two types are the non-degenerate primary spin types characterized by three different eigenvalues, each with one corresponding eigendirection; if the distinction between zero and non-zero eigenvalues  is irrelevant
the joint notation $\{111/0\}$ will be utilized.

If we denote the eigenvalues (without multiplicity) by $X_i$, $i=3\,..\,5$, we get $X=\diag[X_3,X_4]\equiv\diag(X_3,X_4,-X_3-X_4)$ after diagonalization.
For the non-degenerate primary types $\{111/0\}$ we write $X=\diff[X_3,X_4]$ to indicate that $X_i\neq X_j$ for $i\neq j$. We may interchange the $\mm_i$'s such that $X=\diff[X_3,-X_3]$ ($X_5=0$) in the case of spin type $\{110\}$, and $X=X_3\k_3$ in the case of spin type $\{(11)1\}$, where
\begin{eqnarray}
\k_3\equiv\diag\left(1,-\tfrac 12,-\tfrac 12\right)&\leftrightarrow&X_4=X_5=-\tfrac 12 X_3.
\label{kappa3}
\end{eqnarray}
The possible normal forms for $X$ are summarized in table \ref{Table: X normal form}. Notice that for $b_C([\bl])=\pm 2$ the case $\{(000)\}$ should be excluded by definition of boost order (cf. Table~\ref{dim5}).\\

\begin{table}
\begin{tabular}{c|cccc}
Primary spin type& $\{(000)\}$ &$\{(11)1\}$&$\{110\}$&$\{111\}$ \\
\hline
Normal form of $X$&0&$X_3\kappa_3$& $\diff[X_3,-X_3]$&$\diff[X_3,X_4]$
\end{tabular}
\caption{Primary spin types of $[\bl]$ and normal forms for $X$ ($X_3X_4\neq 0$).}\label{Table: X normal form}
\end{table}

Next, if $-1\leq b_C([\bl])\leq 1$ we consider \underline{the vector $x$} and first suppose that $X\neq 0$.
If $x$ is non-zero,  its position relative to the $X$-eigendirections is a Fix$([\bl])$-invariant, and one distinguishes three
qualitatively different cases. Together with the possibility $x=0$ we get four {\em secondary spin types of $[\bl]$}, which we symbolize by
\begin{itemize}
\item $0$: $x=0$;
\item $\para$: $x$ is parallel to a non-degenerate eigendirection;
\item $\bot$: $x$ is orthogonal to a non-degenerate eigendirection, and not coinciding with the other two such directions in the case of primary types $\{111/0\}$;
\item $\g$: $x$ is in `general' position; i.e., none of the above hold.
\end{itemize}
These symbols are added in subscript to the primary spin type to form the {\em (total) spin type of $[\bl]$}. For spin type $\{110\}_\para$ it may be important to indicate whether $x$ is parallel to the $0$-eigendirection or not, and we will symbolize this by $\{110\}_{\para0}$ and $\{110\}_{\para1}$, respectively. Likewise, for spin type $\{110\}_\p$ we will write $\{110\}_{\p0}$ if $x$ is orthogonal to the $0$-eigendirection, and $\{110\}_{\p1}$ if it is not.

In the case $X=0$ the spin type can be $\{(000)\}_\para$ ($x\neq 0$) for $-1\leq b_C([\bl])\leq 1$, and $\{(000)\}_0[R\neq 0]$ ($x=0$) for $b_C[\bl]=0$ (for this case we will add $\Rb=0$ or $\Rb\neq 0$ between square brackets after the spin type symbol whenever this distinction is important).\\

Regarding normal forms, it is possible  and advantageous to order the $\mm_i$'s such that $x_4=0$ (except for spin type $\{111\}_\g$) and  $x_3\neq 0$
whenever $x\neq 0$, instead of taking an arrangement where the $X$-normal forms of table \ref{Table: X normal form} are guaranteed. In fact, this has only implications for spin types $\{111\}_{\para 0}$, $\{110\}_{\bot 0}$ and $\{(11)1\}_\bot$, where the normal forms $X=\diff[0,X_4]$, $X=\diff[X_3,0]$ and, for example, $X=X_5\kappa_5$ should be taken instead, where
\begin{eqnarray}
\k_5\equiv\diag\left(-\tfrac 12,-\tfrac 12,1\right)&\leftrightarrow&X_3=X_4=-\tfrac 12 X_5.
\label{kappa5}
\end{eqnarray}
Additionally, we will take $X=X_5\kappa_5$ for type $\{(11)1\}_\g$ as well.
The resulting normal forms for the highest b.w.\ constituents are summarized in table \ref{Table: Canforms bw zeta}. For $b_C([\bl])=\pm 1$ the case $\{(000)\}_0$ should be excluded by definition of boost order.

\begin{table}[ht]
\begin{tabular}{|c||l|l|c|}
\hline
Sec.\ type & $\begin{matrix}\{111/0\}\\X=\diff[X_3,X_4]\end{matrix}$ & \hspace{2cm} {$\{(11)1\}$} & $\begin{matrix}\{(000)\}\\X=0\end{matrix}$ \\
\hline
\hline
$0$ & \hspace{.5cm}$x=0$  & $ X=X_3\k_3$,\;\;\;  $x=0$ & $x=0$ \\
\hline
$\para$  
& $x=(x_3,0,0)$ & $ X=X_3\k_3$,\;\;\;  $x=(x_3,0,0)$  & $x=(x_3,0,0)$\\
\hline
$\bot$ & $x=(x_3,0
,x_5)$ & $X=X_5\k_5$,\;\;\;  $x=(x_3,0,0)$  & -\\
\hline
$\g$ & $x=(x_3,x_4
,x_5)$ & $X=X_5\k_5$,\;\;\;  $x=(x_3,0,x_5)$& -\\
\hline
\end{tabular}
\caption{Spin types of $[\bl]$ and normal forms for $X$ and $x$ ($-1\leq b_C([\bl])\leq 1$). The scalars $x_i$ are non-zero. For spin types $\{111\}_{\para 0}$ and $\{110\}_{\bot 0}$ we have $X_3=0$, resp.\ $X_4=0$, but in all other cases the scalars $X_i$ are non-zero as well.} 
\label{Table: Canforms bw zeta}
\end{table}

\subsubsection{Weyl spin type at a point}

Since on the one hand Weyl-preferred null
directions may exist, and since there are only a finite number of possible spin types on the other,
it is natural to define the notion of Weyl spin type at a spacetime point $p$. It is possible to
distinguish four cases:\\
(1) If the Weyl tensor is of alignment type {\bf {N}}, {\bf {III}}, {\bf {II}}$_0$ or {\bf {I}}$_0$, then the WAND of maximal alignment order $1-\zeta$ (cf.\ definition (\ref{def zeta})) is unique.\\
(2) If the alignment type is {\bf {D}}, consider two distinct double WANDs $[\bl]\neq [\bn]$. With respect to null frames $(\bl,\bn,\mm_i)$ and $(\bn,\bl,\mm_i)$, the b.w.\ $\zeta=0$ constituents (which are then the only non-zero Weyl components) relate like $(\Sb,\wb,\Rb)$ and $(\Sb,-\wb,\Rb)$, and it immediately follows that the spin types of $[\bn]$ and $[\bl]$ are the same.\\
(3) If the alignment type is {\bf {I}}$_i$, one considers the single WANDs and, for instance, the total ordering
\begin{equation}
(X_1,x_1) < (X_2,x_2) \quad\Leftrightarrow\quad  X_1 < X_2\quad\textrm{or}\quad X_1=X_2,\,x_1<x_2 \label{spin type ordering}
\end{equation}
on the set of possible spin types, where
\begin{eqnarray}
\{(000)\} < \{110\} < \{(11)1\} < \{111\},\qquad 0 \,< \,\,\para\,\, <\, \bot \,<\, \g.\label{Xx ordering}
\end{eqnarray}
(4) If the alignment type is {\bf {G}}, one considers all null directions and, for example, the ordering
\begin{eqnarray}
\{110\} < \{(11)1\} < \{111\} \label{X ordering}
\end{eqnarray}
on the set of possible spin types.

\begin{defn} The {\em Weyl spin type at a spacetime point $p$} is defined to be the spin
type of any maximally aligned null direction in the case of the algebraically special alignment
type {\bf {II}} (and its specializations) and alignment type {\bf {I}}$_0$, and the minimal spin type
of the single WANDs [of all null directions] with respect to the ordering (\ref{Xx ordering})
[(\ref{X ordering})] in the case of alignment type {\bf {I}}$_i$ [{\bf {G}}].
\end{defn}

\begin{rem}
(1) The notion of spin type, introduced here for a 5D Weyl tensor, may be readily
transferred (in principle) to any dimension and any tensor. Regarding the different
alignment types of the Weyl tensor in general dimensions, some spin type subcases have already been pointed out
in earlier work~\cite{Coleyetal04,Ortaggio09,ColHer09}, which will be commented on later.\\
(2) The spin type can be used as a classification tool. In
particular, we may try to determine all spacetimes with given
Ricci-Segre and Weyl alignment-spin types. In this respect, all 5D
Einstein spacetimes ($R_{ab}=\frac{R}{5} g_{ab}$)  of alignment type
{\bf {D}} and spin type $\{(11)1\}_\para$, $\{(11)1\}_0$,
$\{(000)\}_\para$ or $\{(000)\}_0[\Rb\neq 0]$ have been invariantly
classified and partially integrated in \cite{ParradoWyll11}; the
collection of these spin types corresponds to the situation where
the Weyl tensor is isotropic in some spacelike plane, in addition
to the boost isotropy in any plane spanned by double WANDs.
\end{rem}

\subsection{Weyl operator refinement}\label{subsec: Weyl operator geometry}

The Weyl operator $\sf C$ on $\wedge^2T_pM$
is characterized by the list of elementary divisors
\[(x-\l_i)^{m_{ij}},\qquad i=1,\ldots,r,\quad j=1,\ldots,\nu_i,\]
where the $\l_i$ are the distinct, possibly complex, eigenvalues of $\sf C$ and $\nu_i\equiv\dim(E_{\l_i})$ is the dimension of the $\l_i$-eigenspace, which equals the number of Jordan blocks corresponding to $\l_i$ in the Jordan normal form  of $\sf C$. The {\em Segre type} of $\sf C$ is the list of the orders $m_{ij}$ where, for fixed $i$, round brackets are used to enclose them in the case where $\nu_i>1$. For instance, $[(3211)12]$ would indicate that there are 3 distinct eigenvalues, the first one corresponding to four Jordan blocks of dimension 3, 2, 1 and 1, while the other two correspond to one Jordan block each, of dimensions 1 and 2, respectively. The integer  $\sum_{1\leq j\leq\nu_i} m_{ij}$ (equalling 7, 1 and 2 for the respective eigenvalues in the example) is the dimension of the {\em generalized eigenspace} $M^{\l_i}$ corresponding to $\l_i$, which is a $\sf C$-invariant subspace of $\wedge^2T_pM$. A basis of the latter is built by concatenating  $\nu_i$ {\em Jordan normal sequences} (JNSs) 
of the form
\begin{equation}\label{F_j}
\F_j[m_{ij}]\equiv \left(\F_j,{\sf C}_\l(\F_j),\ldots,{\sf C}_\l^{m_{ij}-1}(\F_j)\right),\qquad {\sf C}_\l\equiv {\sf C}-\l\,\,\id_{\wedge^2T_pM}.
\end{equation}
We shall use the notation $W_k^{'\lambda_i}$ for the span of those bivectors $F_j$ for which the length $m_{ij}$ of the corresponding JNS equals $k$, and $p_k(\lambda_i)$ for the invariant $\dim(W_k^{'\lambda_i})$. The concatenation of the $M^{\l_i}$-bases yields a Jordan normal basis (JNB) realizing the Jordan normal form of ${\sf C}$. We refer to appendix A or standard text books on linear algebra (such as \cite{Lang}) for a further discussion, dealing with vector space endomorphisms in general.

In four dimensions (4D), the Segre type classification of the Weyl
bivector operator is fully equivalent to the alignment type
classification, and both reduce to the 6 distinct Petrov types.  In
higher dimensions, however, a particular alignment type can allow
for different Segre types. In the present 5D analysis we shall focus
on the algebraically special alignment type {\bf {II}} (and its
specializations), and regard classification of a certain property of
the Weyl operator as a refinement thereof.

We will treat alignment types {\bf {N}} and {\bf {III}} in full
detail:  we shall deduce the possible Segre types, and write down
the kernel $\Ker(\sf C)$, image $\Im(\sf C)$ and a JNB of $\sf C$.
As a type {\bf {N}} ({\bf {III}}) Weyl operator is nilpotent of
index 2 (3), it suffices to study the possibilities of $\rank(\sf
C)$ (and $\rank({\sf C}^2)$), according to formula (\ref{pk lA});
here and henceforth, we denote $\rank(Z)$ as the rank of a
matrix/linear operator $Z$.  We also present the compatibility of
the Segre and spin type classifications.

Regarding type {\bf {II}} (covering {\bf {II}}$_0$ and {\bf {D}}), we shall emphasize the classification based on $\rank(M)$ and $\rank(\Omega)$,
and on the potential nilpotence of $\sf C$. We shall also determine the possible spin types in the case where $M$ is nilpotent and $\Om$ has a quadruple eigenvalue.
The determination of {\em all} possible Segre types for a given spin type would
involve a straightforward but tedious investigation.
Rather, for illustration, we shall present the complete eigenvalue degeneracies for 
spin types $\{\cdot\}_0$, $\{\cdot\}_\para$ and $\{(11)1\}_\bot$  in
appendix \ref{subsec: type II eigenvalues}.

\section{Type {\bf {N}}}\label{section: type N}

In general dimensions, a type {\bf {N}} Weyl tensor is characterized by having a quadruple WAND $[\bl]$.
With respect to any null frame $(\bl,\bn,\mm_i)$ all components of b.w. greater than $\zeta=-2$
vanish; i.e., only the Weyl constituent $\Hc$ is non-zero:
\begin{equation}\label{type N constituents}
\Hh=0,\quad \nh=\vh=0,\quad \Sb=\wb=\Rb=0,\quad \nc=\vc=0,\quad \Hc\neq 0.
\end{equation}
In fact, the argument given in footnote $^{5}$ shows that $[\bl]$ is the only WAND.

\subsection{Spin types}\label{subsec: type N spin types}

The allowed spin types of the WAND, and thus of the spacetime, are $\{(11)1\}$, $\{110\}$ and $\{111\}$ (these were summed up in section 4.5 of \cite{ColHer09}).
We shall use the normal forms for $X=\Hc$ of table \ref{Table: X normal form}. By an additional boost-normalization, we could naturally take $\Hc=\diff[1,-1]$ in the case of spin type $\{110\}$, and $\Hc_3=\pm \kappa_3$ in the $\{(11)1\}$ case (where $\pm$ is the sign of $X_3$).

\subsection{Weyl operator}\label{subsec: type N Weyl operator}

With (\ref{type N constituents}) and the diagonal normal form of $\Hc$ the Weyl operator takes the form:
\beq\label{WeylOperatorN} {\cal
C}=\begin{bmatrix}
0 & 0 & 0  \\
0 & 0 & 0  \\
\Hc & 0 & 0
\end{bmatrix},\qquad \Hc=\diag(\Hc_3,\Hc_4,\Hc_5).
\eeq
Obviously ${\cal C}^2=0$; i.e., $\sf C$ is nilpotent of index 2 such that its only eigenvalue is 0,
and ${\cal V}\oplus{\cal W}\preceq \Ker({\sf C})$.
Also, ${\sf C}(\U_i)=\Hc_i\V_i$ (no sum over $i$), and by $\Hc_3\Hc_4\neq 0$ it
follows that $2\leq \rank({\sf C})=\rank(\Hc)\leq 3$,
where $\rank({\sf C})=2\Leftrightarrow \Hc_5=0$. From formula (\ref{pk lA}), with
$\l_A=0,\,s(0)=2,\,{\sf N}_A=\sf C$, we get
\[p_2(0)=\rank({\sf C}),\quad p_1(0)=10-2\rank(\sf C).\]
This leads to two possibilities (cf.\ section \ref{subsec: Weyl operator geometry} for the notation):
\begin{enumerate}
\item Segre type $[(2221111)]\,$ $\Leftrightarrow\,\rank({\sf C})=3\,\Leftrightarrow\, \Hc_5\neq 0$, with
\begin{eqnarray*}
&&\Im({\sf C})={\cal V}=\langle\V^3,\V^4,\V^5\rangle,\quad \Ker({\sf C})={\cal V}\oplus{\cal W},\\
&& JNB= (\U_3[2],\U_4[2],\U_5[2],[\W],[\W_{[45]}],[\W_{[53]}],[\W_{[34]}]),
\end{eqnarray*}
corresponding 
to $W_2'^{(0)}={\cal U}$ and $W_1'^{(0)}={\cal W}$;
\item Segre type $[(22111111)]\,$ $\Leftrightarrow\,\rank({\sf C})=2\,\Leftrightarrow\,\Hc_5=0$, with
\begin{eqnarray*}
&& \Im({\sf C})=\langle\V^3,\V^4\rangle,\quad \Ker({\sf C})=\langle\U^5\rangle\oplus{\cal V}\oplus{\cal W},\\
&& JNB= (\U_3[2],\U_4[2],[\U_5],[\V_5],[\W],[\W_{[45]}],[\W_{[53]}],[\W_{[34]}]),
\end{eqnarray*}
corresponding to $W_2'^{(0)}=\langle\U^3,\,\U^4\rangle$ and $W_1'^{(0)}=\langle\U^5\rangle\oplus\langle\V^5\rangle\oplus{\cal W}$.
\end{enumerate}
Here $\U_{i}[2]=[\U^{i},\Hc_i \V_{i}]$ for $\Hc_i\neq 0$ (cf.\ (\ref{F_j})).

\subsection{Intersection of the two refinements}\label{subsec: type N intersection}

The intersection of the spin type and Segre type classifications is trivial
and summarized in table \ref{Table: type N Segre-spin}: case 1 above covers
spin types $\{111\}$ and $\{(11)1\}$, whereas case 2 corresponds precisely to spin type $\{110\}$.

\begin{table}[t]
\begin{tabular}{c|ccc}
& $\{110\}$ & $\{(11)1\}$&$\{111\}$\\
\hline
$[(2221111)]$&-&x&x\\
$[(22111111)]$&x&-&-\\
\end{tabular}
\caption{Type {\bf {N}} Weyl tensors: spin types (columns) for given Segre types (rows). The symbols - and x indicate that the corresponding Segre type is not or is allowed, respectively.}
\label{Table: type N Segre-spin}
\end{table}

\subsection{Comparison with 4D}\label{subsec: type N 4D}

Consider a 4D spacetime which is of
(Weyl-Petrov) type {\bf {N}} at a point. With respect to
any null frame $(\bl,\bn,\mm_3,\mm_4)$ with $[\bl]$ the quadruple,
unique WAND, we formally have (\ref{type N constituents}), so that with respect to a bivector frame
\begin{equation}\label{type N 4D Weyl frame}
(\U_3,\U_4,\W,\W_{[34]},\V_3,\V_4),
\end{equation}
defined as in (\ref{calB def}), the Weyl operator ${\sf C}$ takes the (2+2+2)-block form formally equal to (\ref{WeylOperatorN}). Hence $\sf C$ is nilpotent of index 2. However, $\{11\}$ is the only possible spin type here, where we may take $\Hc=\diag(1,-1)$ after boost-normalization. Also, $\rank({\sf C})=2$ and the only possible Segre type of $\sf C$ is $[(2211)]$. On writing
\begin{eqnarray*}
{\cal V}\equiv \langle\V_3,\V_4\rangle,\quad {\cal W}\equiv
\langle\W,\W_{[34]}\rangle,\quad {\cal U}\equiv \langle
\U_3,\U_4\rangle
\end{eqnarray*}
we have
\begin{eqnarray*}
&&\mbox{Im}({\sf C})={\cal V},\quad \Ker({\sf C})={\cal V}\oplus{\cal W},\\
&& JNB=([\U_3,\V_3],[\U_4,-\V_4],[\W],[\W_{[34]}]).
\end{eqnarray*}
We note that certain four-dimensional results on the Weyl operator (including the rank properties) were given in \cite{Schell61,GolKer61}.

\subsection{Discussion}\label{subsec: type N discussion}

From the above we conclude that {\em for alignment type {\bf {N}} Weyl tensors,
the spin type classification refines the Weyl operator Segre type
classification; the latter coincides with the classification based on the
rank of ${\sf C}$; i.e., on the number of non-zero eigenvalues of the
matrix $\Hc$.}  Although only the $5D$ case has been treated here, it is
clear that these statements still hold in $n+2$ spacetime dimensions:
for $\rank({\sf C})=m\geq 2$ we will have precisely $m$ two-dimensional
and $n-2(m-1)$ one-dimensional Jordan blocks, reflected in the Segre
type; this Segre type is independent of the degeneracies of non-zero
$\Hc$-eigenvalues.  Notice that $m\geq 2$ is due to the tracelessness of
$\Hc$; the case $m=2$ (Segre type $[(2211\cdots 1)]$)
corresponds precisely to the unique spin type $\{1100\cdots 0\}$, while for fixed
$m>2$ $\Hc$-eigenvalue degeneracies become possible and the corresponding
Segre type covers several spin types.

In 4D the matrix $\Hc$ is two-dimensional and $m=2$ is the only possibility, corresponding to spin type $\{11\}$. Most remarkably, {\em for type {\bf {N}}
non-Kundt Einstein spacetimes ($R_{ab}=\frac{R}{D} g_{ab}$) in {\em any} dimensions $D$, the case $m=2$ is the unique possibility} as well, which is due to the compatibility with the Bianchi identities~\cite{Pravdaetal04,Durkeeetal10}. Some explicit examples have been constructed in \cite{OrtPraPra10}.

However, such a constraint does not apply to Kundt Einstein spacetimes. As an illustration, let us recall the homogeneous plane-wave spacetimes,
\[ \d s^2=2\d u(\d v+a_{ij}x^ix^j\d u)+\delta_{ij}\d x^i\d x^j,\]
where $a_{ij}$ is a constant matrix. If the matrix $a_{ij}$ is traceless then this is Ricci-flat, otherwise some pure radiation will be present. In both cases this metric is of Weyl type {\bf {N}} and the eigenvalue type of $a_{ij}$ is directly related to the spin type of the Weyl tensor.

\section{Type {\bf {III}}}\label{section: type III}

In general dimensions, a type {\bf {III}} Weyl tensor is characterized by having a triple WAND $[\bl]$.
With respect to any null frame $(\bl,\bn,\mm_i)$ all components of b.w. greater than $\zeta=-1$ vanish, whereas those of b.w.~-1 are not all zero; in terms of constituents this is
\begin{equation}\label{type III constituents}
\Hh=0,\quad \nh=\vh=0,\quad \Sb=\wb=\Rb=0,\quad (\nc,\vc)\neq (0,0).
\end{equation}
The argument given in footnote $^{5}$ shows that any
other WAND must be single; i.e., $[\bl]$ is the unique triple WAND and there can be no double WANDs.
The existence of a single WAND corresponds to $\Hc=0$, and is symbolized within the full alignment type notation by {\bf {III}}$_i$~\cite{Coleyetal04}.

\subsection{Spin types}\label{subsec: type III spin types}

The allowed spin types of the unique triple WAND, and thus of the spacetime, are the combinations in table \ref{Table: X normal form}, with the exception of $\{(000)\}_0$ (which would yield type {\bf {N}} or {\bf {O}}).
The secondary spin type `0' (i.e, $\vc=0$) was denoted as type {\bf {III}}(a) in \cite{Coleyetal04,Ortaggio09} and {\bf {III}}(A) in \cite{ColHer09}. The primary spin types were also mentioned in \cite{ColHer09}, where $\{(000)\}$ ($\nc=0$) was denoted by {\bf {III}}(B).

\subsection{Weyl operator}\label{subsec: type III Weyl operator}

From (\ref{type III constituents}), and taking the diagonal normal form for $\nc$, the Weyl operator takes the form:
\begin{eqnarray}\label{WeylOperator III}
{\cal C}\equiv\begin{bmatrix}
0 & 0 & 0 \\
\Cc_K{}^t & 0 & 0 \\
\Hc & \Cc_{-K} & 0
\end{bmatrix},\quad
\Cc_{\pm K}\equiv [\pm \Kc\,\,\Lc]=\begin{bmatrix}\mp 2\vc_3&\nc_3&-\vc_5&\vc_4\\
\mp 2\vc_4&\vc_5&\nc_4&-\vc_3\\\mp 2\vc_5&-\vc_4&\vc_3&\nc_5\end{bmatrix}.
\end{eqnarray}
Obviously we have
${\cal C}^3=0$ and thus 0 is the only eigenvalue, just as for
type~{\bf {N}}.  However, contrary to the type~{\bf N} case, a
type {\bf III} Weyl operator satisfies ${\cal C}^2\neq 0$. A proof
hereof, in general dimensions, was given in \cite{Coleyetal04vsi},
Lemma 12; however, let us present a shortcut, specifically for five
dimensions.

\begin{prop} A 5D type {\bf {III}} Weyl operator ${\sf C}$ is nilpotent of index 3; i.e., ${\sf C}^3=0\neq{\sf C}^2$. \label{prop_IIInilp}
\end{prop}
{\bf Proof.} We have
\begin{eqnarray}\label{WeylOperator IIIsquare}
{\cal C}^2\equiv\begin{bmatrix}
0 & 0 & 0 \\
0 & 0 & 0 \\
\Cc_{-K}.\Cc_K{}^t & 0 & 0
\end{bmatrix},
\end{eqnarray}
where a dot denotes matrix multiplication.
Suppose that ${\cal C}^2=0\Leftrightarrow \Cc_{-K}.\Cc_{K}{}^t=0$.
By $\Cc_{\pm K}=[\pm\Kc\,\Lc]$, this is equivalent to
\begin{equation}\label{rank C2 0}
\Lc.\Lc^t=\Kc.\Kc^t.
\end{equation}
This would imply $\rank(\Lc)=\rank(\Lc.\Lc^t)=1$ and hence
$\Lc=ax^t$, with $a,x\in\mathbb{R}^{3\times 1}$. Compatibility with
(\ref{rank C2 0}) then requires $\Lc=\Kc e^t$, with $e$ a unit
vector in $\mathbb{R}^{3\times 1}$.
But then $\Cc_{K}=\Kc[1\,\, e^t]$ such that $2\leq \rank(\Cc_{K})=1$: this is a contradiction.\vrk\\

 The difference between the indices of nilpotence serves as an
easily testable criterion for  distinguishing the alignment
types~{\bf III} and {\bf N}.

Next, from (\ref{WeylOperator III}) it is clear that the order 2
minors~\footnote{An order $p$ minor
$M\left(\begin{smallmatrix}i_1&i_2&\cdots&i_p\\j_1&j_2&\cdots&j_p\end{smallmatrix}\right)$
of a matrix $M\in\mathbb{R}^{m\times n}$, $1\leq i_1<\ldots<i_p\leq
m$, $1\leq j_1<\ldots<j_p\leq n$, is the determinant of the $p\times
p$ submatrix $A$ of $M$ with $A_{kl}=M_{i_kj_l}$, $1\leq k,l\leq
p$.}
$\Cc_K$$\left(\begin{smallmatrix}i_1&i_2\\j_1&j_2\end{smallmatrix}\right)$
of $\Cc_{\pm K}$ cannot all be zero, as this would lead to
$\vc=\nc=0$ and thus type~{\bf N}. Hence,
$\rank(\Cc_{K})=\rank(\Cck)=\rank(\Cckt)$ equals 2 or 3. Moreover,
from the structure of (\ref{WeylOperator III}) we immediately see
that ${\cal V}\precneqq\Ker({\sf C})$ and
\begin{eqnarray*}
4\leq 2\,\rank(\Cc_K)\leq \rank({\sf C})\leq 3+\rank(\Cc_K)\leq 6.
\end{eqnarray*}
Defining the order 3 minors
\begin{eqnarray}
&&d \equiv \det(\Lc)=\nc_3\nc_4\nc_5+\nc_3\vc_3^2+\nc_4\vc_4^2+\nc_5\vc_5^2,\label{ddef}\\
&&d_i \equiv
2[\vc_j\vc_k(\nc_j-\nc_k)-\vc_i(\vc_3^2+\vc_4^2+\vc_5^2+\nc_j\nc_k)],\label{didef}
\end{eqnarray}
we have
\begin{equation}\label{rankC=6}
\rank({\sf C})=6\quad
\Leftrightarrow\quad\rank(C_{K})=3\quad\Leftrightarrow\quad
D^6\equiv d^2+d_3^2+d_4^2+d_5^2\neq 0.
\end{equation}
Conforming to the canonical forms of table \ref{Table: Canforms bw
zeta}, we take $\vc_3\neq 0$ whenever $\vc\neq 0$, while in the case
$\vc=0$ we may permute the $\mm_i$'s such that $\nc_3\nc_4\neq 0$.
Then, the first and second columns of $\Cc_K{}^t$ are always
independent, as well as columns $j$ and $k$ of $\Cc_{-K}$, where
\begin{eqnarray}
&\vc=0\,\,(\nc_3\nc_4\neq 0):& j=2,\,k=3;\label{jk v=0}\\
&\vc\neq 0\,\,(\vc_3\neq 0):& j=1,\,k=4\,\,\mbox{or}\,\,3.\label{jk
v<>0}
\end{eqnarray}
It is then easily seen that only one order 5 minor is needed to
distinguish between $\rank({\sf C})=4$ and $\rank({\sf C})=5$:
\begin{equation}\label{D45def}
\rank({\sf C})=4\quad \Leftrightarrow\quad D^{45}\equiv {\cal
C}\left(\begin{smallmatrix}j+3&k+3&8&9&10\\1&2&3&j+3&k+3\end{smallmatrix}\right)=0.
\end{equation}

Finally, based on formula (\ref{pk lA}) with $\l_A=0,\,s(0)=3,\,{\sf
N}_A=\sf C$, the numbers $p_i(0)$ of Jordan blocks of dimension $i$
are determined by $\rank(\sf C)$ and $\rank({\sf C}^2)$:
\begin{equation}\label{pk III}
p_1(0)=10-2\rank({\sf C})+\rank({\sf C}^2),\qquad p_2(0)=\rank({\sf
C})-2\rank({\sf C}^2),\qquad p_3(0)=\rank({\sf C}^2).
\end{equation}
For $\rank({\sf C})=6$ we get $2\leq \rank({\sf C}^2)\leq 3$ from
$p_1(0)\geq 0$, where
\begin{eqnarray}\label{CK=3 C2=2}
&&\rank({\sf C}^2)=2\quad\Leftrightarrow\quad \Dth\equiv
\det\left(\Cc_{-K}.C_{K}{}^t\right)\equiv d^2-d_3^2-d_4^2-d_5^2=0.
\end{eqnarray}
For $\rank({\sf C})<6$, proposition \ref{prop_IIInilp} and
$p_2(0)\geq 0$ yield $1\leq \rank({\sf C}^2)\leq 2$, where
\begin{eqnarray*}
\rank({\sf C}^2)=1\quad\Leftrightarrow\quad\Dtw \equiv
(\nc_3^2+\vc_4^2+\vc_5^2-4\vc_3^2)(\nc_4^2+\vc^2-5\vc_4^2)-(\vc_5(\nc_3-\nc_4)-5\vc_3\vc_4)^2=0.
\end{eqnarray*}

 This leads to Table~\ref{Table: type III summary}, which
summarizes the possible Segre types that arise in the classification
based on the Weyl operator geometry in the type {\bf III} case. For
further details (like $\Ker({\sf C})$, $\Im({\sf C})$ and the
determination of JNBs in the different cases) we refer to appendix
\ref{subsubsection: type III rank C} and \ref{subsubsection: type
III rank C2}.

\begin{table}[t]
\begin{tabular}{|l||c|l|c|}
\hline
  & $\rank(\sf C^2)=3$ & $\rank(\sf C^2)=2$ & $\rank(\sf C^2)=1$ \\
\hline
\hline
 $\rank({\sf C})=6$ ($D^6\neq 0$) & $[(3331)]$ ($D^{6<}\neq 0$)  & $[(3322)]$ $\;\;$ ($D^{6<}=0$) & - \\
\hline
 $\rank({\sf C})=5$ ($D^6=0\neq D^{45}$) & - & $[(33211)]$ \hfill ($D^{45<}\neq 0$) & $[(32221)]$ \hfill ($D^{45<}=0$) \\
\hline
 $\rank({\sf C})=4$ ($D^6=0=D^{45}$) & - & $[(331111)]$ \hfill ($D^{45<}\neq 0$) & $[(322111)]$ \hfill ($D^{45<}=0$) \\
\hline
\end{tabular}
\caption{Type {\bf {III}} Weyl bivector operators: summary of the (six) possible Segre types and corresponding conditions on relevant determinants (defined in the text).}
\label{Table: type III summary}
\end{table}

\begin{rem}\label{remark type III_i and rank} If the full alignment type
is {\bf {III}}$_i$ ($\Hc=0$ in (\ref{WeylOperator III})) we clearly
have $\rank({\cal C})=2\rank(\Cc_K)=6$ or 4, i.e., $D^{45}$ vanishes
in this case (cf. (\ref{quant v=0}) and (\ref{gener_v_D45})).
\end{rem}

\subsection{Comparison with 4D}\label{subsec: type III 4D}

Consider a 4D, alignment (Weyl-Petrov) type {\bf {III}} Weyl tensor, and let $[\bl]$ be the unique triple WAND. Referring to the decomposition (\ref{C1ijk}) we have $\check{T}_{ijk}=0$ ($\Leftrightarrow \nc=0$) \cite{ColHer09}, such that the only possible spin type is $\{00\}_\para$. By applying a spin and a boost we may set $\vc\equiv(\vc_3,\vc_4)=(0,1)$, while $\Hc$ can be transformed to zero by a null rotation (\ref{null rot l}) about $[\bl]$, such that the full alignment type is {\bf {III}}$_i$~\cite{Stephanibook,ColHer09}. In this gauge
the Weyl operator
${\sf C}$  takes the (2+2+2)-block form 
\begin{equation}
{\cal C}=\begin{bmatrix}
0 & 0 & 0 \\
\Cc_K{}^t & 0 & 0 \\
0 & \Cc_{-K} & 0
\end{bmatrix},\quad
\Cc_K{}^t=\begin{bmatrix}
0&1\\
1&0
\end{bmatrix},\quad \Cc_{-K}=\begin{bmatrix}
0&-1\\
1&0
\end{bmatrix}.
\end{equation}
Obviously ${\sf C}^3=0\neq {\sf C}^2$, $\rank({\sf C})=2\rank({\sf C}^2)=4$, and the only possible Segre type is $[(33)]$ and
\begin{equation*}
\mbox{Im}({\sf C})={\cal W}\oplus{\cal V},\quad \Ker({\sf C})={\cal V},\quad JNB=([\U_3,\W_{[34]},-\V_3],[\U_4,\W,\V_4]).
\end{equation*}

Cf. also \cite{Schell61,GolKer61}.

\subsection{Discussion}\label{subsec: type III discussion}

It is clear that the Weyl operator geometry approach distinguishes
between type {\bf {III}} and type {\bf {N}} (for this purpose it
suffices to consider $\rank({\sf C})$, equivalent to specifying the
number of eigenvectors, or simply the index of nilpotence of ${\sf
C}$).  However, for type {\bf {III}} there exist  different spin
types that are indistinguishable from the Segre type viewpoint (and
vice versa). We refer to appendix \ref{subsec: type III
intersection} for a full discussion. Table \ref{Table:  type III
spin-Segre} summarizes the relation between the spin type and Segre
type refinement schemes for 5D alignment type {\bf {III}} Weyl
tensors.  Just as for type {\bf {N}}, a certain Segre type covers
several spin types, and different spin types may allow for exactly
the same list of Segre types.  However, except for spin types
$\{(11)1\}_0$, $\{111\}_0$ and $\{(000)\}_\para$, such a list does
not contain a unique element any more (as was the case for type {\bf
{N}}).

Even before studying more general Weyl types (and/or going to higher
dimensions), it is thus already clear from the above 5D type {\bf
{III}} analysis that the two schemes are essentially independent and
represent substantial refinements.  This is in contrast to the 4D
case, where type {\bf {III}} exactly corresponds to a Segre type
$[(33)]$ Weyl operator and allows for a single spin type
($\{00\}_\para$); the same happens for the 4D types {\bf {II}} and
{\bf {I}} (see below).

Finally, we mention that some of the discussed features readily
generalize to $n+2$-dimensional type {\bf {III}} Weyl tensors.  In particular,
the index of nilpotence for ${\sf C}$ is still 3~\cite{Coleyetal04vsi}
such that formula (\ref{pk III}) still holds, with 10 replaced by
$(n+2)(n+1)/2$.  Also, $\rank({\sf C})\leq 2n,\,\rank({\sf C}^2)\leq n$ a
priori, but the lower bounds would need a detailed analysis.  Constraints on type {\bf III} Ricci-flat spacetimes were derived in \cite{Pravdaetal04} using the Bianchi identities. Some examples of type {\bf III} Ricci-flat/Einstein spacetimes were constructed in \cite{OrtPraPra10}, where the ``asymptotic'' behavior of the Weyl tensor was also discussed.

\section{Type {\bf {II}}}\label{section: type II}

In general dimensions, a (primary) type {\bf {II}} Weyl tensor is characterized by having a
double WAND $[\bl]$. With respect to any null frame $(\bl,\bn,\mm_i)$  all components
of b.w. greater than $\zeta=0$ vanish, whereas those of b.w.~0 are not all zero; in terms of constituents
this is:
\begin{equation}\label{type II constituents}
\Hh=0,\quad \nh=\vh=0,\quad (\Sb,\wb,\Rb)\neq (0,0,0).
\end{equation}
If the (full) type is D, and if $\bn$ is a second double WAND, then the components of b.w.
less than $0$ also vanish:
\begin{equation}\label{type D constituents}
\nc=\vc=0,\quad \Hc=0.
\end{equation}

\subsection{Spin types}\label{subsec: type II spin types}

The allowed spin types of any double WAND, and thus of the spacetime
(cf.\ section \ref{subsec: spin type}), are the combinations in
table \ref{Table: X normal form}, with the exception of
$\{(000)\}_0[\Rb=0]$. In \cite{Coleyetal04,Ortaggio09,ColHer09} the
secondary spin type `0' (i.e, $\wb=0$) was denoted by subtype {\bf
{II}}(d), the primary spin type $\{(000)\}$ ($\Sb=0$) by {\bf
{II}}(b), and the case $\Rb=0$ by {\bf {II}}(a). In general
dimensions the subtype {\bf {II}}(c) is defined by equation
(\ref{Cijkl=0}); however, this is identically satisfied in five
dimensions.

\subsection{Weyl operator}\label{subsec: type II Weyl operator}

With (\ref{type II constituents}) the block representation of the Weyl operator reduces to
\begin{eqnarray}\label{WeylOperator II}
{\cal C}\equiv\begin{bmatrix}
M^t & 0 & 0 \\
\Cc_K{}^t & \Omega & 0 \\
\Hc & \Cc_{-K} & M
\end{bmatrix}.
\end{eqnarray}
The various submatrices are defined in table \ref{dim5} and (\ref{Omega_etc.}), and
we work with the diagonal normal form for $\Sb$, where
\[\Rb_i\equiv\Sb_{i}+\tfrac 13 \Rb,\]
with $\Sb_i$ being the diagonal entries (i.e., eigenvalues) of $\Sb$.
Then the diagonal block entries $M$ and $\Omega$ take the form
\begin{eqnarray}
M=\begin{bmatrix}-\tfrac{\Rb_3}{2}&-\tfrac{\wb_5}{2}&\tfrac{\wb_4}{2}\\
\tfrac{\wb_5}{2}&-\tfrac{\Rb_4}{2}&-\tfrac{\wb_3}{2}\\-\tfrac{\wb_4}{2}&\tfrac{\wb_3}{2}&-\tfrac{\Rb_5}{2}\end{bmatrix},\quad
\Omega=\begin{bmatrix}\tfrac{\Rb}{2}&-\wb_3&-\wb_4&-\wb_5\\
\wb_3&\tfrac{\Rb}{2}-\Rb_3&0&0\\\wb_4&0&\tfrac{\Rb}{2}-\Rb_4&0\\
\wb_5&0&0&\tfrac{\Rb}{2}-\Rb_5.
\end{bmatrix}.
\end{eqnarray}
We have
\begin{equation}\label{M=0 iff Om=0}
M=0\quad\Leftrightarrow\quad\Omega=0\quad\Leftrightarrow \quad\Rb_i=\wb_i=0,
\end{equation}
which is {\em not allowed for type {\bf II}}. Also notice that $M$, being the sum of an antisymmetric and a diagonalized symmetric part, may represent any $3\times 3$ matrix and thus may take any Jordan normal form, while this is not the case for $\Omega$.

We shall start by showing that a 5D, primary alignment type {\bf
{II}} Weyl operator ${\sf C}$ can be nilpotent, in contrast with the
4D case (cf.\ section \ref{subsec:  type II 4D}).  We will then
study its classification from different perspectives:  Segre type of
$M$ or $\Om$, $\rank(M)$ and $\rank(\Om)$, and spin type.  More
precisely, we will exemplify the Segre type classifications by
treating particular cases and determining all possibilities for the
other properties.  We then give the conditions for when the ranks
$M$ and $\Om$ attain a given value, and comment on the relation with
$\rank({\sf C})$.  We also describe the kernels and images of $M$
and $\Om$, and pay special attention to the type {\bf{ D}} subcase.
Finally, in appendix \ref{subsec: type II eigenvalues} we present
the possible eigenvalue degeneracies for spin types $\{\cdot\}_0$,
$\{\cdot\}_\para$ and $\{(11)1\}_\bot$ and comment on the more
general spin types.

We shall use an extended Segre type notation to indicate the Jordan
block dimensions (elementary divisors) of $M$, $\Om$ and ${\sf C}$
(cf.\ section \ref{subsec:  Weyl operator geometry}).  If there is a
single eigenvalue zero we indicate this by writing 0 instead of 1.
For a multiple zero eigenvalue, or if we want to emphasize a
specific eigenvalue, we write this value followed by the
corresponding Jordan block dimensions between round brackets.  For a
pair of single complex eigenvalues we use $Z\Zc$ or $a\pm i b(Z\Zc)$
(and, e.g., $(Z21)(\Zc21)$ or $a+ib(Z21),a-ib(\Zc 21)$ if both
correspond to one Jordan block of dimension 2 and one of dimension
1). $W\Wc$ indicates a different pair of complex eigenvalues (for
instance, ${\sf C}[\Rb(2211),\pm 3 i\omega(Z\Zc),10]$ indicates that
${\sf C}$ has one eigenvalue $\Rb$ with Jordan block dimensions 2,
2, 1, 1, one pair of single complex eigenvalues $\pm 3 i\omega$, one
single zero and one single non-zero eigenvalue).

\subsubsection{Nilpotence of ${\sf C}$}

\label{subsubsec: type II C nilpotent}

The lower-triangular block structure of a type {\bf {II}} Weyl operator
(\ref{WeylOperator II}) is preserved by taking powers ${\cal C}^k$, which
have $((M^k)^t,\Omega^k,M^k)$ on the diagonal.  Hence, if
$M^{k_0}=\Omega^{k_0}=0$ then ${\sf C}^{k_0}$ is (at most) a type {\bf {III}}
operator, and thus ${\sf C}^{3k_0}=0$.  Therefore, {\em ${\sf C}$ is nilpotent
if and only if both $\sf M$ and $\sf\Omega$ are nilpotent}.

The characteristic polynomials of $M$ and $\Omega$ are:
\beq
&&k_M(x)=x^3-\s^1_M x^2+\s^2_M x-\s^3_M,\label{kM}\\
&&k_\Omega(x)=x^4-\s^1_\Omega x^3+\s^2_\Omega x^2-\s^3_\Omega x+\s^4_\Omega,\label{kOm}
\eeq
where
\beq
&-2\,\s^1_M=&-2\,\mbox{Tr}\,M=\s^1_\Omega=\mbox{Tr}\,\Omega=\Rb=\Rb_3+\Rb_4+\Rb_5,\label{s1M}\\
&4\s^2_M=&\s^2_\Omega=\Rb_4\Rb_5+\Rb_5\Rb_3+\Rb_3\Rb_4+\wb_3^2+\wb_4^2+\wb_5^2,\label{s2M}\\
&-8\s^3_M=&-8\det(M)=\Rb_3\Rb_4\Rb_5+\Rb_3\wb_3^2+\Rb_4\wb_4^2+\Rb_5\wb_5^2,\label{s3M}\\
&\s^3_\Omega=&-8\s^3_M+\tfrac 14 (\Rb-2\Rb_3)(\Rb-2\Rb_4)(\Rb-2\Rb_5),\label{s3Om}\\
&4\s^4_\Omega=&4\det(\Om)=(\Rb-2\Rb_4)(\Rb-2\Rb_5)\wb_3^2+(\Rb-2\Rb_5)(\Rb-2\Rb_3)\wb_4^2+(\Rb-2\Rb_3)(\Rb-2\Rb_4)\wb_5^2\nonumber\\
&&+\tfrac 14 \Rb(\Rb-2\Rb_3)(\Rb-2\Rb_4)(\Rb-2\Rb_5).\label{s4Om}
\eeq
For further purpose, notice that these coefficients are linear polynomials in
$\wb_3^2$, $\wb_4^2$ and $\wb_5^2$. The matrices $M$ and $\Omega$ are nilpotent
{\em if and only if} (abbreviated {\em iff} henceforward)
all their eigenvalues are zero, which is equivalent to the vanishing of all $\s^i_M$ and
$\s^i_\Omega$. From $\s^1_M=\s^3_\Omega+8\s^3_M=0$ it follows that $\Rb_4=0=\Rb_3+\Rb_5$
(by suitable axis permutation), and then  from $\s^2_M=\s^3_M=\s^4_\Omega=0$ we obtain
\beq
 \Rb_4=0, \qquad \Rb_3=-\Rb_5\neq 0, \qquad 2\wb_3^2=2\wb_5^2=\Rb_5^2 , \qquad \wb_4=0.
 \label{C nilpotent}
\eeq
Hence, {\em ${\sf C}$ is nilpotent iff (\ref{C nilpotent}) is
satisfied.} Notice that the spin type is then $\{110\}_{\bot0}[\Rb=0]$.  We  easily find that $M^3=0\neq M^2$ and $\Omega^3=0\neq
\Omega^2$ in this case, such that the Segre types of $M$ and $\Omega$ are
$M[0(3)]$ and $\Omega[0(31)]$.  Thus, the value of $k_0$ above equals 3, and
the operator ${\sf C}$ will be generically nilpotent of index $9$, but
lower indices $\geq 3$ can occur.  In particular, we clearly have that {\em if a
5D, alignment type D Weyl bivector operator is nilpotent, its index of
nilpotence is 3, the Segre type being ${\sf C}[(3331)]$}.  Thus, in this
case and from this viewpoint, it is undistinguishable from a generic type
{\bf {III}} Weyl operator. Let us discuss an easily
testable criterion distinguishing between these two situations.

From lemma 8 in \cite{Coleyetal04vsi} we know that any type {\bf
{D}} (and, in fact, any primary type {\bf II}) Weyl tensor (in
arbitrary dimensions) must have a non-zero polynomial invariant,
whereas all such  invariants necessarily vanish in the type {\bf
{III}} (or {\bf N}) case.~\footnote{Indeed, this is true for any
tensor: the alignment theorem and the VSI corollary in
\cite{Her11-align} state that a tensor has only vanishing polynomial
invariants if and only if it is of type {\bf III}, or simpler.} The
polynomial invariants of the Weyl tensor $C_{abcd}$ are (by
definition) traces of {\em curvature operators} built from
$C_{abcd}$~\cite{HerCol10-op}. An example of a first order operator
is the aforementioned operator ${\sf C}$ acting on bivector space.
An example of a second order operator is $C^{abcd}C_{defg}$, acting
on the space of contravariant 3-tensors. Remarkably, there is
another natural first order operator associated to the Weyl tensor,
which does not seem to have been considered in the literature before
(not even in the 4D case). This is the following operator, acting on
the space $S^2 (T_p M)$ of {\em symmetric} two-tensors~\footnote{It
can be shown that the same operator acting on bivector space is
equivalent to ${\sf C}$.}:
\begin{equation}\label{Cs}
{\sf C}^s: Z^{ab}=Z^{(ab)}\mapsto C^{a}{}_{c}{}^b{}_d Z^{cd} .
\end{equation}

Now, both ${\sf C}$ and ${\sf C}^s$ are nilpotent in the type {\bf
III} (or {\bf N}) case, and it is easy to verify, in an {\em ad hoc}
manner in the present 5D context, that in the proper type {\bf II}
case where ${\sf C}$ is nilpotent (i.e., when (\ref{C nilpotent})
holds), then ${\sf C}^s$ {\em cannot} be nilpotent at the same time.
However, let us first explicitly prove the following more general result,
slightly strengthening lemma 8 in \cite{Coleyetal04vsi} and valid in
{\em any} dimension:
\begin{prop} Suppose that a rank 4 tensor
$T_{abcd}$ has symmetries $T_{abcd}=-T_{bacd}=T_{cdab}$, is of
primary type {\bf II} or more special and has nilpotent associated
operators ${\sf T}$ and ${\sf T}^s$. Then $T_{abcd}$ is necessarily
of type {\bf III} or {\bf N}, i.e., a frame exists in which all
components of b.w.\ 0, 1 and 2 vanish.
\end{prop}
{\em Note.} For a tensor $T_{abcd}$ satisfying the mentioned
symmetries we can indeed define the operators ${\sf T}$ and ${\sf
T}^s$ on $\Lambda^2 T_pM$ and $S^2(T_pM)$ by formally replacing
$C_{abcd}$ by $T_{abcd}$ in (\ref{Cop}) and (\ref{Cs}),
respectively. A null frame (\ref{5D null frame}) of $T_pM$, where
the indices $i$ now run from 3 to the dimension $(n+2)$ of the spacetime, induce a basis (\ref{calB def}) of
$\Lambda^2 T_pM$ wrt which ${\sf T}$ takes a $3\times 3$ block
matrix representation ${\cal T}$ as in (\ref{WeylOperator}), where
we will use the same symbols as there but with a tilde decoration.
On the other hand, (\ref{5D null frame}) induces the basis
\begin{eqnarray}
\label{S2 frame -21} &&{\bf P}^{ab}={\mbold n^{a}}{\mbold n}^{b},
\qquad \quad{\bf Q}_i^{ab}=\sqrt{2}\,{\mbold n^{(a}}{\bf m}_i^{b)},\\
\label{S2 frame 0}&&{\bf O}^{ab}=\sqrt{2}\,{\mbold n^{(a}}{\mbold
\ell}^{b)}, \quad {\bf O}^{ab}_{ij}=\sqrt{2}\,{\bf m}_i^{(a}{\bf
m}_j^{b)}\;(i<j), \quad {\bf
O}^{ab}_{ii}={\bf m}_i^{a}{\bf m}_i^{b}\;(\text{no sum over $i$}),\\
\label{S2 frame 12}&& {\bf R}_i^{ab}=\sqrt{2}\,{\mbold
\ell^{(a}}{\bf m}_i^{b)}, \qquad \quad {\bf S}^{ab}={\mbold
\ell^{a}}{\mbold \ell}^{b}
\end{eqnarray}
of $S^2(T_p M)$. Notice that the boost orders of the tensors
${\bf P}$, ${\bf Q}_i$, $({\bf O},{\bf O}_{ij})$, ${\bf R}_i$ and
${\bf S}$  along $\bl$ are -2,-1,0,1 and 2, respectively;
accordingly, wrt such a frame the operator ${\sf T}^s$ takes a
$5\times 5$ block matrix representation ${\cal T}^s$.
\begin{proof}
Since $T_{abcd}$ is of primary type {\bf {II}} or more special, a
null frame exists in which all its components of strictly positive
b.w.\ (1 or 2) vanish, such that the block  matrices ${\cal T}$ and
${\cal T}^s$ are lower triangular.
Hence, the nilpotence of ${\sf T}$ and ${\sf T}^s$
is equivalent to the nilpotence of all block entries on the
diagonals of ${\cal T}$ and ${\cal T}^s$. Consider first the middle
(b.w.\ 0) block of ${\cal T}^s$, corresponding to (\ref{S2 frame
0}). It is easy to check that it is symmetric,~\footnote{The
$\sqrt{2}$ normalization factors in (\ref{S2 frame 0}) have been
introduced for this reason. Essentially this normalization implies
that the dual basis vectors of the (\ref{S2 frame 0}) frame vectors
are the metric dual ones.} and thus should vanish (since a symmetric
matrix is diagonalizable). In particular we get ${\bf
O}^{ab}T_{acbd}{\bf O}^{cd}=-T_{0101}=0$, and ${\bf
O}^{ab}T_{acbd}{\bf O}_{ij}^{cd}=\tilde M_{ij}+\tilde M_{ji}=0$ (for
all $i$ and $j$) implying that the matrix with entries $\tilde
M_{ij}:=T_{1i0j}$ is antisymmetric. Next, consider the first block
on the diagonal of ${\cal T}$ and the second block on the diagonal
of ${\cal T}^s$: these have entries $2({\bf V}_i)_{ab}{\sf T}({\bf
U}_j)^{ab}=-\tilde M_{ij}$ and $({\bf R}_i)_{ab}{\sf T}^s({\bf
Q}_j)^{ab}=T_{01ij}-\tilde M_{ij}$. Thus they are antisymmetric,
whence diagonalizable and 0. This gives $T_{01ij}=T_{1i0j}=0$.
Finally, the middle block $\tilde\Omega$ of ${\cal T}$ is now
symmetric and thus should also vanish; this gives $T_{ijkl}=0$.
Hence, all b.w.\ 0 components of $T_{abcd}$ also vanish in the
considered frame, and thus the tensor is of type {\bf III} or {\bf
N}.
\end{proof}

In particular we thus conclude that \emph{if a proper type {\bf {II}}
Weyl tensor has a nilpotent Weyl operator ${\sf C}$, then the
operator ${\sf C}^s$ has non-zero eigenvalues, and consequently
non-zero polynomial invariants (contrary to any type {\bf III}
tensor).}

\subsubsection{Case of a nilpotent $\sf M$}

\label{subsubsec: type II}

In order to illustrate the classification of a type {\bf {II}} operator ${\sf C}$
from the viewpoint of the extended Segre type of $M$, we shall determine
the possible spin types and the extended Segre types of $\Omega$, in the
case of a nilpotent $M$; i.e., when the extended Segre type of $M$ is
either $M[0(3)]$ ($M^3=0\neq M^2\leftrightarrow\rank(M)=2$) or $M[0(21)]$
($M^2=0\neq M\leftrightarrow\rank(M)=1$).

The matrix $M\neq 0$ is nilpotent iff all of the eigenvalues are zero. This happens iff
\begin{equation}\label{M nilpot general}
\s^1_M=0,\qquad\s^2_M=0,\qquad\s^3_M=0.
\end{equation}
The first equation gives $\Rb=0$,
so that $\Rb_i=\Sb_i$, and the second equation together with (\ref{s2M}) implies that  spin type $\{(000)\}_\para$ (all $\Rb_i=0$) is not allowed.
Therefore, when $\sf M$ is nilpotent at least two $\Rb_i$ must be non-zero, and at most
two of them can coincide. Hence, without loss of generality, we may assume that $\Rb_3\neq \Rb_5$.
We can then solve $\s^2_M=\s^3_M=0$ for $\wb_3^2$ and $\wb_5^2$, yielding the {\em necessary and sufficient conditions for $\sf M$ being nilpotent}:
\beq
 \Rb_4=-\Rb_3-\Rb_5, \qquad \wb_3^2=\frac{\Rb_5^3-\wb_4^2(\Rb_3+2\Rb_5)}{\Rb_5-\Rb_3} , \qquad \wb_5^2=\frac{\Rb_3^3-\wb_4^2(2\Rb_3+\Rb_5)}{\Rb_3-\Rb_5}
 \label{Mnilp} .
\eeq

To determine which spin types are allowed we use the normal
forms of table \ref{Table: Canforms bw zeta} with $(X,x)=(\Sb,\wb)$ and proceed by increasing
the number of zero $\wb$-components.
The generic spin type is $\{111/0\}_\g$, corresponding to $\wb_3\wb_4\wb_5\neq 0$.
If $\wb_4=0\neq \wb_3\wb_5$ we generically have spin type $\{111\}_\bot$; the special case of spin type $\{(11)1\}_g$ corresponds to the subcase
\begin{equation}\label{(11)1g}
\Rb_3=\Rb_4=-\tfrac 12 \Rb_5,\qquad \wb_4=0,\qquad 3\wb_3^2=24\wb_5^2=2\Rb_5^2
\end{equation}
of (\ref{Mnilp}), while $\{110\}_\bot$ leads to $\{110\}_{\bot0}$ and gives back (\ref{C nilpotent}).
If $\wb_4=\wb_5=0\neq \wb_3$, we  find from (\ref{Mnilp}):
\beq
 \Rb_3=0, \qquad \Rb_4=-\Rb_5\neq 0, \qquad \wb_4=\wb_5=0 , \qquad \wb_3^2=\Rb_5^2.
 \label{M2=0}
\eeq
It can be readily verified that this subcase corresponds {\em precisely} to the case
where the nilpotence index of $M$ is 2 (Segre type $M[0(21)]$), whereas in all other
cases this index is 3 (Segre type $M[0(3)]$). Notice that (\ref{M2=0}) is a subcase
of spin type $\{110\}_{\para0}[\Rb=0]$. Finally, $\wb=0$ leads to a diagonal, and hence
non-nilpotent, $M$.

To determine the possible Segre types of $\Omega$, we first determine the possible multiplicities and nature (real or complex) of the eigenvalues. This is most easily done by means of the discriminant sequence of (\ref{kOm})~\cite{YangXia,ColHer11}. Given (\ref{Mnilp}) the coefficients (\ref{s1M})-(\ref{s4Om}) reduce to
\begin{eqnarray}
&&\s^\Omega_1=\s^\Omega_2=0,\label{s12 Om}\\
&&\s^\Omega_3=-2\Rb_3\Rb_4\Rb_5,\label{s3Om M nilp}\\
&&\s^\Omega_4=(\Rb_3-\Rb_4)(\Rb_5-\Rb_4)\wb_4^2-(\Rb_3^2+\Rb_5^2)\Rb_4^2,\label{s4Om M nilp}
\end{eqnarray}
where $\Rb_4=-\Rb_3-\Rb_5$, and the discriminant sequence list becomes
\begin{equation}
[4, 0, -36(\s^\Omega_3)^2, D_4],\qquad D_4 = 256(\s^\Omega_4)^3-27(\s^\Omega_3)^4.\label{Dseq4}
\end{equation}
Here $D_4$ is the classical discriminant of the degree 4 polynomial $k_\Omega(x)$ (given (\ref{s12 Om})) which vanishes iff  the latter has multiple roots.
If $\s^\Omega_3=\s^\Omega_4=0$ (i.e., $0$ is a multiple root) we are in the case (\ref{C nilpotent}) of a nilpotent ${\sf C}$.
In all other cases it follows that the so called revised sign list (see \cite{YangXia,ColHer11})
must have at least one sign switch, such that $\Omega$ has at least one pair of complex eigenvalues.
If $D_4>0$ there are two (differing) complex pairs (Segre type $\Om[Z\Zc W\Wc]$), while for
$D_4\leq 0$ there are two additional real eigenvalues, which coincide iff $D_4=0$. In this
last case one can further show that this real (non-zero) eigenvalue must correspond to a
two-dimensional Jordan block (i.e., the Segre type of $\Omega$ must be $[Z\Zc 2]$); if $D_4<0$
we either have $[Z\Zc 11]\,(\s^\Omega_4\neq 0)$ or $[Z\Zc 10]\,(\s^\Omega_4= 0\neq \s^\Omega_3)$.
In the latter case we have $\rank(\Omega)=3$, while $\rank(\Omega)=2$ in the nilpotent case
(\ref{C nilpotent}); in all other cases we have $\rank(\Omega)=4$.

Regarding the compatibility of the spin types and the Segre types of
$\Omega$, we first observe from (\ref{Dseq4}) that $D_4<0$ is implied by
$\s^\Omega_4<0$.  Primary spin type $\{110\}$ is equivalent to
$\s^\Omega_3=0$ (see (\ref{s3Om M nilp})), and we may take
$\Rb_4=\Rb_3+\Rb_5=0$ by permuting the axes; moreover, if $\wb_4\neq 0$ we
then see from (\ref{s4Om M nilp}) that $\s^\Omega_4<0$.  The case
$\wb_4=0\neq\Rb_4$ immediately yields $\s^\Omega_4<0$ as well.  It
follows that $D_4\geq 0$ is only consistent with the most generic spin
type $\{111\}_\g$ and the spin type $\{110\}_{\bot0}$ ($\wb_4=0=\Rb_4$)
corresponding to nilpotent ${\sf C}$; in all other case we have $D_4<0$
and thus Segre type $\Omega[Z\Zc 11]$.

A summary of the allowed spin types and Segre types of $M$ and $\Om$, and their compatibility, in the case of a nilpotent $M$ is given in table \ref{Table: type II M nilp}.

\begin{table}[t]
\begin{tabular}{|c|c|c|c|c|}
\hline
Spin types&  Segre type of $M$ &$\rank(M)$&$\Om$ Segre type&$\rank(\Omega)$\\
\hline
\hline
$\{110\}_{\para 0}$ &  $[0(21)]$ &1& $[Z\Zc 11]$&4 \\
\hline
$\{110\}_{\bot 0}$ &  $[0(3)]$ &2& $[0(31)]$&2 \\
\hline
$\{(11)1\}_\g,\,
\{111\}_\bot,\,\{110\}_\g$  & $[0(3)]$&2 & $[Z\Zc 11]$ &4\\
\hline
$\{111\}_g$& $[0(3)]$&2 & $[Z\Zc 11],\,[Z\Zc W\Wc],\,[Z\Zc 2]$ &4\\
&&&$[Z\Zc 10]$&3\\
\hline
\end{tabular}
\caption{Type {\bf {II}} Weyl tensors: allowed spin types, and
corresponding Segre types and ranks of $M$ and $\Om$, in the case of
a {\em nilpotent matrix $M$}. Spin types $\{110\}_{\bot 0}$,
$\{(11)1\}_\g$ and $\{110\}_{\para 0}$ correspond to the special
cases (\ref{C nilpotent}), (\ref{(11)1g}) and (\ref{M2=0}) of the
characterizing equations (\ref{Mnilp}), respectively. {In all cases
$\Rb=0$.}} \label{Table: type II M nilp}
\end{table}

\subsubsection{Case of a quadruple $\sf \Omega$-eigenvalue}

\label{subsubsec: type II Omega nilpotent}

Given the analysis of the previous paragraph, we now illustrate the
classification of a type {\bf {II}} operator ${\sf C}$ from the viewpoint of the
extended Segre type of $\Omega$, by working out the case where $\Omega$
has a quadruple eigenvalue $x_0$.  This comprises the subcase $x_0=0$ of a
nilpotent $\Omega$.  Strikingly, we will see that the spin type
classification forms a pure refinement of the $\Om$-Segre type
classification (in the sense that a certain spin type is compatible with
at most one $\Omega$-Segre type) and that the extended Segre type of $M$
is almost always $M[Z\Zc 1]$.

The characteristic polynomial $k_\Om(x)$ of the matrix $\Omega$ has a quadruple root, namely 
\begin{equation}
x_0=\tfrac{1}{4}\s^\Om_1=\tfrac 14 \Rb,
\end{equation}
if and only if
\beq
\s_2^\Om=\tfrac 38 \Rb^2,\qquad \s_3^\Om=\tfrac {1}{16} \Rb^3,\qquad \s_4^\Om=\tfrac {1}{256}\Rb^4.\label{s234 spec}
\eeq
By (\ref{s1M}), (\ref{s2M}) and (\ref{s3Om}) it follows that the $\s^M_i$ can be written in terms of the $\Rb_i$ only:
\beq\label{sMs}
\s^M_1=-\tfrac 12 \Rb,\qquad \s^M_2=\tfrac{3}{32}\Rb^2,\qquad \s^M_3=\tfrac {1}{32}(F-\tfrac{1}{4}\Rb^3),
\eeq
where
\beq
F\equiv (\Rb-2\Rb_3)(\Rb-2\Rb_4)(\Rb-2\Rb_5),\qquad \Rb=\Rb_3+\Rb_4+\Rb_5.\label{Fdef}
\eeq
Therefore, the cases where $k_M(x)$ has a root 0 or $x_0$  respectively correspond to
\begin{equation}\label{F conds}
F=\tfrac 14\Rb^3,\qquad F=\tfrac 52\Rb^3,
\end{equation}
and they occur simultaneously iff ${\sf C}$ is nilpotent. Also, the discriminant sequence of a general degree 3 polynomial (\ref{kM}), with arbitrary $\s^i_M$, is
\begin{equation}
[3,F_3,D_3],\qquad F_3\equiv 2(\s^M_1)^2-6\s^M_2,\qquad 27 D_3\equiv -[27\s^M_3-\s_M^1(9\s^M_2-2(\s_M^1)^2)]^2+4F_3^3.\label{Dseq3 gen}
\end{equation}
In the considered situation of a type {\bf {II}} Weyl operator with a quadruple $\Om$-eigenvalue, (\ref{s1M})-(\ref{s2M}) and the first equation of (\ref{s234 spec}) imply $F_3=-\tfrac{1}{16}\Rb^2\leq 0$, such that $D_3\leq 0$ from (\ref{Dseq3 gen}), where $F_3<0\Rightarrow D_3<0$. It follows that {\em either ${\sf C}$ is nilpotent or $k_M(x)$ has a pair of complex roots.} The nilpotent case is characterized here by $\Rb=\Rb_3\Rb_4\Rb_5=0$. Regarding the latter case we will indicate the respective subcases in (\ref{F conds}) by $M[Z\Zc0]$ and $M[Z\Zc1_c]$, and otherwise write $M[Z\Zc1]$.

Just as in the case of a nilpotent $M$, we can show that $\wb=0$ and spin type $\{(000)\}_\para$ are not allowed. The assumptions of $\wb_3\neq 0=\wb_4=\wb_5$ {\em or} primary spin type $\{(11)1\}$ both lead to the case (up to permutation of the axes):
\beq\label{Om(211)}
\wb_4=\wb_5=0,\qquad \tfrac 12\Rb_3=\Rb_4=\Rb_5=\wb_3\qquad (x_0=\wb_3),
\eeq
which is a subcase of spin type $\{(11)1\}_{\para}[\Rb\neq 0]$. It can be verified that this is the only case where $\Om$ has Jordan blocks of dimension at most 2, and that we have extended Segre types $\Om[\wb_3(211)]$ and
$M[-\tfrac{1\pm i}{2}(Z\Zc),-\wb_3(1)]$
and thus ${\sf C}[(ZZ)(\Zc\Zc)(11)(1111)]$. In all other cases,
given (\ref{s2M}), (\ref{s3Om}) and (\ref{s4Om}),  (\ref{s234 spec}) can be solved for the $\wb_i^2$, yielding:
\begin{equation}\label{Om(4)}
\wb_3^2=\frac{\left(\Rb_3-\tfrac 14\Rb\right)^4}{(\Rb_3-\Rb_4)(\Rb_3-\Rb_5)} , \quad \wb_4^2=\frac{\left(\Rb_4-\tfrac 14\Rb\right)^4}{(\Rb_4-\Rb_5)(\Rb_4-\Rb_3)} , \quad \wb_5^2=\frac{\left(\Rb_5-\tfrac 14\Rb\right)^4}{(\Rb_5-\Rb_3)(\Rb_5-\Rb_4)}.
\end{equation}
By substituting this into $\left(\Om-\tfrac 14 \Rb\right)^3$ we find that Segre
type $\Om[(31)])$ corresponds precisely to spin type $\{111/0\}_\bot$ (i.e., the situation
where exactly one $\wb$-component vanishes); in agreement with table \ref{Table: Canforms bw zeta} this is for
\begin{equation}\label{wb4=0 cond}
\wb_4=0\neq \wb_3\wb_5\quad\Leftrightarrow\quad 4\Rb_4-\Rb=0\neq (4\Rb_3-\Rb)(4\Rb_5-\Rb).
\end{equation}
The case $\Rb=\Rb_4=0$ is equivalent to spin type $\{110\}_{\bot 0}$ and
gives the case (\ref{C nilpotent}) of nilpotent ${\sf C}$; for $\Rb\neq
0$ one necessarily has spin type $\{111\}_\bot$, (i.e., $\{110\}_{\bot
1}$ is not allowed) and extended Segre type $M[Z\Zc 1]$ (i.e.,
(\ref{wb4=0 cond}) is incompatible with (\ref{Fdef})-(\ref{F conds})).
Finally, the Segre type $\Om[4]$ allows for the spin types
$\{110\}_\g[\Rb\neq 0]$, $\{111\}_\g[\Rb=0]$ and $\{111\}_\g$, where the
second one gives a {\em nilpotent $\Om$} and only the last one is compatible with
either of the equations in (\ref{F conds}).

A summary of the allowed spin types and Segre types of $\Om$ and $M$, and their compatibility, in the case of a quadruple $\Om$-eigenvalue is given in table \ref{Table: type II Om quadr}.

\begin{table}[t]
\begin{tabular}{|c|c|c|c|c|}
\hline
Spin types&  $\Om$ Segre type  &$\rank(\Om)$&$M$ Segre type&$\rank(M)$\\
\hline
\hline
$\{(11)1\}_{\para}[\Rb\neq 0]$ &  $[(211)]$ &4& $[Z\Zc 1]$&3 \\
\hline
$\{110\}_{\bot 0}[\Rb=0]$ &  $[0(31)]$ &2& $[0(3)]$&2 \\
\hline
$\{111\}_\bot[\Rb\neq 0]$  & $[(31)]$&4 & $[Z\Zc 1]$ &3\\
\hline
$\{111\}_\g[\Rb=0]$  & $[0(4)]$&3 & $[Z\Zc 1]$ &3\\
\hline
$\{110\}_\g[\Rb\neq 0]$  & $[4]$&4 & $[Z\Zc 1]$ &3\\
\hline
$\{111\}_\g[\Rb\neq 0]$  & $[4]$&4 & $[Z\Zc 1]$, $[Z\Zc 1_c]$ &3\\
&&&$[Z\Zc 0]$&2\\
\hline
\end{tabular}
\caption{Type {\bf {II}} Weyl tensors: allowed spin types, and corresponding Segre types and ranks of $\Om$ and $M$, in the case where {\em $\Om$ has a quadruple eigenvalue}. The second row corresponds the the case of a nilpotent~${\sf C}$.}
\label{Table: type II Om quadr}
\end{table}

\subsubsection{Classification based on $\rank(M)$ and $\rank(\Om)$}\label{subsubsec: type II rank M}

The classification by the ranks of $M$ and $\Om$, and their intersection, gives a rather course subclassification of an alignment type {\bf {II}} Weyl tensor, but in combination with the determination of the (kernel and) image of ${\sf C}$ may be useful for particular purposes.

For example, the image has a particular implication for the holonomy group of the spacetime under consideration. The infinitesimal generators of the holonomy group are spanned by:
\[ R_{abcd}X^cY^d, \quad R_{abcd;e_1}X^cY^dZ^{e_1},\quad ...\]
for all vectors $X, ~Y,~Z,...\in T_pM$, through the isomorphsim $\bar\iota: \wedge^2T^*_pM\mapsto {\mf o}(1,n-1)$, where  ${\mf o}(1,n-1)$ is the Lie algebra of the Lorentz group. The map $\bar\iota$ is explicitly given by raising an index: $X_{ab}\mapsto X^a_{~b}$. Similarly, by lowering an index we have an isomorphism ${\iota}: \wedge^2T_pM\mapsto {\mf o}(1,n-1)$. In particular, this implies that the image of the Riemann bivector operator generates a vector subspace of $\mf{h}\mf{o}\mf{l}$ (the Lie algebra of the holonomy group); i.e.,
\[{\iota}(\mathrm{Im}({\sf R}))\subset \mf{h}\mf{o}\mf{l}.\]
In the case of a Ricci-flat spacetime, this implies that
${\iota}(\mathrm{Im}({\sf C}))\subset \mf{h}\mf{o}\mf{l}$, and thus we can
obtain a minimal dimension for the holonomy algebra by considering the rank of ${\sf C}$.
Indeed, if the ${\iota}(\mathrm{Im}({\sf C}))$ does not close as a Lie algebra,
we can consider the algebra it generates (which must also be in $\mf{h}\mf{o}\mf{l}$). For example, since the dimension of the Lorentz group is 10, we immediately get that the vacuum cases where $\rho({\sf C})=10$ generate the whole Lorentz group (=holonomy group). Indeed, since there are no 9 or 8-dimensional proper subgroups of the Lorentz group, we also have that  $\rho({\sf C})\geq 8$ must have the full Lorentz group as its holonomy group. Thus to conclude, the image $\Im({\sf C})$ generates a subalgebra of the infinitesimal holonomy algebra.
The allowed spin types in each case can be computed.

The generic case, of course, corresponds to having $\rank(M)=3$ and
$\rank(\Omega)=4$.  We will now work out all possible special cases where
lower rank combinations are allowed.  From (\ref{M=0 iff Om=0}) we have
$\rank(M)\neq 0 \neq\rank(\Omega)$ and cases of zero-rank matrices can
thus be ruled out from the start.  Moreover, by considering the order~2
minors of $\Om$ it is easily shown that $\rank(\Omega)=1$ is not
allowed either.  As a final preliminary remark we mention that $\rank(\Om)=2$
(all order 3 $\Om$-minors being zero) leads to the two cases (\ref{2_2b})
and (\ref{2_2a}) below, for which $\rank(M)=2$.

We now work out the classification based on $\rank(M)$, and its intersection with that of $\rank(\Om)$.

\paragraph{\underline{$\rank(M)=3$}} This is the case of a generic matrix $M$, corresponding to
\begin{equation}\label{detM}
-8\det(M)=\Rb_3\Rb_4\Rb_5+\Rb_3\wb_3^2+\Rb_4\wb_4^2+\Rb_5\wb_5^2\neq 0.
\end{equation}
If $\Omega$ is also generic then $\rank(\Omega)=4$. The case $\rank(\Omega)<4$ necessarily
gives $\rank(\Omega)=3$ by the above remark. This happens if $\det(\Om)=0$
(i.e., if (\ref{s4Om}) vanishes while (\ref{detM}) is valid), and occurs for either
\beqn
 & & w_3^2=\frac{\Rb_4+\Rb_5-\Rb_3}{4(\Rb_4-\Rb_5-\Rb_3)(\Rb_4-\Rb_5+\Rb_3)}{\cal P} , \nonumber \label{3_3a} \\
 & & {\cal P}= \Rb_3[\Rb_3^2+\Rb_3(\Rb_4+\Rb_5)-(\Rb_4-\Rb_5)^2+4(\wb_3^2+\wb_4^2)]-\Rb_5[\Rb_5^2-\Rb_4\Rb_5-\Rb_4^2+4(\wb_4^2-\wb_5^2)] \nonumber \\
 & & \qquad\qquad\qquad\qquad {}-\Rb_4[\Rb_4^2-4(\wb_4^2-\wb_5^2)] ,
\eeqn
or
\beq
 \Rb_3=0=\Rb_4-\Rb_5 .
 \label{3_3b}
\eeq
In the latter case (\ref{detM}) reduces to $\Rb_5(\wb_4^2+\wb_5^2)\neq 0$.
It is understood that other cases can be obtained from these two by simple axis-permutations (hereafter we will not mention such possibilities any further).
The case of nilpotent $\Om$, corresponding to (\ref{Om(4)}) with $\Rb=0\neq \Rb_3\Rb_4\Rb_5$,
is a subcase of (\ref{3_3a}) (cf.\ also table \ref{Table: type II Om quadr}).

\paragraph{\underline{$\rank(M)=1$}} This is the case where all order 2 minors of $M$ vanish, which happens iff
\beq
 \Rb_3=0, \qquad \wb_4=\wb_5=0, \qquad \wb_3^2=-\Rb_4\Rb_5,\quad (\Rb_4,\Rb_5)\neq (0,0).
 \label{rankM=1}
\eeq
We have $\s^M_3=\s^M_2=0$, and generically the zero-$M$ eigenvalue is double
($\s^M_1\sim\Rb\neq 0$).
It becomes triple (i.e., $\sf M$ is nilpotent of index 2) for $\Rb=\Rb_4+\Rb_5=0$ (cf.~(\ref{M2=0})).
Notice that $\Rb_4\neq\Rb_5$, and since $16\det(\Omega)=-(\Rb_4-\Rb_5)^4$ by (\ref{s4Om}) we have
$\rank(\Omega)=4$.

\paragraph{\underline{$\rank(M)=2$}} By setting $\s^M_3=\det(M)=0$ to zero we find all cases
with $\rank(M)<3$, where we just have to exclude (\ref{rankM=1}) and its permutations
to get $\rank(M)=2$ (this is implicitly understood in the following). Generically,
$M$ has a single zero eigenvalue. This comprises the case where all $R_i$'s coincide;
by (\ref{detM}) it then follows that all of the $R_i$'s are zero, which is the case (\ref{2_2b}) below.
Otherwise, it is  always possible to solve $\s^M_3=0$ for one $\wb_i^2$,
and we may assume $\Rb_3\neq \Rb_5$. The zero eigenvalue becomes at least double
iff $\s^M_3=\s^M_2=0$, which is the case iff
\beq\label{wb35}
 \wb_3^2=\frac{-\Rb_5^2(\Rb_3+\Rb_4)+\wb_4^2(\Rb_4-\Rb_5)}{\Rb_5-\Rb_3} ,
 \qquad \wb_5^2=\frac{-\Rb_3^2(\Rb_4+\Rb_5)+\wb_4^2(\Rb_4-\Rb_3)}{\Rb_3-\Rb_5}
\eeq
(note that different subcases have already been discussed in the tables).
There is a triple zero eigenvalue iff, in addition, $\s^M_1=0$, which is the case (\ref{Mnilp}) of a nilpotent $\sf M$ of index 3; the last three rows of table \ref{Table: type II M nilp} produce examples fitting in the subsequent discussion of $\rank(\Om)$.

Generically, we have $\rank(\Omega)=4$. For $\rank(\Omega)<4$ we have two possibilities.
If (\ref{3_3b}) holds then $\det(\Om)=0$ automatically,
and $\det(M)=0\Leftrightarrow \Rb_5(\wb_4^2+\wb_5^2)=0$, which leads to
(\ref{2_2b}) and a subcase of (\ref{2_2a}). If (\ref{3_3b}) or its permutations do not hold,
we necessarily have (e.g., $(\Rb_4-\Rb_5)(\Rb_3-\Rb_4-\Rb_5)\neq 0$) and the conditions $\det(M)=0=\det(\Omega)$ are solved simultaneously by
\beqn
 & & \wb_4^2=\frac{(\Rb_3-\Rb_4+\Rb_5)^2}{\Rb_4-\Rb_5}\left[\frac{\Rb_5-\Rb_3}{(\Rb_3-\Rb_4-\Rb_5)^2}\wb_3^2+\frac{1}{4}\Rb_5\right] , \nonumber \label{2_3} \\
 & & \wb_5^2=\frac{(\Rb_3+\Rb_4-\Rb_5)^2}{\Rb_4-\Rb_5}\left[\frac{\Rb_3-\Rb_4}{(\Rb_3-\Rb_4-\Rb_5)^2}\wb_3^2-\frac{1}{4}\Rb_4\right] .
\eeqn
This generically corresponds to $\rank(\Omega)=3$.

The case $\rank(\Omega)=2$ arises for two different choices of the parameters.
\begin{enumerate}

\item The first possibility is
\beq
  \Rb_i=0 , \quad i=3,4,5 \label{2_2b} .
\eeq
This is precisely spin type $\{(000)\}_\para[\Rb=0]$. In this case the extended Segre types are $M[\pm i\omega(Z\Zc),0]$, $\Om[\pm i\omega(Z\Zc),0(11)]$ and thus ${\sf C}[i\omega(Z111),-i\omega(\Zc111),0(1111)]$.

\item The second possibility is
\beq
  \Rb_5=\Rb_4-\Rb_3 , \qquad \wb_4=0, \qquad \wb_5^2=\frac{\Rb_3(\Rb_3\Rb_4-\Rb_4^2-\wb_3^2)}{\Rb_4-\Rb_3} , \label{2_2a}
\eeq
which can be understood as a subcase of~(\ref{2_3}). The further subcase
hereof
\beq
 \Rb_4=0, \qquad \wb_3^2=\tfrac{1}{2}\Rb_3^2
,
 \label{IInilpotent}
\eeq
is precisely the case  (\ref{C nilpotent}) of a nilpotent ${\sf C}$ (cf.\ also table \ref{Table: type II M nilp}).
\end{enumerate}

A summary for the possible relative ranks of $M$ and $\Omega$ is given in table \ref{Table: type II M/Omega}.

\begin{table}[t]
\begin{tabular}{|l||c|c|c|c|}
\hline
  & $\rank(\Omega)=4$ & $\rank(\Omega)=3$ & $\rank(\Omega)=2$ & $\rank(\Omega)=1$ \\
\hline
\hline
 $\rank(M)=3$ $\Leftrightarrow$ (\ref{detM}) & generic case  & (\ref{3_3a}) or (\ref{3_3b}) & - & - \\
\hline
 $\rank(M)=2$ & (\ref{w3_M2}) & (\ref{2_3}) & (\ref{2_2b}) or (\ref{2_2a}) & - \\
\hline
 $\rank(M)=1$  & (\ref{rankM=1}) & - & - & - \\
\hline
\end{tabular}
\caption{Weyl operator geometry for type {\bf {II}} spacetimes: possible relative ranks of $M$ and $\Omega$ and equation numbers of the corresponding conditions (up to  permutations of the axes).}
\label{Table: type II M/Omega}
\end{table}

\subsubsection{\Ker({\sf M}) and $\Ker({\sf\Omega})$}\label{subsubsec: type II kernels M/Omega}

From (\ref{WeylOperator II}) it is easy to see that $\Ker({\sf M})\preceq\Ker({\sf C})$.
Of course, in the generic case ${\sf M}$ and ${\sf\Omega}$ have full rank (i.e., $\rank({\sf M})=3$ and $\rank({\sf\Omega})=4$), so that $\Ker({\sf M})=\{0\}$ and $\Ker({\sf \Omega})=\{0\}$. The spin type is generically $\{111\}_g[\Rb\neq0]$, but many special subcases are possible.
In particular, we have the type $\{(000)\}_\para[\Rb\neq0]$ if $\Rb_3=\Rb_4=\Rb_5$,
 or $\{(000)\}_0[\Rb\neq0]$ if, additionally, $\wb=0$. For these two simple types the
 conditions $\rank({\sf M})=3$ and $\rank({\sf\Omega})=4$ are, in fact, necessary.

The more special cases of table~\ref{Table: type II M/Omega} are now discussed.

\paragraph{\underline{$\rank(M)=3$, $\rank({\Omega})=3$}}

This case is defined by either (\ref{3_3a}) or (\ref{3_3b}). Clearly $\Ker({\sf M})=\{0\}$ here.

\begin{enumerate}

\item When (\ref{3_3a}) is satisfied we find
that
\beqn
 \Ker({\sf\Omega})=\langle\W^+\rangle , \quad \W^+ = & & [(\Rb_4-\Rb_5)^2-\Rb_3^2]\W+2(\Rb_4-\Rb_5+\Rb_3)\wb_4\W_{[53]} \nonumber \\
 & & {}+2(\Rb_5+\Rb_3-\Rb_4)\wb_5\W_{[34]}+\sqrt{\frac{{\cal P}[\Rb_3^2-(\Rb_4-\Rb_5)^2]}{\Rb_3-\Rb_4-\Rb_5}}\W_{[45]} .
 \label{33}
\eeqn
The spin type is generically $\{111\}_g[\Rb\neq0]$, but many special subcases are possible.
\item If, instead, (\ref{3_3b}) holds we get
\beq
 \Ker({\sf\Omega})=\langle\W^*\rangle , \qquad \W^* = -\wb_5\W_{[53]}+\wb_4\W_{[34]} .
\eeq
Here the spin type is $\{(11)1\}_g[\Rb\neq0]$; it specializes to $\{(11)1\}_\p[\Rb\neq0]$ if $\wb_3=0$.
\end{enumerate}

\paragraph{\underline{$\rank(M)=2$, $\rank(\Omega)=4$}} It is easy to see that if
$\Rb_3=\Rb_4=\Rb_5=0$ then $\rank(\Omega)=2$; therefore, here we can assume that
the $\Rb_i$ are not all identically zero. Let us take, for definiteness, $\Rb_3\neq0$. Then, from~(\ref{detM}) the condition $\det(M)=0$ gives
\beq
 \wb_3^2=-\frac{\Rb_3\Rb_4\Rb_5+\Rb_4\wb_4^2+\Rb_5\wb_5^2}{\Rb_3} .
 \label{w3_M2}
\eeq

It is now straightforward to see that
\beq
 \Ker({\sf M})=\langle\V^+\rangle , \qquad \V^+=\frac{\Rb_4\wb_4^2+\Rb_5\wb_5^2}{\Rb_3}\V_3-(\Rb_5\wb_5+\wb_3\wb_4)\V_4+(\Rb_4\wb_4-\wb_3\wb_5)\V_5 ,
 \label{24}
\eeq
where we substitute $\wb_3$ from~(\ref{w3_M2}). Again,  the spin type is generically $\{111\}_g[\Rb\neq0]$, but many special subcases are possible.

\paragraph{\underline{$\rank(M)=2$, $\rank(\Omega)=3$}}

Using~(\ref{2_3}) we can easily compute the  generators of the one-dimensional
spaces $\Ker({\sf M})$ and $\Ker({\sf \Omega})$; the expressions are rather
long and non-illuminating, and are therefore omitted.

The spin type is in general $\{111\}_g[\Rb\neq0]$. This specializes
to $\{(11)1\}_g[\Rb\neq0]$ (but in a non-canonical frame) if $\Rb_4=\Rb_3$, or
to $\{111\}_g[\Rb=0]$ if $\Rb_5=-\Rb_3-\Rb_4$. When $\wb_3=0$ we get the type $\{111\}_\p[\Rb\neq0]$, which becomes $\{111\}_\p[\Rb=0]$ if, additionally, $\Rb_5=-\Rb_3-\Rb_4$. Further, the type is $\{110\}_g[\Rb\neq0]$ if $\Rb_5=2\Rb_3-\Rb_4$; this becomes $\{110\}_{\p0}[\Rb\neq0]$ with the further condition $\wb_3=0$, or $\{110\}_{\p1}[\Rb\neq0]$ for $\wb_4=0$.

\paragraph{\underline{$\rank(M)=2$, $\rank(\Omega)=2$}}

There are two different possibilities corresponding to this situation.

\begin{enumerate}

\item  When~(\ref{2_2b}) holds we get
\beqn
 & & \Ker({\sf M})=\langle\V^*\rangle , \qquad \V^*=\wb_3\V_3+\wb_4\V_4+\wb_5\V_5 , \label{22b} \\
 & & \Ker({\sf \Omega})=\langle\W^{*1},\W^{*2}\rangle , \qquad
                \W^{*1}=-\wb_5\W_{[45]}+\wb_3\W_{[34]} , \quad \W^{*2}=-\wb_4\W_{[45]}+\wb_3\W_{[53]} . \nonumber
\eeqn

This corresponds univocally to spin type $\{(000)\}_\para[\Rb=0]$.

\item  Assuming, instead, that~(\ref{2_2a}) is satisfied we obtain
\beqn
 & & \Ker({\sf M})=\langle\V^\times\rangle , \qquad \V^\times=\wb_5(\Rb_3-\Rb_4)\V_3-\Rb_3(\Rb_3-\Rb_4)\V_4+\wb_3\Rb_3\V_5 ,  \label{22a} \\
 & &  \Ker({\sf \Omega})=\langle\W_{[53]},\W^\times\rangle , \qquad
                \W^\times=\wb_5(\Rb_3-\Rb_4)\W+\wb_3\wb_5\W_{[45]}+(\Rb_3\Rb_4-\Rb_4^2-\wb_3^2)\W_{[34]} , \nonumber
\eeqn
where it is understood that $\wb_5$ is as in~(\ref{2_2a}).
We also recall the special subcase~(\ref{IInilpotent}) where both $\sf M$ and $\sf\Omega$
are nilpotent.

The spin type is in general $\{111\}_\p[\Rb\neq0]$. It becomes
$\{(11)1\}_\para[\Rb\neq0]$ when $\Rb_3=0$ (and $\{(11)1\}_0[\Rb\neq0]$ if,
additionally, $\wb_3=0$); this is the situation, for instance, for the Kerr black string (where $\wb_3=0$ in the equatorial plane). In addition, we  may  have the type $\{110\}_{\p0}[\Rb=0]$ for $\Rb_4=0$ (and $\{110\}_{0}[\Rb=0]$ if also $\wb_3=0$), which includes the case when $\sf C$ is nilpotent; or $\{111\}_{\para}[\Rb\neq0]$ if, instead, $\wb_3=0$.

\end{enumerate}

\paragraph{\underline{$\rank(M)=1$}} The conditions for this to occur were already given
in~(\ref{rankM=1}) (and correspond to a specialization of ~(\ref{w3_M2}) above). Now we have (without loss of generality we can assume $\Rb_4\neq0$)
\beq
 \Ker({\sf M})=\langle\V_3,\V^0\rangle , \qquad \V^0=-\frac{\wb_3}{\Rb_4}\V_4+\V_5 ,\label{14}\\
 \Im({\sf M})=\langle-\Rb_4\V_4+\wb_3\V_5\rangle ,
\eeq
while $\Ker({\sf \Omega})=\{0\}$. If $\Rb\neq 0$ then ${\sf M}$ also admits a non-zero eigenvalue $-\Rb/2=(\wb_3^2-\Rb_4^2)/(2\Rb_4)$, with eigendirection $\Im({\sf M})$.
If $\Rb=0$ then ${\sf M}$ is nilpotent and $\Im({\sf M})$ coincides with the subspace $\langle\V^0\rangle$ of $\Ker({\sf M})$.

Here the spin type is in general $\{111\}_\para[\Rb\neq0]$.  It
specializes to $\{(11)1\}_0[\Rb\neq0]$ for $\Rb_5=0$, and to
$\{110\}_{\para0}[\Rb=0]$ for $\Rb_5=-\Rb_4$ (in which case $\sf M$ is
nilpotent).  The Segre types of $M$, $\Om$ and ${\sf C}$ are easy to
determine (some cases are presented below).

\subsubsection{$\rank({\sf C})$ in the various subcases}\label{subsubsec: type II rank C}

We have
\beq
 6\le 2\rank(M)+\rank(\Omega)\le\rank({\sf C})\le\mbox{min}\{\rank(M)+7,\rank(\Omega)+6\}\le 10 .
\eeq
All inequalities are a priori clear from (\ref{WeylOperator II}), except
for $6\le 2\rank(M)+\rank(\Omega)$ which follows from the detailed discussion above.
Considering the various previous subcases we thus obtain the possibilities summarized
in Table~\ref{Table:type_II_rank_C}. In general, the precise value of $\rank({\sf C})$  also
depends on the negative b.w. components.


\begin{table}[t]
\begin{tabular}{|l||c|c|c|}
\hline
  & $\rank(\Omega)=4$ & $\rank(\Omega)=3$ & $\rank(\Omega)=2$ \\
\hline
\hline
 $\rank(M)=3$  & $\rank({\sf C})=10$  & $\rank({\sf C})=9$ & - \\
\hline
 $\rank(M)=2$ & $8\le\rank({\sf C})\le 9$ & $7\le\rank({\sf C})\le 9$ & $6\le\rank({\sf C})\le 8$ \\
\hline
 $\rank(M)=1$  & $6\le\rank({\sf C})\le 8$ & - & - \\
\hline
\end{tabular}
\caption{Possible values of $\rank({\sf C})$ for all permitted combinations of values of $\rank(M)$ and $\rank(\Omega)$ in type {\bf {II}} spacetimes.}
\label{Table:type_II_rank_C}
\end{table}

\subsection{Type {\bf {D}} spacetimes}\label{subsec: type D}

Type {\bf {II}} spacetimes specialize to type {\bf {D}} when $\Cc_K{}^t=\Hc=\Cc_{-K}=0$ in~(\ref{WeylOperator II}).
In this case, ${\cal U}$, ${\cal W}$ and ${\cal V}$ are all invariant subspaces
(recall also~(\ref{p_U})--(\ref{p_V})) and the equality $\rank({\sf C})=2\rank(M)+\rank(\Omega)$
holds.  Therefore, in each case $\rank({\sf C})$ sticks to the lower value given
in Table~\ref{Table:type_II_rank_C} and we can explicitly present the Kernel
and Image of ${\sf C}$, as follows:


\paragraph{\underline{$\rank(M)=3$, $\rank(\Omega)=4$}} Here $\rank({\sf C})=10$, so that
\beq
 \Ker({\sf C})=\{0\}, \qquad \Im({\sf C})={\cal U}\oplus{\cal W}\oplus{\cal V} .
\eeq
Both the Schwarzschild-Tangherlini and Myers-Perry solutions belong to this class.

\paragraph{\underline{$\rank(M)=3$, $\rank(\Omega)=3$}} Here $\rank({\sf C})=9$, and we have
\beq
 \Ker({\sf C})=\langle\W^+\rangle , \qquad \Im({\sf C})={\cal U}\oplus\Im({\sf\Omega}|_{{\cal W}\setminus\langle\W^+\rangle})\oplus{\cal V} .
\eeq
with $\W^+$ defined as in~(\ref{33}).

\paragraph{\underline{$\rank(M)=2$, $\rank(\Omega)=4$}} Here $\rank({\sf C})=8$ and
\beq
 \Ker({\sf C})=\langle\V^+\rangle\oplus\langle\U^+\rangle , \qquad \Im({\sf C})=\Im({\sf M}|_{{\cal V}\setminus\langle\V^+\rangle})\oplus{\cal W}
 \oplus\Im({\sf M}^t|_{{\cal U}\setminus\langle\U^+\rangle}) .
\eeq
with $\V^+$ as in~(\ref{24}) and $\U^+$ defined analogously (i.e., $\Ker({\sf M}^t)=\langle\U^+\rangle$).

\paragraph{\underline{$\rank(M)=2$, $\rank(\Omega)=3$}} Now $\rank({\sf C})=7$ and
\beq
 & & \Ker({\sf C})=\Ker({\sf M})\oplus\Ker({\sf M}^t)\oplus\Ker({\sf\Omega}) , \nonumber \\
 & & \Im({\sf C})=\Im({\sf M}|_{{\cal V}\setminus\Ker({\sf M})})\oplus\Im({\sf\Omega}|_{{\cal W}\setminus\Ker({\sf\Omega})})
 \oplus\Im({\sf M}^t|_{{\cal U}\setminus\Ker({\sf M}^t)}) ,
\eeq
 with long expressions for $\Ker({\sf M})$, $\Ker({\sf M}^t)$ and $\Ker({\sf\Omega})$, that we omit.

\paragraph{\underline{$\rank(M)=2$, $\rank(\Omega)=2$}} Now $\rank({\sf C})=6$ and there are two possibilities, so that either
\beq
 & & \Ker({\sf C})=\langle\V^*\rangle\oplus\langle\U^*\rangle\oplus\langle\W^{*1},\W^{*2}\rangle , \nonumber \\
 & &  \Im({\sf C})=\Im({\sf M}|_{{\cal V}\setminus\langle\V^*\rangle})\oplus\Im({\sf\Omega}|_{{\cal W}\setminus\langle\W^{*1},\W^{*2}\rangle})\oplus\Im({\sf M}^t|_{{\cal U}\setminus\langle\U^*\rangle}) ,
\eeq
with the definitions of~(\ref{22b}) (and $\Ker({\sf M}^t)=\langle\U^*\rangle$), or
\beq
 & & \Ker({\sf C})=\langle\V^\times\rangle\oplus\langle\U^\times\rangle\oplus\langle\W_{[53]},\W^\times\rangle , \nonumber \\
 & &  \Im({\sf C})=\Im({\sf M}|_{{\cal V}\setminus\langle\V^\times\rangle})\oplus\Im({\sf\Omega}|_{{\cal W}\setminus\langle\W_{[53]},\W^\times\rangle})
 \oplus\Im({\sf M}^t|_{{\cal U}\setminus\langle\U^\times\rangle}) ,
\eeq
with the definitions of~(\ref{22a}) (and $\Ker({\sf M}^t)=\langle\U^\times\rangle$). For instance, the Kerr black string belongs to the latter class (see the text after (\ref{22a})).

\paragraph{\underline{$\rank(M)=1$, $\rank(\Omega)=4$}} Also in this case $\rank({\sf C})=6$, but now
\beq
 \Ker({\sf C})=\langle\V_5,\V^0\rangle\oplus\langle\U_5,\U^0\rangle , \qquad \Im({\sf C})=\Im({\sf M}|_{{\cal V}\setminus\langle\V_5,\V^0\rangle})\oplus{\cal W}\oplus\Im({\sf M}^t|_{{\cal U}\setminus\langle\U_5,\U^0\rangle}) ,
\eeq with $\V^0$ as in~(\ref{14}) and $\U^0$ defined analogously
(i.e., $\Ker({\sf M}^t)=\langle\U_5,\U^0\rangle$). Cf.~(\ref{14})
for the explicit form of $\Im({\sf M})$.

\subsection{Comparison with 4D}\label{subsec: type II 4D}

For $n=2$ we have $\Sb=0$ in addition to (\ref{Cijkl=0}). Thus the only possible spin types for a 4D
primary alignment type {\bf {II}} (i.e., Petrov type {\bf {II}} or {\bf {D}}) Weyl bivector operator
are $\{(00)\}_\para[R\neq 0]$,  $\{(00)\}_\para[R=0]$ and  $\{(00)\}_0[R\neq 0]$.
The first one is the generic case. In the type {\bf {D}} subcase, in  the second and third cases the Weyl tensor is dubbed purely magnetic and purely electric, respectively (see also section \ref{sec:PEPM}).~\footnote{Observe that $\Rb$ and $A_{34}$ correspond, in the Newman-Penrose notation, to the real and imaginary part of $\Psi_2$, respectively.}

The $2\times 2$ matrices $M$ and $\Omega$ take the form~\cite{ColHer09}:
\begin{equation}
M=\begin{bmatrix}-\tfrac 14 \Rb&-\tfrac 12 A_{34}\\\tfrac 12 A_{34}&-\tfrac 14 \Rb\end{bmatrix},\qquad \Om=\begin{bmatrix}\tfrac 12 \Rb&- A_{34}\\ A_{34}&\tfrac 12 \Rb\end{bmatrix},
\end{equation}
with respective eigenvalues $-\tfrac 14 (\Rb\pm 2i A_{34})$ and $\tfrac 12 (\Rb\pm 2i A_{34})$ . Thus the eigenvalue degeneracies are 
\beq
\mbox{Spin type}\,\{(00)\}_\para:&&M[Z\Zc],\quad\Om[W\Wc],\quad {\sf C}[(ZZ)(\Zc\Zc)(W\Wc)];\\
\mbox{Spin type}\,\{(00)\}_0:&&M[(11)],\quad\Om[(11)],\quad {\sf C}[(1111)(11)].
\eeq
For Petrov type {\bf {D}} this also gives the extended Segre types for ${\sf C}$; for Petrov type {\bf {II}} the extended Segre types are ${\sf C}[Z[2]\Zc[2],W\Wc]$ and ${\sf C}[(22)(11)]$, respectively. Notice that {\em neither $M$ nor $\Om$ can be nilpotent}, basically due to the fact that they are the sum of an antisymmetric matrix and a multiple of the identity (instead of a general symmetric matrix).


\subsection{Discussion}\label{subsec: type II discussion}

We have studied the (refined) algebraic classification of the
5D primary alignment type
{\bf {II}} Weyl operator ${\sf C}$
from the perspectives of  Segre type (of $M$ or $\Om$, and also of $\rank(M)$ and
$\rank(\Om)$) and spin type.  In particular, we have
described the Segre type classifications by treating
particular cases, and then determined
other properties such as the values of the
ranks of $M$ and $\Om$,  the
relation with $\rank({\sf C})$, and the kernels
and images of $M$ and $\Om$, within each particular case.

The nilpotence of ${\sf C}$ is of particular interest.
We found that the type {\bf {II}} Weyl
operator ${\sf C}$ is nilpotent if and only
if both $\sf M$ and $\sf\Omega$ are nilpotent (and
the Segre types of $M$ and $\Omega$ are $M[0(3)]$ and
$\Omega[0(31)]$, respectively),
and the operator ${\sf C}$ is generically nilpotent of
index $9$ (but lower indices can occur).
In  order to illustrate the classification of a type {\bf {II}}
operator ${\sf C}$ we determined the possible spin types and
the extended Segre types of $\Omega$ in the case of a
nilpotent $M$;
a summary of the allowed spin types and Segre types of $M$
and $\Om$, and their compatibility,  was given in table \ref{Table:  type II M nilp}.

The classification of a type {\bf {II}} operator ${\sf C}$
from the viewpoint of the extended Segre type of $\Omega$ was also illustrated
in the case where $\Omega$ has a quadruple
eigenvalue.  Remarkably, we found that the spin
type classification forms a pure refinement of the
$\Om$-Segre type classification
and that the extended Segre type of $M$ is almost always
$M[Z\Zc 1]$.
A summary of the allowed spin types and Segre types of $\Om$
and $M$, and their compatibility, in the case of a quadruple
$\Om$-eigenvalue was given in table \ref{Table:  type II Om
quadr}.

We then focussed attention on the special
type {\bf {D}} subcase of
type {\bf {II}} spacetimes.  In
this case,
the Kernel and Image of ${\sf C}$ were explicitly presented,
and the possible values of $\rank({\sf C})$ for all
permitted combinations of values of $\rank(M)$ and
$\rank(\Omega)$ in type {\bf {II}} spacetimes were given (see table~\ref{Table:type_II_rank_C}).

In appendix \ref{subsec: type II eigenvalues} we discuss the
classification of a type {\bf {II}} Weyl operator based on its spin
type. For spin types $\{\cdot\}_0$, $\{\cdot\}_\para$ and
$\{(11)1\}_\bot$ we present the  degeneracies in the eigenvalue
spectra of ${\sf M}$, ${\sf \Omega}$ and ${\sf C}$ and briefly
comment on the more general spin types. As was the case for type
{\bf {III}}, a certain spin type gives rise to several possible
eigenvalue degeneracies, opposed to the situation in 4D.

Finally, let us point out that several other properties of type {\bf
{II}}/{\bf D} Einstein spacetimes in higher (in particular, five)
dimensions have been studied in \cite{PraPraOrt07,Durkee09}.


\section{Types {\bf {I}} and {\bf {G}}}

\label{section: type IG}

In general dimensions, a type {\bf {I}} Weyl tensor is characterized by having
a single WAND $[\bl]$ and no multiple WANDs. With respect to any null frame $(\bl,\bn,\mm_i)$
all components of b.w. greater than ~$\zeta=1$ vanish, whereas those of b.w.~1 are not all zero:
\begin{equation}\label{type I constituents}
\Hh=0,\quad (\nh,\vh)\neq (0,0).
\end{equation}
If this is the unique single WAND the type is {\bf {I}}$_0$,
otherwise the type is {\bf {I}}$_{i}$. If there are no WANDs at all,
the Weyl type is {\bf {G}}.

In 4D type {\bf {G}} does not occur (i.e., WANDs always exist), and if the type is {\bf {I}} then there
are exactly four WANDs (such that the type is automatically {\bf {I}}$_i$; this is the so called algebraically general case
in 4D). In higher than four dimensions, however, type {\bf {G}}
is generic~\cite{Milsonetal05}
and type {\bf {I}} is algebraically special with respect to type {\bf {G}}.
However, in many
applications the presence of a {\em multiple} WAND is important, and thus the distinction
between types {\bf {I}}/{\bf {G}} and types {\bf {II}}/{\bf {III}}/{\bf {N}} also appears to be significant.

We further note that type {\bf {I}}(A) \cite{ColHer09}
(i.e., type {\bf {I}} (a) of \cite{Coleyetal04}) corresponds to the collection of the
spin types I$\{\cdot\}_0$ (i.e., $\vh=0$), and  type {\bf {I}}(B) to I$\{(000)\}_\para$
($\nh=0$).  See, e.g., \cite{PraPra05,GodRea09} for examples of type {\bf {I}}/{\bf {G}} (vacuum) spacetimes in five-dimensions.

\subsection{The ``electric'' and ``magnetic'' parts of the Weyl operator}\label{sec:EMparts}
For the type {\bf {I}}/{\bf {G}} case there is another  split which may be useful \cite{HOW}.
The split can be done for any of the types, but it is probably most useful for the
type {\bf {I}}/{\bf {G}} cases (and also possibly for type {\bf {D}}). The split utilises the existence of a \emph{Cartan involution} of the general linear group.

Consider the full Lorentz group $G=O(1,4)$. Let $K\cong O(4)$ be a maximal compact
subgroup of $O(1,4)$. Then there exists a unique Cartan involution $\theta$ of $G$ with
the following properties \cite{RS90}: (i) $\theta$ is invariant under the adjoint
action of $K$: $Ad_{K}(\theta)=\theta$. (ii) $O(1,4)$ is $\theta$-stable. (iii)
$\theta$ is the following automorphism of the Lie algebra $\mf{g}\mf{l}(n,\mb{R})$:
$X\mapsto -X^*$, where $^*$ denotes the adjoint (which is equal to the transpose
here since the coefficients
are real).

If $\theta_1$ and
$\theta_2$ are two such Cartan involutions of $G$, associated with maximal compact subgroups $K_1$ and $K_2$, then there exists a
$g\in G$ such that $\theta_1=\mathrm{Int}(g)\theta_2\mathrm{Int}(g^{-1})$,
where $\mathrm{Int}(g)$ is the inner automorphism by $g$.
By a slight abuse of notation, we will denote any representation of
$\theta$ simply by $\theta$.

First, let us  consider the case when $\theta:T_pM\mapsto T_pM$.
The above-mentioned requirements enable us to choose a unit time-like vector
${\bf u}$ that is $K$-invariant, and let us then choose the $\theta$ which has ${\bf u}$
 as an eigenvector. Therefore, in the orthonormal basis,
$\{{\bf u},{\bf m}_{i=2,...,5}\}$, we have the matrix representation:
\[ \theta=(\theta^a_{~b})=\diag (-1,1,1,1,1).\]
We note that $\theta^2={\sf 1}$. This consequently picks out a
special time-like direction and any other $\theta_2$ is related to a
(different) time-like vector ${\bf u}_2$.

Through the tensor structure of bivector space, we can let $\theta$
act on bivector space; explicitly, $\theta: F^{ab}\mapsto
\theta^a_{~c}\theta^b_{~d}F^{cd}$. By choosing the $\theta$ adapted
to the time-like vector ${\bf u}=({\mbold\ell}+{\mbold
n})/\sqrt{2}$, we get the matrix representation of $\theta$ acting
on bivector space in a $(3+4+3)$-block form relative to the basis
(\ref{calB bivector basis})-(\ref{calB def}):
\begin{eqnarray}
\theta=\begin{bmatrix}
0 & 0 & {\sf 1} \\
0 & \eta & 0 \\
{ {\sf 1} } & 0 & 0
\end{bmatrix}, \quad \eta=\diag (-1,1,1,1).
\end{eqnarray}
This will then act on the Weyl operator ${\sf C}$ through conjugation $\theta{\sf C}\theta$.

Since $\theta^2=1$, the eigenvalues of $\theta$ are $\pm 1$. We can
thus project  the operator ${\sf C}$ along  the eigenspaces of
$\theta$: \beq {\sf C}={\sf C}_++{\sf C}_-, \quad \text{where}\quad
{\sf C}_+=\frac 12({\sf C}+\theta{\sf C}\theta), \quad {\sf
C}_-=\frac 12({\sf C}-\theta{\sf C}\theta). \eeq Using the Weyl
operator (\ref{WeylOperator}) we can then compute the components of
the matrix representation of ${\sf C}_{\pm}$ relative to (\ref{calB
bivector basis})-(\ref{calB def}): \beq \mathcal{C}_+=
\begin{bmatrix}
\frac 12(M+M^t) & \frac 12(\Kh+\Kc) & \frac 12(\Lh+\Lc) & \frac 12(\Hh+\Hc) \\
\frac 12(\Kh^t+\Kc^t) &-\Phi& 0 &-\frac 12(\Kh^t+\Kc^t)\\
\frac 12(\Lh^t+\Lc^t) & 0 & \Hb & \frac 12(\Lh^t+\Lc^t) \\
\frac 12(\Hh+\Hc) & -\frac 12(\Kh+\Kc) & \frac 12(\Lh+\Lc) & \frac 12(M+M^t)
\end{bmatrix}\label{C+}
\\
\mathcal{C}_-=
\begin{bmatrix}
\frac 12(M-M^t) & \frac 12(\Kh-\Kc) & \frac 12(\Lh-\Lc) & \frac 12(\Hh-\Hc) \\
-\frac 12(\Kh^t-\Kc^t) &0 & -A^t &-\frac 12(\Kh^t-\Kc^t)\\
-\frac 12(\Lh^t-\Lc^t) & A & 0 & \frac 12(\Lh^t-\Lc^t) \\
-\frac 12(\Hh-\Hc) & \frac 12(\Kh-\Kc) & -\frac 12(\Lh-\Lc) & \frac 12(M^t-M)
\end{bmatrix}.\label{C-}
\eeq Therefore, we can see that the components  ${\sf C}_{\pm}$ are
the symmetric and anti-symmetric parts of the Weyl operator with
respect to the Euclidean metric on bivector space. In 4D these
components are referred to as the electric and magnetic parts of the
Weyl tensor. In \cite{HOW} these were defined as the
higher-dimensional electric and magnetic part of the Weyl tensors;
thus, henceforth we will refer to the component  ${\sf C}_+$ as the
\emph{electric part} of the Weyl operator (tensor), while  ${\sf
C}_-$ will be referred to as the \emph{magnetic part}. Note that, as
in 4D, these parts depend on the choice of a time-like vector ${\bf
u}$ (and their representation with respect to a different time-like
vector ${\bf u}_2$ will change accordingly).

For type {\bf {I}}/{\bf {G}} Weyl tensors it is cumbersome to say anything general
about their eigenvalue structure; however, for purely electric or purely magnetic Weyl operators, we have the following \cite{HOW}:
\begin{thm} A purely electric (PE, ${\sf
C}_-=0$) Weyl operator has only real eigenvalues. A purely magnetic
(PM, ${\sf C}_+=0$) Weyl operator has at least 2 zero eigenvalues
while the remaining eigenvalues are purely imaginary.
\end{thm}
This can be seen more easily if we switch to an orthonormal frame
(see \cite{ColHer09}). Then using a $(4+6)$-block form:
\beq
\mathcal{C}_+=\begin{bmatrix}
S_1 & 0 \\ 0 & S_2
\end{bmatrix},
 \quad \mathcal{C}_-=\begin{bmatrix}
0 & T \\ -T^t & 0
\end{bmatrix}, \quad S_1,~S_2 ~~\text{symmetric}, ~T \text{ a $4\times 6$ matrix}.
\eeq
In the purely electric case, the eigenvalues are the eigenvalues of $S_1$ and $S_2$, which are clearly real. In the purely magnetic case, we note that the matrix $T$ can be decomposed (using the singular value decomposition) as $T=g_1Dg_2$, where $g_1$ and $g_2$ are $SO(4)$ and $SO(6)$ matrices, respectively, and $D$ is a diagonal $4\times 6$ matrix $D=\diag(\lambda_1,\lambda_2,\lambda_3,\lambda_4)$. Consequently, a purely magnetic Weyl operator has eigenvalues $\{0,0,\pm i\lambda_1,\pm i \lambda_2,\pm i \lambda_3,\pm i \lambda_4\}$.

From the vanishing of the diagonal in (\ref{C-}) we see, in
particular, that the b.w.\ 0
part of a PE Weyl tensor has spin type $\{...\}_0$. The fact that PE
implies only real eigenvalues is illustrated, for example, by the
classification of the eigenvalue structure of the 5D type {\bf D}
Weyl operator in appendix \ref{subsec: type II eigenvalues}. On the
other hand, the converse is not true -- these are only
\emph{necessary} criteria, not sufficient; indeed, among
(\ref{App172})--(\ref{AppLast}) in Appendix \ref{app: spin
w3<>0=w4=w5}, one can find cases where the Weyl operator has only
real eigenvalues (i.e., only 1 and 0 occurring), which thus
constitute counterexamples to the converse (since $\wb\neq 0$ and
hence not PE).

For a PM Weyl tensor the b.w.\ 0 part  has spin type
$\{(000)\}_\parallel[\Rb=0]$ (the diagonal of (\ref{C+}) vanishes)
which does indeed give purely imaginary eigenvalues, as illustrated by
(\ref{AppPM}) in Appendix \ref{app: spin w3<>0=w4=w5} in the type
{\bf D} case. Also for the PM case the converse is not true, as is
illustrated by (\ref{AppIm}) in Appendix \ref{app: spin w3<>0=w4=w5}
which has purely imaginary eigenvalues but is not PM.

\subsection{Exact purely electric/magnetic solutions}\label{sec:PEPM}
In 4D there have been many studies attempting to classify purely
electric or magnetic solutions with various sources. There is a
wealth of  4D purely electric Weyl spacetimes; e.g., all static
spacetimes  (or, more generally, all those admitting a shear- and
vorticity-free timelike vector field), all spacetimes with
spherical, hyperbolic or planar symmetry, all Bianchi type I
spacetimes, and the Schwarzschild, C and G{\"o}del metrics.
However, only a very limited number of 4D purely magnetic Weyl
spacetimes are known to exist. It has even been conjectured that
purely magnetic vacuum or dust spacetimes do not exist (and this has
been proved under quite mild conditions, but not in general; see,
e.g.,
\cite{Hall73,Haddow95,VandenBergh03a,VandenBergh03b,Wylleman06}).
For an extensive review of known 4D purely magnetic Weyl solutions
see \cite{WyllVdB06}; see also \cite{Loz07,LozWyll11} for two more
recent contributions. All known purely magnetic metrics are
algebraically general (Petrov type {\bf I}), except for the locally
rotationally symmetric metrics given in \cite{LozCarm03,VdBWyll06},
which are of Petrov type {\bf D}.

There are also many examples of purely electric spacetimes in 5D,
including static spacetimes, spacetimes with an $\mb{R}^4$ spatial
translational invariance, spherically symmetric spacetimes, and
spacetimes with $SO(2)\times SO(2)$ isotropy or with spatial
isotropy $H\supset SO(2)\times SO(2)$ \cite{ColHer09}. Remarkably,
with our general definition of Weyl electric and magnetic parts and
of PE/PM spacetimes, many of the generic examples mentioned above
can be lifted to 5 or higher dimensions -- see \cite{HOW} for full
details. It would be useful to classify all such solutions. In
addition, the purely magnetic spacetimes are much harder to find.
The only purely magnetic spacetimes known so far in higher than 4
dimensions are conformally related to a two-parameter family of {\em
Riemann purely magnetic} spacetimes; these are all type {\bf I}$_i$,
see \cite{HOW}. A complete algebraic classification may be helpful
in the search for new exact 5D solutions.


\section{Conclusions and Discussion}\label{section: Conclusion}

In this paper we have presented a
refinement of the  null alignment classification of the Weyl
tensor of a 5D spacetime based on
the notion of spin type of the
components of highest boost weight and
the Segre types of the Weyl operator acting on bivector space
(and we have examined the intersection between the two
(sub)classifications).  We have presented a full treatment for types {\bf {N}} and
{\bf {III}}, and illustrated the classification from different
viewpoints (Segre type, rank, spin type) for types {\bf {II}} and {\bf {D}},
paying particular attention to possible nilpotence,  since this is a completely new feature of higher dimensions.  We also briefly discussed
alignment types  {\bf {I}} and {\bf {G}}.

In future work we shall develop the algebraic classification further. In
particular, it is hoped that
canonical forms can be determined explicitly in each algebraic subcase. The analysis
may be used to study particular spacetimes of special interest in detail; for
example, stationary (static) spacetimes and
warped product spacetimes.
In particular, we could attempt to classify  and analyse all vacuum Einstein type {\bf {III}}
spacetimes in 5D. We also note that the algebraic techniques may be of
use in other applications, since the
analysis is independent of any field equations.

This work is timely because of the
recent interest in the study of
general relativity (GR) in higher dimensions and, in
particular, in higher dimensional black holes \cite{EmpRea08}, motivated, in part,
by supergravity, string theory and the gauge-gravity
correspondence.
Indeed, even at the classical level gravity
in higher dimensions exhibits a much richer structure than in
4D. For example, there is no unique black hole
solution in higher dimensions.  In fact, there now exist a
number of different asymptotically flat, higher-dimensional
vacuum black hole solutions \cite{EmpRea08},
including Myers-Perry black holes \cite{MyePer86}, black rings
\cite{EmpRea02prl,PomSen06}, and various solutions with multiple
horizons (e.g., \cite{ElvFig07,ElvRod08}).

Since the algebraic classification of spacetimes has
 played such a crucial role in understanding exact solutions in 4D,
it is likely to play an important role also in higher
 dimensions.  However,
compared to 4D, the algebraic types defined by
the higher-dimensional alignment classification are rather
broad and it has proven more difficult to derive general
results.  Therefore, it is important to develop  more
refined algebraic classifications, and it is hoped that the work
in this paper will prove useful
in the search and analysis of exact higher dimensional (and, in particular, 5D)
black hole solutions.

It would also be useful to obtain a more constructive way
of accessing the invariant classification information in higher
dimensions.  For example, in \cite{ColHer11} {\em {discriminants}}
were used to study the
necessary conditions for the Weyl
curvature operator (and hence the higher dimensional Weyl
tensor) to be of algebraic type {\bf II} or {\bf D} in terms of simple scalar
polynomial curvature invariants.
In particular,  the Sorkin-Gross-Perry soliton, the supersymmetric black
ring, the doubly-spinning black ring, and a number of other higher
dimensional spacetimes were  investigated using
discriminant techniques.~{\footnote {Indeed, it was found that
the Sorkin-Gross-Perry soliton spacetimes and  the 5D supersymmetric
rotating black holes are
of type {\bf I} or {\bf G}, while  the doubly spinning black ring can only
be of type {\bf II} or more special at
the horizon~\cite{Coleyetal11} (in fact, it is more generally known that at Killing horizons the Weyl type must be {\bf II} or more special, at least under some assumptions on the matter content \cite{LewPaw05,PraPraOrt07}).}}

\section*{Acknowledgments}

A. C. was supported, in part, by NSERC. M.O. has been
supported by research plan RVO: 67985840 and research grant GA\v CR
P203/10/0749. L.W.\ has been supported by an Yggdrasil mobility
grant No 211109 to Stavanger University, a BOF research grant of
Ghent University, and a FWO mobility grant No V4.356.10N to Utrecht
University, where parts of this work were performed. M.O.\ and L.W.\
are grateful to the Department of Mathematics and Statistics,
Dalhousie University, for hospitality during the initial stages of
this work.


\appendix

\renewcommand{\theequation}{A\arabic{equation}}
\setcounter{equation}{0}

\section{Jordan normal structure and projectors}\label{appendix: JNB}


Let $V$ be the complexification of a real vector space of finite dimension $m$, and let ${\sf T}$ be a linear operator (endomorphism) on  $V$. For any  $\l\in\mathbb{C}$, define
\begin{equation}\label{Tlambda}
{\sf T}_\l\equiv {\sf T}-\l\,\id_V,
\end{equation}
with $\id_V$ the identity transformation on $V$, and
\begin{equation}
W_k^{(\l)} \equiv \Ker({\sf T}_\l^k),\qquad k\in\mathbb{N}.
\end{equation}
Notice that $W_0^{(\l)}=\{0\}$. One can show that
a smallest natural number $s=s(\l)$, $0\leq s\leq m$, exists such that for all $k\in\mathbb{N}$:
\begin{equation}\label{Wseq}
W_0^{(\l)}\precneqq  W_1^{(\l)}\precneqq \ldots \precneqq W_{s-1}^{(\l)}\precneqq W_{s}^{(\l)}=W_{s+k}^{(\l)}.
\end{equation}
The complex number $\l$ is an {\em eigenvalue} of ${\sf T}$ if and only if $s(\l)\geq 1$, i.e., it is a solution of the characteristic equation $\det({\cal T}-x {\bf 1}_m)=0$ of ${\sf T}$, where ${\bf 1}_m$ is the unit matrix of dimension $m$ and ${\cal T}$ is the representation matrix of ${\sf T}$ with respect to any basis of $V$. In this case, $E_\l\equiv W_1^{(\l)}\neq \{0\}$ and $M^{(\l)}\equiv  W_{s}^{(\l)}$ are called the {\em eigenspace}, respectively, {\em generalized eigenspace} of ${\sf T}$ corresponding to $\l$. Since ${\sf T}_\l\left(W_{k+1}^{(\l)}\right)\preceq W_k^{(\l)}$, the number $s(\l)$ is the index of nilpotence of the restriction of ${\sf T}_\l$ to $M^{(\l)}$.
If $\{\l_1,\ldots,\l_r\}$ is the set of different eigenvalues of ${\sf T}$, then
\begin{equation}\label{V=Mplus}
V=\bigoplus_{A=1}^r M^{(\l_A)}.
\end{equation}

Let $\l$ be an eigenvalue of ${\sf T}$, take an arbitrary complementary subspace $W_s'^{(\l)}$ of $W_{s-1}^{(\l)}$ in $M^{(\l)}=W_{s}^{(\l)}$:
\begin{equation*}
W_{s}^{(\l)}=W_{s-1}^{(\l)}\oplus W_s{''}^{(\l)},
\end{equation*}
and put $W_s''^{(\l)}=W_s'^{(\l)}$ (for consistency with the next pair of equations).
By induction, define for $k=1,\,\ldots,\,s-1$ an arbitrary complementary subspace $W_{s-k}'^{(\l)}$ of $W_{s-k-1}^{(\l)}\oplus{\sf T}_\l(W''^{(\l)}_{s-k+1})$ in $W_{s-k}^{(\l)}$ via
\begin{eqnarray*}
&&W_{s-k}^{(\l)}= W_{s-k-1}^{(\l)}\oplus{\sf T}_\l(W''^{(\l)}_{s-k+1})\oplus W_{s-k}'^{(\l)},\\
&&W_{s-k}''^{(\l)}\equiv {\sf T}_\l(W''^{(\l)}_{s-k+1})\oplus W_{s-k}'^{(\l)} .
\end{eqnarray*}
By construction, the dimensions $p_k=p_k(\l)$ of $W_{k}'^{(\l)}$ are independent of the choice of $W_{k}'^{(\l)}$. Let $(\eb_{i_k}^{\l,k})_{i_k=1..p_k}$ be an arbitrary basis of $W_k'^{(\l)}$, $k=1,\,..\,s$. A sequence $\eb_{i_k}^{\l}[k]$ of the form
\begin{equation}\label{JNS}
\eb_{i_k}^{\l}[k]\equiv \left(\eb_{i_k}^{\l,k},{\sf T}_\l(\eb_{i_k}^{\l,k}),\ldots,{\sf T}_\l^{k-1}(\eb_{i_k}^{\l,k})\right)
\end{equation}
is called a {\em Jordan normal sequence} (JNS) of length $k$, corresponding to $\l$.
It follows that a basis of $M^{(\l)}$ is given by
\begin{equation}\label{JNBpre}
\bigcup_{k=1}^s\bigcup_{i_k=1}^{p_k} \eb_{i_k}^{\l}[k].
\end{equation}
Notice that, since $\eb_{i_k}^{\l,k}\in W_k^{(\l)}=\Ker({\sf T}_\l^k)$, the last element of a JNS, and only this one, is an element of $\Ker({\sf T}_\l)=E_{\l}$, i.e., it is an eigenvector of ${\sf T}$ corresponding to $\l$.
Combining (\ref{V=Mplus})-(\ref{JNBpre}) yields a basis of $V$, which is called a {\em Jordan normal basis} (JNB) for ${\sf T}$.
The matrix representation of ${\sf T}$ with respect to the JNB has a block-diagonal form, called the {\em Jordan normal form} of ${\sf T}$; for a given eigenvalue $\l$ the corresponding JNSs of length $k$ give rise to $p_k(\l)$ {\em Jordan blocks} of dimension $k$ ($1\leq k\leq s(\l)$):
\begin{equation}
B_\l[k]\equiv\begin{bmatrix}
\lambda & 0 & 0& \cdots & 0 \\
1 & \lambda & 0& \ddots  & \vdots \\
0      & 1 &\lambda& \ddots & 0 \\
\vdots & \ddots     &\ddots& \ddots & 0 \\
0    &     \hdots &0   &  1      & \lambda
\end{bmatrix}.
\end{equation}
It follows that the algebraic multiplicity of the eigenvalue $\l$ (i.e., the power of $x-\l$ in the characteristic polynomial), $\dim(M^{(\l)})$, and the geometric multiplicity, $\dim(E_\l)$, are given by
\begin{equation}\label{alggeom}
\dim(M^{(\l)})=\sum_{k=1}^{s(\l)} k\,p_k(\l),\quad \dim(E_\l)=\sum_{k=1}^{s(\l)} p_k(\l),
\end{equation}
the latter being the total number of Jordan blocks corresponding to the eigenvalue $\l$.
In particular, $\dim(\Ker({\sf T}))$ equals the total number $\sum_{k=1}^s p_k(0)=\dim(E_0)$ of blocks corresponding to $\l=0$. Also, blocks of dimension 1 give rise to a diagonal block of ${\sf T}$.

Consider a particular element $\l_A$ of the set $\{\l_B\}$ of eigenvalues of ${\sf T}$. As $s(\l_A)$ is the index of nilpotence of ${\sf T}_{\l_A}$, it follows that $\bot_A$ defined by
\begin{eqnarray}\label{botA}
\bot_A\equiv {\sf \id_V}-\left(\id_V-\widetilde{\bot}_A\right)^{s(\l_A)},\qquad \widetilde{\bot}_A\equiv \frac{\prod_{B\neq A}{\sf T}_{\l_B}^{s(\l_B)}}{\prod_{B\neq A}(\lambda_A-\lambda_B)^{s(\l_B)}}
\end{eqnarray}
is the projection operator on $M^{(\l_A)}$. Clearly, $\id_V=\sum_{A=1}^r\bot_A$, and
one can use these projectors to decompose the operator ${\sf T}$:
\begin{eqnarray}\label{decomp}
{\sf T}={\sf N}_{\sf T}+\sum_{A=1}^r\lambda_A\bot_A.
\end{eqnarray}
This defines the nilpotent operator ${\sf N}_{\sf T}$ which contains all the information not
encapsulated in the eigenvalues $\lambda_A$. The operator
\begin{equation}\label{def NA}
{\sf N}_A\equiv \bot_A{\sf N}_{\sf T}={\sf N}_{\sf T}\bot_A
\end{equation}
has index of nilpotence $s(\l_A)$, and since $\bot_A\bot_B=\delta_{AB}\bot_A$ and
\begin{equation}\label{Ndef}
{\sf N}_{\sf T}=\sum_{A=1}^r {\sf N}_A
\end{equation}
it follows that ${\sf N}_{\sf T}$ has index of nilpotence $\max_A\,s(\l_A)$.
Combining (\ref{decomp}) and (\ref{Ndef}), we get the orthogonal decomposition:
\begin{equation}\label{Torth}
{\sf T}=\sum_{A=1}^r {\sf T}_A,\qquad {\sf T}_A\equiv {\sf N}_A+\l_A\bot_A.
\end{equation}
Notice that for all $A$, $\bot_A$ and ${\sf N}_A$ are elements of the commutative algebra $\mathbb{Q}(\l_1,\ldots,\l_r)[{\sf T}]$, where $\mathbb{Q}(\l_1,\ldots,\l_r)$ is the field of rational functions in the eigenvalues $\l_A$ over $\mathbb{Q}$. Since the polynomial invariant $\mbox{tr}(T^l)$ of ${\sf T}$ equals the sum of the $l^{th}$ powers of the eigenvalues, viz.\ $\sum_{A=1}^{r}\dim(M^{(\l)})\l_A^l$, and since the coefficients of the characteristic equation of ${\sf T}$ are combinations of the latter via Newton's identities, the eigenvalues are uniquely determined by the polynomial invariants. It follows that $\bot_A$ and ${\sf N}_A$ can be determined (in principle) by taking powers of ${\sf T}$ and traces thereof.

Finally, given the eigenvalues of ${\sf T}$ its Jordan normal form
can be easily determined by calculating $\rank({\sf N}_A^j)$,
$j=0,\ldots,s(\l_A)-1$, with $\rank({\sf N}_A^0)\equiv
\dim(M^{(\l_A)})$. Indeed, for $1\leq i\leq k$ we have
$\rank([B_{\l_A}[k]-\l_A{\bf 1}_k]^j)=k-j$. Hence,
\begin{equation}\label{rank NAj}
\rank({\sf N}_A^j)=\sum_{k=j+1}^{s(\l_A)} (k-j)p_k(\l_A),\qquad j=0,\ldots,s(\l_A)-1.
\end{equation}
This gives a linear system of $s(\l_A)$ equations in the same number of unknowns $p_k(\l_A)$~\footnote{Notice that (\ref{rank NAj}), $j=0$ is the first relation in (\ref{alggeom}), while subtracting (\ref{rank NAj}), $j=1$ from it gives the second relation.}, which can be readily solved to give
\begin{equation}\label{pk lA}
p_k(\l_A)=\rank({\sf N}_A^{k-1})-2\rank({\sf N}_A^{k})+\rank({\sf N}_A^{k+1}),\qquad k=1,\ldots,s(\l_A).
\end{equation}
Given the eigenvalues, ${\sf N}_A$ can be represented with respect to any basis, and the ranks of its powers can be easily determined by computing determinants.

\section{Details of the type III Weyl operator
classification}\label{app: type III}

\renewcommand{\theequation}{B\arabic{equation}}
\setcounter{equation}{0}

\subsection{Classification based on $\rank({\sf
C})$}\label{subsubsection: type III rank C}

Let $(ijk)$ be a cyclic permutation of $(345)$. We symbolize the
order 2 minors
$\Cc_K$$\left(\begin{smallmatrix}i_1&i_2\\j_1&j_2\end{smallmatrix}\right)$
of $\Cc_{K}$ - up to sign - by (no summation over repeated indices)
\begin{eqnarray}
&&\a_i\equiv 2(\vc_j^2+\vc_k^2),\quad \b_i\equiv \vc_i^2+\nc_j\nc_k,\label{aibi}\\
&&\gamma_{i+}=\vc_j\vc_k+\nc_i\vc_i,\quad \gamma_{i-}=\vc_j\vc_k-\nc_i\vc_i,\\
&&\delta_{i+}=2(\vc_j\vc_k+\nc_j\vc_i),\quad
\delta_{i-}=2(\vc_j\vc_k-\nc_k\vc_i).\label{deltai}
\end{eqnarray}
Adding to $\Cc_{-K}$ its $i^{th}$ row and taking the determinant we
arrive at the identity
\begin{equation}\label{identity}
2\vc_i d+\nc_id_i-\vc_kd_j+\vc_jd_k=0.
\end{equation}
Referring to (\ref{p_U}-\ref{p_V}) we have
\begin{equation}\label{direct sum}
\rank({\sf C})=3+\rank(\Cc_K)\,\Leftrightarrow\, \rank({\sf
C})=\;6\;\mbox{or}\;5\;\Leftrightarrow\;\Im({\sf C})=\Im({\sf
C}|_{\cal U})\oplus\Im(\Cck).
\end{equation}
When $\rank({\sf C})=4$ ($\Rightarrow \rank(\Cc_K)=2$) the last sum
is no longer direct (see also~(\ref{Im4}) below). This leads to the
following discussion (see (\ref{ddef})-(\ref{D45def}) for
definitions).

\begin{itemize}
\item \underline{$\rank({\sf C})=6$}\,\,\, $\Leftrightarrow\,\,\rank(C_{K})=3\,\,\Leftrightarrow\,\, D^6\neq
0$. In this case $\Cckt$, and hence ${\sf C}|_{\cal U}$, is
one-to-one while $\Cck$ maps onto ${\cal V}$. Hence, by (\ref{direct
sum}):
\begin{eqnarray}
&&\Im({\sf C})=\langle{\sf C}(\U_3),{\sf C}(\U_4),{\sf C}(\U_5)\rangle\oplus{\cal V}=\Im(\Cckt)\oplus{\cal V},\nonumber\\
&&\Im(\Cckt)=\langle\Cckt(\U_3),\Cckt(\U_4),\Cckt(\U_5)\rangle.
\label{Immax specify}
\end{eqnarray}
As (\ref{identity}) is the $i^{th}$ component of the vector relation
$\Cc_{-K}Y=0$, where
\begin{equation}\label{Ydef}
Y\equiv [d\;d_3\;d_4\;d_5]^t\quad\leftrightarrow\quad\Y\equiv
d\W+\sum_{i=3}^5 d_i\W^{[jk]},
\end{equation}
we have that
\begin{equation}\label{Kermax}
\Ker({\sf C})=\Ker(\Cck)\oplus{\cal
V},\qquad\Ker(\Cck)=\langle\Y\rangle.
\end{equation}

\item \underline{$\rank({\sf C})<6$}\,\,\, $\Leftrightarrow\,\,\rank(C_{K})=2\,\,\Leftrightarrow\,\, D^6=0$.

With the choice (\ref{jk v=0})-(\ref{jk v<>0}), for this case we
have that:
\begin{eqnarray}
&&\label{ImCKt}\Im(\Cckt)=\langle \Cckt(\U_3),\Cckt(\U_4)\rangle,\quad \Ker(\Cckt)\equiv\langle\U^+\rangle,\\
&&\label{ImC-K}\Im(\Cc_{-K})=\langle(\Cc_{-K})_{*j},\,(\Cc_{-K})_{*k}\rangle,\quad
\Ker(\Cck)\equiv\langle\Y^1,\Y^2\rangle. \label{Y1Y1}
\end{eqnarray}
Here and in general, $N_{*i}$ ($N_{i*}$) stands for the $i^{th}$
column (row) of a matrix $N$. One has $\U^+\in{\cal U}$,
$\U^+\notin\langle\U_3,\U_4\rangle$ and ${\sf C}(\U^+)\in{\cal V}$,
and
\begin{eqnarray*}
\rank({\sf C})=4\quad\Leftrightarrow\quad D^{45}=0\quad
\Leftrightarrow\quad {\sf C}(\U^+)\,\in\,\Im(\Cck).
\end{eqnarray*}
In summary, we have that
\begin{itemize}
\item[$(i)$] $\rank({\sf C})=5\,\,\Leftrightarrow\,\,D^6=0\neq D^{45}$, where, with the aid of (\ref{direct sum}):
\begin{eqnarray}
&&\Im({\sf C})=\langle{\sf C}(\U_3),{\sf C}(\U_4),{\sf C}(\U^+)\rangle\oplus\Im(\Cck)=\Im(\Cckt)\oplus{\cal V},\nonumber\\
&&\Ker({\sf C})=\Ker(\Cck)\oplus{\cal V}.\label{Im5}
\end{eqnarray}
\item[$(ii)$] $\rank({\sf C})=4\,\,\Leftrightarrow\,\,D^6=0=D^{45}$, where
\begin{eqnarray}
&&\Im({\sf C})=\langle{\sf C}(\U_3),{\sf C}(\U_4)\rangle\oplus\Im(\Cck),\nonumber\\
&& \Ker({\sf C})=\langle\U^++\W^+\rangle\oplus\Ker(\Cck)\oplus{\cal
V}. \label{Im4}
\end{eqnarray}
\end{itemize}
In Sec.\ \ref{subsec: type III intersection} we will  give explicit
expressions for $\Y^1$, $\Y^2$, $\W^+$ (which all belong to $\cal
W$), $\U^+$ and $D^{45}$, in the separate cases $\vc=0$ and $\vc\neq
0$.
\end{itemize}

\subsection{Subclassification based on $\rank({\sf C}^2)$}\label{subsubsection: type III rank C2}

Referring to (\ref{WeylOperator IIIsquare}) we explicitly have
\begin{equation}\label{C2 explicit}
\Cc_{-K}.\Cc_{K}{}^t=
\begin{bmatrix}
\nc_3^2+\vc^2-5\vc_3^2&\vc_5(\nc_3-\nc_4)-5\vc_3\vc_4&\vc_4(\nc_5-\nc_3)-5\vc_5\vc_3\\
\vc_5(\nc_3-\nc_4)-5\vc_3\vc_4&\nc_4^2+\vc^2-5\vc_4^2&\vc_3(\nc_4-\nc_5)-5\vc_4\vc_5\\
\vc_4(\nc_5-\nc_3)-5\vc_5\vc_3&\vc_3(\nc_4-\nc_5)-5\vc_4\vc_5&\nc_5^2+\vc^2-5\vc_5^2
\end{bmatrix}.
\end{equation}
Referring to (\ref{p_U})-(\ref{p_V}), the restriction of the map
$\Cck$ to $\Im(\Cckt)$ induces the isomorphism of vector spaces
\begin{equation}\label{Im C^2}
\Im({\sf C}^2)\cong \frac{\Im(\Cckt)}{\Im(\Cckt)\cap \Ker(\Cck)},
\end{equation}
such that
\begin{eqnarray}\label{rank CK C2}
\rank(\Cc_K)-\rank({\sf C}^2)=\dim(\Im(\Cckt)\cap\Ker(\Cck)).
\end{eqnarray}
For $\rank(\Cc_{K})=3$ we have $\dim(\Im(\Cckt)\cap\Ker(\Cck))\leq
\dim(\Ker(\Cck))=1$; regarding $\rank(\Cc_{K})=2$, and in view of
(\ref{Im C^2}), proposition \ref{prop_IIInilp} essentially states
that the 2-dimensional vector spaces $\Ker(\Cck)$ and $\Im(\Cckt)$
do not coincide. It follows that the difference
$\rank(\Cc_K)-\rank({\sf C}^2)$ in (\ref{rank CK C2}) is either 0 or
1, the latter case characterized by the existence of a (unique)
bivector direction $\langle\U^0\rangle$, $\U^0\in{\cal U}$,
satisfying
\begin{eqnarray}
0\neq \Cckt(\U^0)\in\Ker(\Cck),\quad\mbox{i.e.,}\quad {\sf
C}(\U^0)\notin{\cal V},\quad {\sf C}^2(\U^0)=\Cck(\Cckt(\U^0))=0.
\end{eqnarray}
This reconfirms the possible Segre types mentioned in table
\ref{Table: type III summary}. For fixed value of $\rank(\sf C)$ the
first Segre type, corresponding to the highest value of $\rank({\sf
C}^2)$, is the generic case. This leads to the following discussion
(using also the corresponding previous results of
subsection~\ref{subsubsection: type III rank C}).

\begin{itemize}
\item \underline{$\rank({\sf C})=6\Leftrightarrow \rank(\Cc_K)=3$.}

In this case $\rank({\sf C}^2)$ is either 3 or 2, where (cf.\
(\ref{Kermax}))
\begin{eqnarray}\label{CK=3 C2=2}
&&\rank({\sf C}^2)=2\quad\Leftrightarrow\quad
\Dth=0\quad\Leftrightarrow\quad
\langle\Cckt(\U^0)\rangle=\langle\Y\rangle.
\end{eqnarray}
Here the determinant $\Dth$ is given by
\begin{eqnarray}
\Dth&\equiv& \det\left(\Cc_{-K}.C_{K}{}^t\right)
=\det\left(\begin{bmatrix}({\Cc_{-K}})_{1*}\\ ({\Cc_{-K}})_{2*}\\
({\Cc_{-K}})_{3*}\end{bmatrix}.
\begin{bmatrix}(C_{K}{}^t)_{*1}\,(C_{K}{}^t)_{*2}\,(C_{K}{}^t)_{*3}\end{bmatrix}\right)\nonumber\\
&=&d^2-d_3^2-d_4^2-d_5^2\label{D6<def}\\
&=&\det\left(\left[Y\;\;(\Cc_K{}^t)_{*1}\,\,(\Cc_K{}^t)_{*2}\,\,(\Cc_K{}^t)_{*3}\right]\right),\nonumber
\end{eqnarray}
with $d$, $d_i$ and $Y$ as defined in (\ref{ddef}), (\ref{didef})
and (\ref{Ydef}). For later use, we notice that
\begin{equation}\label{d=0}
\rank({\sf C})=6,\,\,d=0\quad\Rightarrow\quad D^{6<}\neq 0.
\end{equation}

\begin{itemize}

\item In the case $\rank({\sf C}^2)=2$ $\leftrightarrow$  Segre type $[(3322)]$,
we have $W_2^{(0)}\equiv\Ker({\sf C}^2)={\cal W}\oplus{\cal
V}\oplus\langle \U^0\rangle$. If we take $\U'_3,\,\U'_4$ such that
${\cal U}=\langle\U'_3,\,\U'_4,\U^0\rangle$ and $\W^*\in{\cal
W}\setminus\Im(\Cckt)$, a Jordan basis is given by
\begin{equation}\label{JNB 3 spec}
JNB=(\U'_3[3],\U'_4[3],\U^0[2],\W^{*}[2]) ,
\end{equation}
corresponding to $W_3'^{(0)}=\langle\U'_3,\,\U'_4\rangle$ and
$W_2'^{(0)}=\langle\U^0,\,\W^{*}\rangle$.~\footnote{We have used the
result $\Im(\Cckt)=\langle \Cckt(\U'_3),\,\Cckt(\U'_4),\,\Y\rangle$,
which follows from (\ref{Immax specify}) and (\ref{CK=3 C2=2}).}

\item In the generic case $\rank({\sf C}^2)=3$ $\leftrightarrow$  Segre type $[(3331)]$, we have $W_2^{(0)}\equiv\Ker({\sf C}^2)={\cal W}\oplus{\cal V}$ and 
\begin{equation}\label{JNB 3 reg}
JNB=(\U_3[3],\U_{4}[3],\U_5[3],\Y[1]),
\end{equation}
corresponding to $W_3'^{(0)}={\cal U}$ and
$W_1'^{(0)}=\langle\Y\rangle$.

\end{itemize}

\item \underline{$\rank({\sf C})<6\Leftrightarrow \rank(\Cc_K)=2$.}

We choose $\V^*\in{\cal V}\setminus\Im(\Cck)$ and write
\begin{eqnarray}\label{JNB 2 X}
\X\equiv \left\{\begin{matrix}\U^+[2],&\rank({\sf C})=5;\\
(\U^++\W^+)[1],\V^*[1],&\rank({\sf C})=4.\end{matrix}\right.
\end{eqnarray}
In this case $\rank({\sf C}^2)$ is either 2 or 1. With the choice
(\ref{jk v=0})-(\ref{jk v<>0}) we have
\begin{eqnarray}\label{CK=2 C2=1}
&&\rank({\sf C}^2)=1\quad\Leftrightarrow\quad
\Dtw=0\quad\Leftrightarrow\quad
\Cckt(\U^0)\in\langle\Y^1,\Y^2\rangle,
\end{eqnarray}
where the determinant $\Dtw$ is given by
\begin{eqnarray*}
\Dtw&\equiv& \det\left(\begin{bmatrix}({\Cc_{-K}})_{1*}\\
({\Cc_{-K}})_{2*}\end{bmatrix}.
\begin{bmatrix}(C_{K}{}^t)_{*1}\, (C_{K}{}^t)_{*2}\end{bmatrix}\right)\\
&=&\beta_5^2+\gamma_{3+}^2+\gamma_{4-}^2-\a_5^2-\delta_{3+}^2-\delta_{4-}^2\\
&=&(\nc_3^2+\vc^2-5\vc_3^2)(\nc_4^2+\vc^2-5\vc_4^2)-(\vc_5(\nc_3-\nc_4)-5\vc_3\vc_4)^2\\
&\sim&\det\left(\left[Y^1\,\,Y^2\,\, (C_{K}{}^t)_{*1}\,\,
(C_{K}{}^t)_{*2}\right]\right).
\end{eqnarray*}
Here definitions~(\ref{aibi})--(\ref{deltai}) and (\ref{Y1Y1}) have
been used, as well as the correspondence $Y^1\leftrightarrow\Y^1$
and $Y^2\leftrightarrow\Y^2$ between column vectors in
$\mathbb{R}^{4\times 1}$ and elements of ${\cal W}$, as in
(\ref{Ydef}).

\begin{itemize}
 \item In the case $\rank({\sf C}^2)=1$ we have
$W_2^{(0)}\equiv\Ker({\sf C}^2)={\cal W}\oplus{\cal V}\oplus\langle
\U^+\rangle\oplus\langle \U^0\rangle$. If we take $\U'_3$ such that
${\cal U}=\langle\U'_3,\U^0,\U^+)$, and
\begin{equation*}
\Y^*\in\Ker(\Cck)\setminus \langle \Cckt(\U^0)\rangle,\quad
\W^*\in{\cal W}\setminus(\Im \Cckt+\Ker(\Cck))={\cal
W}\setminus(\langle\Y^1,\,\Y^2\rangle\oplus\langle
\Cckt(\U'_3)\rangle),
\end{equation*}
a Jordan basis is then given by
\begin{equation}\label{JNB 2 spec}
JNB=(\U'_3[3],\U^{0}[2],\W^{*}[2],\X,\Y^*[1]) ,
\end{equation}
corresponding for $\rank(\sf C)=5$ to
$W_3'^{(0)}=\langle\U'_3\rangle$,
$W_2'^{(0)}=\langle\U^0,\,\W^{*},\,\U^+\rangle$,
$W_1'^{(0)}=\langle\Y^*\rangle$, and for $\rank(\sf C)=4$ to
$W_3'^{(0)}=\langle\U'_3\rangle$,
$W_2'^{(0)}=\langle\U^0,\,\W^{*}\rangle$,
$W_1'^{(0)}=\langle\U^++\W^+,\,\V^*,\,\Y^*\rangle$.

\item In the generic case $\rank(C^2)=2$ we have $W_2^{(0)}\equiv\Ker({\sf C}^2)={\cal W}\oplus{\cal V}\oplus\langle \U^+\rangle$. A Jordan basis is
\begin{equation}\label{JNB 2 reg}
JNB=(\U_{3}[3],\U_{4}[3],\X,\Y^1[1],\Y^2[1]),
\end{equation}
corresponding for $\rank(\sf C)=5$ to
$W_3'^{(0)}=\langle\U_3,\,\U_4\rangle$,
$W_2'^{(0)}=\langle\U^+\rangle$,
$W_1'^{(0)}=\langle\Y^1,\,\Y^2\rangle$, and for $\rank(\sf C)=4$ to
$W_3'^{(0)}=\langle\U_3,\,\U_4\rangle$,
$W_1'^{(0)}=\langle\U^++\W^+,\,\V^*,\,\Y^1,\,\Y^2\rangle$.


\end{itemize}

\end{itemize}

\subsection{Intersection of the two refinements}\label{subsec: type III intersection}

In the subsequent analysis we will use the normal forms for
$(X,x)=(\nc,\vc)$ of table \ref{Table: X normal form}, which is
compatible with (\ref{jk v<>0}) when $\vc\neq 0$. Referring to these
forms, it is advantageous to distinguish between the cases $\vc=0$,
$\vc_3\neq 0=\vc_4=\vc_5$, $\vc_3\vc_5\neq 0=\vc_4$ and
$\vc_3\vc_4\vc_5\neq 0$. For $\vc=0$ we will work with the choice
(\ref{jk v=0}).

\subsubsection{Case $\vc=0$ (spin types $\{\cdot\}_0$)}

\label{subsubsec: type III vc=0}

This case is characterized by $\W\in\Ker(\Cck)\precneqq\Ker({\sf
C})$ or diagonal $\Lc$, cf.\ (\ref{WeylOperator III}). Regarding
$\Im(\Cckt)$, notice that we may replace $\Cckt(\U_3)$ by
$\W_{[45]}$ etc.\ in (\ref{Immax specify}) and (\ref{ImCKt}); thus
$\Im(\Cckt)=\langle\nc_3\W_{[45]},$
$\nc_4\W_{[53]},\,\nc_5\W_{[34]}\rangle$ in a unified form for all
values of $\rank({\sf C})$.

Regarding $\rank(\sf C)$ all $d_i$'s vanish, such that by
(\ref{rankC=6}):
\[D^6=d^2=(\nc_3\nc_4\nc_5)^2=\det(\Lc)^2.\]
Thus, with the choice (\ref{jk v=0}):
\[\rank(C)=6\quad\Leftrightarrow\quad \nc_5\neq 0\qquad \rightarrow \quad \langle\Y\rangle=\langle\W\rangle.\]
This precisely covers the primary spin types $\{(11)1\}$ and
$\{111\}$, while $\rank({\sf C})<6$ is equivalent to primary type
$\{110\}$. In the latter case we get
$\Im(\Cck)=\langle\V_3,\V_4\rangle$ and
\begin{eqnarray}\label{quant v=0}
\langle\Y^1,\,\Y^2\rangle=\langle\W,\,\W_{[34]}\rangle,\qquad \U^+=\U^5, \qquad D^{45}=\b_5^2\Hc_{55}. 
\end{eqnarray}
Thus
\begin{eqnarray}
&&\rank({\sf C})=5\quad \Leftrightarrow\quad \nc_5=0\neq\Hc_{55};\label{vc=0 rank 5}\\
&&\rank({\sf C})=4\quad \Leftrightarrow\quad \nc_5=0=\Hc_{55} \qquad
\rightarrow \quad \W^+=0.\label{vc=0 rank 4}
\end{eqnarray}

Regarding $\rank({\sf C}^2)$ we obtain
\[
\rank({\sf
C}^2)=\rank(\diag(\nc_3^2,\nc_4^2,\nc_5^2))=\rank(\diag(\nc_3,\nc_4,\nc_5))=\rank(\Lc)=\rank(\Cc_K),
\]
such that {\em for fixed value of $\rank(\sf C)$ the Segre type is
always the generic one.} Referring to (\ref{rank CK C2}) this tells
that $\Ker(\Cck)$ and $\Im(\Cckt)$ only have the zero vector in
common (as is readily checked), and is also in accordance with
\[D^{6<}=d^2\neq0\,[\rank({\sf C})=6],\quad D^{45<}=\b_5^2\neq0\,[\rank({\sf C})<6].\]

\subsubsection{Specifications for $\vc\neq 0$}

\label{subsubsec: type III vc<>0}

With the choice (\ref{jk v<>0}), and referring to (\ref{rankC=6}) and
identity (\ref{identity}), it is easily seen that
\begin{equation}\label{D6 vc<>0}
\rank({\sf C})<6\quad\Leftrightarrow\quad
d_3=d_4=0\,\,\mbox{or}\,\,d_3=d_5=0,
\end{equation}
respectively. None of these conditions are automatically satisfied;
i.e., the generic situation for {\em all} $\vc\neq 0$ spin types is
$\rank({\sf C})=6$. Regarding $\rank({\sf C})<6$, the choice
$j=1,\,k=4k$ naturally leads to the explicit expressions
\begin{equation}\label{Y1Y2}
\langle\Y^1,\Y^2\rangle=\langle\a_5\W_{[45]}+\delta_{4-}\W_{[34]}-\gamma_{3+}\W,\,\a_5\W_{[53]}+\delta_{3+}\W_{[34]}-\gamma_{4-}\W\rangle
\end{equation}
and
\begin{eqnarray}
\U^+&=&\delta_{4+}\U_3+\delta_{3-}\U_4-\a_5\U_5,\\
\W^+&=&\left(-{\delta_{4+}}\vc_3\Hc_{33}-{\delta_{3-}}\vc_4\Hc_{44}+2(\nc_5(\vc_3^2-\vc_4^2)-2\vc_3\vc_4\vc_5){\Hc_{34}}\right)\frac{\W}{\a_5}\nonumber\\
&&+2\left(-{\delta_{4+}}\vc_4\Hc_{33}+{\delta_{3-}}\vc_3\Hc_{44}+2(\vc_5(\vc_3^2-\vc_4^2)+2\vc_3\vc_4\nc_5){\Hc_{34}}\right)\frac{\W_{[34]}}{\a_5}\nonumber\\
&&+\left(\vc_3{\W}+2\vc_4\W_{[34]}\right)\Hc_{35}+\left(\vc_4{\W}-2\vc_3\W_{[34]}\right)\Hc_{45} , \\
-D^{45}&=&\delta_{4+}^2\Hc_{33}
+\delta_{3-}^2\Hc_{44}+\a_5^2\Hc_{55}+2\delta_{3-}\delta_{4+}\Hc_{34}-2\delta_{4+}\a_5\Hc_{35}-2\delta_{3-}\a_5\Hc_{45}.
\label{gener_v_D45}
\end{eqnarray}
In the cases where $\rank({\sf C}^2)=\rank(C_K)-1$, the bivector
$\U^0$ can be determined from $\Cckt(\U^0)=\Y$ by Cramer's method
(case $\rank({\sf C})=6,\,D^{6<}=0$) or from $a_3{\sf
C}^2(\U_3)+a_4{\sf C}^2(\U_4)=0$ (case $\rank({\sf
C})<6,\,D^{45<}=0$).

\subsubsection{Case $\vc_3\neq 0=\vc_4=\vc_5$ (spin types $\{\cdot\}_\para$ and $\{(11)1\}_\bot$)}

\label{subsubsec: type III vc3<>0}

Regarding $\rank(\sf C)$ we have
\begin{eqnarray}
d_4=d_5=0,\quad d_3=-2\vc_3(\vc_3^2+\nc_4\nc_5),\quad
d=\nc_3(\vc_3^2+\nc_4\nc_5),
    \label{d_d3_2.3.1}
\end{eqnarray}
such that (cf.\ (\ref{D6 vc<>0}))
\begin{eqnarray}\label{parallel <6 cond}
\rank({\sf C})<6\quad\Leftrightarrow \quad\vc_3^2+\nc_4\nc_5=0.
\end{eqnarray}
This condition is only compatible with $\nc_4\nc_5\neq0$ and
$\nc_4\neq\nc_5$, and thus occurs in the following cases:
\begin{eqnarray}
&\{110\}_{\para 0},&\quad \nc_3=0,\,\nc_5=-\nc_4=\pm\vc_3;\label{a1}\\
&\{111\}_\para,&\quad  \vc_3^2+\nc_4\nc_5=0;\label{a2}\\
&\{(11)1\}_{\bot},&\quad \nc_5=-2\nc_3=-2\nc_4,\quad
\vc_3^2=2\nc_3^2.\label{a3}
\end{eqnarray}
Then (\ref{Y1Y2}) and (\ref{gener_v_D45}) reduce to
\begin{eqnarray*}
&&\langle\Y^1,\Y^2\rangle=\langle-\nc_3{\W}+2\vc_3\W_{[45]},\,
\nc_4\W_{[34]}+\vc_3\W_{[53]}\rangle,\\
&&{D^{45}}=4\nc_4\nc_5^2(\nc_5\Hc_{44}-\nc_4\Hc_{55}+2\vc_3\Hc_{45}).
\end{eqnarray*}
Thus, for the cases (\ref{a1})-(\ref{a3}) we have  $\rank({\sf
C})=4$ if and only if
\begin{equation}\label{D45 v3<>0}
\nc_5\Hc_{44}-\nc_4\Hc_{55}+2\vc_3\Hc_{45}=0,
\end{equation}
else $\rank({\sf C})=5$ (cf.\ also remark \ref{remark type III_i and
rank}).
In all cases different from (\ref{a1})-(\ref{a3}) we have $\rank(\sf
C)=6$ and, from (\ref{Ydef}) with (\ref{d_d3_2.3.1}):
\[\langle\Y\rangle=\langle\nc_3\W-2\vc_3\W_{[45]}\rangle.\]

Regarding $\rank({\sf C}^2)$ the governing determinants reduce to
\begin{eqnarray*}
D^{6<}=(\nc_3^2-4\vc_3^2)(\vc_3^2+\nc_4\nc_5)^2,\qquad
D^{45<}=(\nc_3^2-4\vc_3^2)(\nc_4^2+\vc_3^2),
\end{eqnarray*}
such that, also in view of (\ref{parallel <6 cond}):
\begin{eqnarray}\label{n3=2v3}
\rank({\sf C}^2)=\rank(\Cc_K)-1\quad \Leftrightarrow\quad \nc_3=\pm
2\vc_3.
\end{eqnarray}
In conjunction with (\ref{C2 explicit}) it follows that $\U^0=\U_3$
in this case. If $\rank({\sf C})=6$, (\ref{n3=2v3})
$\leftrightarrow\rank({\sf C}^2)=2$ is possible for spin types
$\{(11)1\}_\para$, $\{110\}_{\para 1}$, $\{111\}_\para$ and
$\{(11)1\}_\bot$, but not for $\{(000)\}_\para$ and $\{110\}_{\para
0}$. Regarding the $\rank({\sf C})<6$ cases (\ref{a1})-(\ref{a3})
only (\ref{a2}) allows for the subcase (\ref{n3=2v3}) giving
$\rank({\sf C}^2)=1$, namely
\begin{equation}\label{spec a2}
\vc_3=\pm \frac{\nc_3}{2},\quad \nc_4=-\frac{1+\sqrt{2}}{2}\nc_3,\quad \nc_5=-\frac{1-\sqrt{2}}{2}\nc_3.
\end{equation}

\subsubsection{Case $\vc_3\vc_5\neq 0=\vc_4$ (spin types $\{111/0\}_\bot$ and $\{(11)1\}_\g$)}

\label{subsubsec: type III vc4=0}

From (\ref{didef}) we have \[d_4=2\vc_3\vc_5(\nc_5-\nc_3)\neq 0\]
such that {\em $\rank(C)=6$}. It is readily checked that $D^{6<}=0$
(see (\ref{D6<def})) admits solutions for both spin types, leading
to subcases where  $\rank({\sf C}^2)=2$.

\subsubsection{Case $\vc_3\vc_4\vc_5\neq 0$ (spin type $\{111/0\}_\g$)}

\label{subsubsec: type III generic}

First we take $\{110\}_\g$, with $\nc_5=0$ and $\nc_4=-\nc_3\neq 0$.
As a necessary condition for $\rank({\sf C})<6$ we obtain from
(\ref{ddef}):
\begin{equation}\label{110 d=0}
d=0\quad\Leftrightarrow\quad \vc_3^2=\vc_4^2.
\end{equation}
By possibly reflecting the $\mm_i$'s we may assume that the
$\vc_i$'s are positive. Then we obtain
\begin{equation}
\rank({\sf C})<6\quad\Leftrightarrow\quad\vc_3=\vc_4, \quad \nc_3=
-\frac{2\vc_3^2+\vc_5^2}{\vc_5}.\label{rankC<6 110g}
\end{equation}
Moreover, if $\rank({\sf C})<6$ it is readily computed that the
condition for having $\rank({\sf C^2})=1$ reads
\begin{equation}
D^{45<}=0\quad\Leftrightarrow\quad \nc_3=-5\vc_5,\quad
\vc_3=\sqrt{2}\vc_5,\label{rank C2 1 110g}
\end{equation}
while the condition for $\rank({\sf C})=4$ follows
from~(\ref{gener_v_D45}) and can always be solved for one of the
$\Hc_{ij}$.

In the case $\rank({\sf C})=6$, the condition for $\rank({\sf
C^2})=2$ is $D^{6<}=0$, which reads
\begin{equation}\label{rankC=6 C2=2}
\vc_5^2\nc_3^4+4\vc_3\vc_4\vc_5\nc_3^3-\left[\left(\frac{\vc_3^2}{2}+\frac{\vc_4^2}{2}+\vc_5^2\right)^2+\vc_5^4-5\vc_3^2\vc_4^2\right]\nc_3^2+(\vc_3^2+\vc_4^2+\vc_5^2)^3=0.
\end{equation}
Relations (\ref{d=0}) and (\ref{110 d=0}) imply that there are no
solutions to this equation with $\vc_3^2=\vc_4^2$. However, an
infinity of solutions exists: it is readily verified
that there are, e.g., two real solutions $\nc_3/\vc_5$ when
$\vc_3=\vc_5$ and $\vc_4=3\vc_5$. We conclude that {\em for spin
type III$\{110\}_\g$ all six Segre types of table \ref{Table: type
III summary} are possible.}


{\em The same conclusion is valid, a fortiori, for III$\{111\}_\g$}.
Notice from (\ref{ddef}) that $\vc_3=\vc_4=\vc_5$ implies
$d=\nc_3\nc_4\nc_5\neq 0$ (recall $\nc_3+\nc_4+\nc_5=0$) and thus
$\rank({\sf C})=6$; there are, e.g., two real solutions
$\nc_3/\vc_5$ to $D^{6<}=0$ for the subcase $\nc_4=2\nc_3$ thereof.
For $\vc_3=\vc_4(\neq \pm\vc_5)$, we have  that 
\begin{eqnarray*}
\rank({\sf C})<6\quad\Leftrightarrow\quad \nc_3-\nc_4=3\vc_5,  \quad \nc_ 5=\pm\sqrt{4\vc_3^2+5\vc_5^2},\\
\end{eqnarray*}
while $D^{45<}=0$ if, in addition, $\vc_5^2=7\vc_3^2$.\vspace{.2cm}

Table \ref{Table:  type III spin-Segre} summarizes the relation
between the spin type and Segre type refinement schemes for 5D
alignment type {\bf {III}} Weyl tensors.

\begin{table}[t]
\begin{tabular}{|c||cc|cccc|}
\hline
& $[(3331)]$& $[(3322)]$&$[(33211)]$&$[(32221)]$&$[(331111)]$&$[(322111)]$\\
\hline \hline
$\{110\}_0$ & - & - & (\ref{vc=0 rank 5}) &- &  (\ref{vc=0 rank 4})& -\\
$\{(11)1\}_0$ & x & - & - &- & -& -\\
$\{111\}_0$ & x & - & - &- & -& -\\
\hline
$\{(000)\}_\para$ & x & - & - &- & -& -\\
$\{(11)1\}_\para$ & G & (\ref{n3=2v3}) & - &- & -& -\\
$\{110\}_{\para 1}$ & G & (\ref{n3=2v3}) & - &- & -& -\\
$\{110\}_{\para 0}$ & G & - & (\ref{a1}) &- & (\ref{a1}), (\ref{D45 v3<>0}) & -\\
$\{111\}_\para$ & G & (\ref{n3=2v3}) & (\ref{a2}) & (\ref{a2}), (\ref{n3=2v3}) & (\ref{a2}), (\ref{D45 v3<>0})& (\ref{a2}), (\ref{D45 v3<>0}), (\ref{n3=2v3})\\
$\{(11)1\}_\bot$ & G & (\ref{n3=2v3}) & (\ref{a3}) &- & (\ref{a3}), (\ref{D45 v3<>0})& -\\
\hline
$\{111/0\}_\bot$ & G & $D^{6<}=0$ & - &- & -& -\\
$\{(11)1\}_\g$ & G & $D^{6<}=0$ & - &- & -& -\\
\hline
$\{110\}_\g$ & G & (\ref{rankC=6 C2=2}) & (\ref{rankC<6 110g}) & (\ref{rankC<6 110g}), (\ref{rank C2 1 110g}) & (\ref{rankC<6 110g}), $D^{45}=0$ & (\ref{rankC<6 110g}), (\ref{rank C2 1 110g}), $D^{45}=0$\\
$\{111\}_\g$ & G & $D^{6<}=0$ & $D^6=0$ & $D^6=D^{6<}=0$ & $D^6=D^{45=0}$ & $D^6=D^{6<}=D^{45}=0$ \\
\hline
\end{tabular}
\caption{Type {\bf {III}} Weyl tensors: possible  Segre types
(columns) for given spin types (rows). The normal forms for
$(X,x)=(\nc,\vc)$ of table \ref{Table: X normal form} are used to
define the spin types. The symbols - and x indicate that the
corresponding Segre type is either not allowed or is unique,
respectively. If $[(3331)]$ is the generic case (but other Segre
types are allowed) this is denoted by G, where it is understood that
the relations in columns 2 and 3 of the same row are not satisfied;
for Segre type $[(3322)]$  the relation in column 3 is not
satisfied. If a spin type allows for a Segre type corresponding to
$\rank({\sf C})<6$ (columns 3 to 6), the Segre type is defined by
the mentioned relation(s) and the negation of the remaining
relations in columns 3 to 6 of the same row; e.g., spin type
$\{111\}_\para$ will be of Segre type $[(32221)]$ if (\ref{a2}) and
(\ref{n3=2v3}) hold (being equivalent to (\ref{spec a2})), but not
(\ref{D45 v3<>0}). In this manner, the relations written down for
spin type $\{111\}_\g$ are the general Segre type defining
relations; for more degenerate spin types these reduce to more
specific conditions, as mentioned in the text and indicated here.}
\label{Table: type III spin-Segre}
\end{table}

\section{Spin types versus eigenvalues for Weyl type II}\label{subsec: type II eigenvalues}

\renewcommand{\theequation}{C\arabic{equation}}
\setcounter{equation}{0}

In this section we shall illustrate the classification of a type {\bf {II}} Weyl operator
based on its spin type.  We shall indicate the {\em degeneracies in the
eigenvalue spectra} of ${\sf M}$, ${\sf \Omega}$ and ${\sf C}$, and
indicate a zero eigenvalue by 0.  So, for instance, ${\sf
C}\{(000)(1111)(11)1\}$ indicates that ${\sf C}$ has one single, one
double and one quadruple non-zero eigenvalue, while 0 is a triple
eigenvalue. In particular, we present the case $\wb=0$ and comment on
the more general situations, where the subdivision is based on the number
of zero components in the normal forms of table \ref{Table:  Canforms bw
zeta}.  

\subsection{\bf Case $\wb=0$.} The spin type can be any of the $\{\ldots\}_0$
types. The matrices  $M$ and $\Omega$ are  diagonal so that their
eigenvalues can be immediately read off; namely,
$[-\Rb_3/2,-\Rb_4/2,-\Rb_5/2]$ for $M$ and
$[\Rb/2,\Rb/2-\Rb_3,\Rb/2-\Rb_4,\Rb/2-\Rb_5]$ for $\Om$. Obviously,
none of these can be nilpotent in this case (unless they vanish
identically, leading to type {\bf {III}} spacetimes). We also
observe that the combination $\rank(M)=2$, $\rank(\Omega)=3$ is not
permitted here. Let us remark that, in the case of type {\bf D}, the
presently considered case $\wb=0$ is of some interest since it
defines a {\em purely electric} type {\bf D} Weyl tensor -- several
properties and explicit examples of such spacetimes are known
\cite{HOW} (see also section~\ref{section: type IG}).

\paragraph{\underline{$\{(000)\}_0[\Rb\neq0]$}} This type arises when $\Rb_3=\Rb_4=\Rb_5=\Rb/3\neq0$.
Clearly we have $M: \{(111)\}$ (or $M: \{-\Rb/6[3]\}$), $\Omega: \{(111)1\}$
(or $\Omega: \{\Rb/6[3],\Rb/2[1]\}$), so that
\be
 C: \{(111111)(111)1\} ,
\ee
or $C: \{-\Rb/6[6],\Rb/6[3],\Rb/2[1]\}$. { We observe that in the case of type {\bf D} this is the spin type of the five-dimensional Schwarzschild-Tangherlini solution (cf. subsection~6.4 of \cite{PraPraOrt07}) and, more generally, of any Robinson-Trautman spacetime (not necessarily empty) \cite{PodOrt06}.

\paragraph{\underline{$\{(11)1\}_0[\Rb\neq0]$}} This occurs for $\Rb_5=\Rb_4$
(up to permutations of the axes). In general we have $M: \{(11)1\}$, $\Omega: \{(11)11\}$, so that
\be
 C: \{(1111)(11)(11)11\} .
\ee
 It is interesting to observe that, for type {\bf D} Einstein spacetimes,
the presence of a non-geodetic multiple WAND is equivalent to the spacetime having this spin type, with $\Rb_5=\Rb_4=-\Rb_3$ (see Proposition 9 in
\cite{PraPraOrt07} and \cite{DurRea09,ParradoWyll11}). All such spacetimes have in fact been found in \cite{DurRea09}, where it was also shown that in any dimensions an Einstein spacetime with a non-geodetic multiple WAND must be of type {\bf D}.

If $\Rb_3=2\Rb_4$ the second (non-degenerate) eigenvalue of $\Omega$ turns into 0,
so that $C: \{(1111)(11)(11)10\}$. There may be more special types in the following cases.

\begin{enumerate}

\item For $\Rb_3=0$, $M: \{(11)0\}$  ($M: \{-\Rb/4[2],0[1]\}$), $\Omega: \{(11)(00)\}$ ($\Omega: \{\Rb/2[2],0[2]\}$), so that
\be\label{blackstring_eqplane}
 C: \{(1111)(0000)(11)\} .
\ee
In the case of type {\bf D} this is, in particular, the spin type of Schwarzschild black strings (and Kerr black strings on the equatorial plane).

\item For $\Rb_4=0$, $M: \{(00)1\}$ ($M: \{0[2],-\Rb/2[1]\}$), $\Omega: \{(111)1\}$ ($\Omega: \{\Rb/2[3],-\Rb/2[1]\}$), so that
\be
 C: \{(0000)(111)(111)\} ,
\ee
or $C: \{0[4],\Rb/2[3],-\Rb/2[3]\}$.

\end{enumerate}

\paragraph{\underline{$\{(11)1\}_0[\Rb=0]$}} This can be seen as a subcase of
the previous spin type, and occurs for $\Rb_5=\Rb_4$, $\Rb_3=-2\Rb_4$. We find $M: \{(11)1\}$ and $\Omega: \{(11)10\}$, so that
\be
 C: \{(1111)(11)(11)10\} .
\ee

\paragraph{\underline{$\{111/0\}_0[\Rb\neq0]$}} This is the case of a generic matrix $\Sb$,
and in general we have $M: \{111\}$, $\Omega: \{1111\}$ and hence
\be
 C: \{(11)(11)(11)1111\} .
\ee
A single eigenvalue of $\Omega$ can be zero for special values of the $\Rb_i$ (e.g., for $\Rb_5=\Rb_3+\Rb_4$; the case $\Rb_5=-\Rb_3-\Rb_4$ corresponds to $\Rb=0$ and is thus discussed separately below).

Further degeneracies in the spectrum of $C$ can still occur when  $\Omega$ has degenerate eigenvalues, or when $M$ and $\Omega$ have some common eigenvalues, as we now discuss.

\begin{enumerate}

\item $\Omega$ has a degenerate eigenvalue iff $\Rb_5=0$ (up to axis permutation) so that $M: \{110\}$, $\Omega: \{(11)11\}$ ($M: \{-\Rb_3/2[1],-\Rb_4/2[1],0[1]\}$ and $\Omega: \{\Rb/2[2],(\Rb_3-\Rb_4)/2[1],(\Rb_4-\Rb_3)/2[1]\}$) and
\be
 C: \{(11)(11)(11)(00)11\} .
\ee

\item For $\Rb_5=-\Rb_4$, $M: \{111\}$ and $\Omega: \{1111\}$ (more precisely, $M: \{-\Rb_3/2[1],-\Rb_4/2[1],\Rb_4/2[1]\}$ and $\Omega: \{\Rb_3/2[1],-\Rb_3/2[1],(\Rb_3/2-\Rb_4)[1],(\Rb_3/2+\Rb_4)[1]\}$), so that
\be
 C: \{(111)(11)(11)111\} .
\ee

\item For $\Rb_5=-(\Rb_3+\Rb_4)/2$, $M: \{111\}$ and $\Omega: \{1111\}$ (more precisely, $M: \{-\Rb_3/2[1],-\Rb_4/2[1],(\Rb_3+\Rb_4)/4[1]\}$ and $\Omega: \{(\Rb_3+\Rb_4)/4[1],(\Rb_4-3\Rb_3)/4[1],(\Rb_3-3\Rb_4)/4[1],3(\Rb_3+\Rb_4)/4[1])$), so that, again
\be
 C: \{(111)(11)(11)111\} .
\ee

\item For $\Rb_5=(\Rb_3-\Rb_4)/2$, $M: \{111\}$ and $\Omega: \{1111\}$ (more precisely, $M: \{-\Rb_3/2[1],-\Rb_4/2[1],(-\Rb_3+\Rb_4)/4[1]\}$ and $\Omega: \{(3\Rb_3+\Rb_4)/4[1],(\Rb_4-\Rb_3)/4[1],3(\Rb_3-\Rb_4)/4[1],(\Rb_3+3\Rb_4)/4[1])$), and again
\be
 C: \{(111)(11)(11)111\} .
\ee

\end{enumerate}

\paragraph{\underline{$\{111\}_0[\Rb=0]$}} This case arises for $\Rb_5=-\Rb_3-\Rb_4$.
In general $M$ and $\Omega$ have no common eigenvalues,  and $\Omega$ has no multiple eigenvalues, so that
\be
 C: \{(11)(11)(11)1110\} .
\ee

However, if $\Rb_4=-\Rb_3$ then $M: \{110\}$ ($M: \{-\Rb_3/2[1],\Rb_3/2[1],0[1]\}$) and
$\Omega: \{(00)11\}$ ($\Omega: \{0[2],-\Rb_3[1],\Rb_3[1])$), and hence
\be
 C: \{(0000)(11)(11)11\} .
\ee


\subsection{\bf Case $\wb_3\neq0=\wb_4=\wb_5$.} It can easily  be seen that the combination $\rank(M)=2$,
$\rank(\Omega)=3$ is still not permitted here, and $\sf\Omega$ cannot be nilpotent.


\label{app: spin w3<>0=w4=w5}


\label{subsec_II100}


\paragraph{\underline{$\{111/0\}_\para[\Rb\neq0]$}} This is the spin type defined by the conditions $\wb_3\neq0=\wb_4=\wb_5$ when no further restrictions apply (special subtypes are discussed in detail below). It is generically $\{111\}_\para$, although also type $\{110\}_\para$ shows up in special instances described below.  While we focus here mostly on the case $\Rb\neq0$, some special subcases with $\Rb=0$ will also be mentioned, when they arise naturally.

The eigenvalues of $M$ and $\Omega$ are \beqn
 & & \lambda_M=\left\{-\frac{\Rb_3}{2},\frac{1}{4}\left(-\Rb_4-\Rb_5\pm\sqrt{(\Rb_4-\Rb_5)^2-4\wb_3^2}\right)\right\} , \nonumber \label{w3_eigenvalues} \\
 & & \lambda_\Omega=\left\{\frac{1}{2}(\Rb_3+\Rb_4-\Rb_5),\frac{1}{2}(\Rb_3-\Rb_4+\Rb_5),
                \frac{1}{2}\left(\Rb_4+\Rb_5\pm\sqrt{\Rb_3^2-4\wb_3^2}\right)\right\} .
\eeqn

Let us remark that our analysis is always restricted to Weyl tensors
with real components only. Note, however, that in certain ranges of
the parameters $M$ and $\Omega$ may admit a pair of complex
conjugate eigenvectors. When that happens the corresponding pair
$\{11\}$ should be replaced by $\{X\bar X\}$. {For brevity, we shall
sometimes omit this distinction at intermediate steps (whenever it
depends on the value of an arbitrary parameter), and we shall be
fully explicit only in the summarizing results for the Segre type of
$C$. Thus, with these conventions we generically have $M: \{111\}$
and $\Omega: \{1111\}$.}

An eigenvalue of $M$ vanishes for $\Rb_3=0$ or
$\wb_3^2=-\Rb_4\Rb_5$, while $\Omega$ has a zero eigenvalue for
$\Rb_5=\pm\Rb_3+\Rb_4$. It easily follows that $M$ is nilpotent
(with $M^2=0$) iff $\Rb_3=0=\Rb_4+\Rb_5$ (so that $\Rb=0$) with
$\wb_3^2=\Rb_4^2$, corresponding to a subcase of the case
$\rank(M)=1$, $\rank(\Omega)=4$, cf.~(\ref{rankM=1}) (up to a
trivial axis permutation). As already mentioned, $\Omega$ can not be
of type $\{(0000)\}$ here. Possible repeated eigenvalues are now
discussed.

\begin{enumerate}

\item For $2\wb_3=\Rb_4-\Rb_5$,  $M: \{(11)1\}$, $\Omega: \{1111\}$, and
\be
 C: \{(1111)(11)1111\} ,
\label{App172}\ee where the last pair becomes $\{X\bar X\}$ if
$(\Rb_4-\Rb_5)^2>\Rb_3^2$.

\begin{enumerate}

\item If, additionally, $2\Rb_3=\Rb_4+\Rb_5$ then $M: \{(111)\}$, $\Omega: \{1111\}$, and
\be
 C: \{(111111)1111\} ,
\ee where the last pair becomes $\{X\bar X\}$ if
$(\Rb_4-3\Rb_5)(\Rb_5-3\Rb_4)<0$. We observe that the spin type
becomes $\{110\}_\para[\Rb\neq0]$ here. Note also that for
$\Rb_5=-\Rb_4$ one gets that $M: \{(000)\}$ is {\em nilpotent} (as
already mentioned above), $\Omega: \{11X\bar X\}$ and $C:
\{(000000)11X\bar X\}\}$ (with $\Rb=0$ in this case).

\begin{enumerate}

\item With the further condition $\Rb_5=3\Rb_4$ one gets $M: \{-\Rb_4[3]\}$, $\Omega: \{2\Rb_4[3],0\}$, so that
\be
 C: \{(111111)(111)0\} .
\ee

\item If, instead, $\Rb_5=0$ one gets $M: \{(111)\}$, $\Omega: \{11X\bar X\}$ with one common eigenvalue, hence
\be
 C: \{(1111111)1X\bar X\} .
\ee

\end{enumerate}

\item If $4\Rb_3\Rb_4=5\Rb_4^2-2\Rb_4\Rb_5+\Rb_5^2$ then $M: \{(11)1\}$, $\Omega: \{(11)11\}$, therefore
\be
 C: \{(1111)(11)(11)11\} .
\ee

\begin{enumerate}

\item If $\Rb_5=-3\Rb_4$ then $M$ and $\Omega$ share their double eigenvalue and one single eigenvalue, so that
\be
 C: \{(111111)(111)1\} .
\ee The spin type is $\{110\}_\para[\Rb\neq0]$ here.

\item If $\Rb_5=9\Rb_4$, the double $M$-eigenvalue equals a single $\Omega$-eigenvalue, hence
\be
 C: \{(11111)(11)(11)1\} .
\ee The spin type is $\{110\}_\para[\Rb\neq0]$ here.

\end{enumerate}

\item If $\Rb_3=\Rb_4-\Rb_5$ then $M: \{(11)1\}$, $\Omega: \{(11)10\}$, so that
\be
 C: \{(1111)(11)(11)10\} .
\ee

\begin{enumerate}

\item If $\Rb_5=-\Rb_4$ the double eigenvalues of $M$ and $\Omega$ vanish, thus giving
\be
 C: \{(0000000)(11)1\} .
\ee

\item If $\Rb_4=0$ the double eigenvalue of $\Omega$ equals the single eigenvalue of $M$, therefore
\be
 C: \{(1111)(1111)10\} .
\ee The spin type is $\{110\}_\para[\Rb=0]$.

\item If $3\Rb_5=5\Rb_4$ the single non-zero eigenvalue of $\Omega$ equals the double eigenvalue of $M$, which implies
\be
 C: \{(11111)(11)(11)0\} .
\ee

\end{enumerate}

\item If $2\Rb_3=\Rb_4-\Rb_5$, $M: \{(11)1\}$, $\Omega: \{11X\bar X\}$, and the single eigenvalue of $M$ is also an eigenvalue of $\Omega$, hence
\be
 C: \{(1111)(111)1X\bar X\} ,
\ee which reduces to $C: \{(0000)(111)1X\bar X\}$ if $\Rb_5=-\Rb_4$.

When, additionally, $\Rb_5=2\Rb_4$, then also the double eigenvalue
of $M$ is an eigenvalue of $\Omega$, so that \be
 C: \{(11111)(111)X\bar X\} .
\ee

\item If $(\Rb_4+\Rb_5)\Rb_3=-(\Rb_4^2+\Rb_5^2)$, $M: \{(11)1\}$, $\Omega: \{1111\}$, and the single eigenvalue of $M$ is also an eigenvalue of $\Omega$, hence
\be
 C: \{(1111)(111)111\} .
\ee

\item If $2\Rb_3=\Rb_4-3\Rb_5$, $M: \{(11)1\}$, $\Omega: \{1111\}$, and the double eigenvalue of $M$ is also an eigenvalue of $\Omega$, hence
\be
 C: \{(11111)(11)111\} ,
\ee where the last pair becomes $\{X\bar X\}$ if
$(\Rb_4+\Rb_5)(5\Rb_5-3\Rb_4)<0$.

\item If $2\Rb_3=\pm\sqrt{13(\Rb_4^2+\Rb_5^2)+10\Rb_4\Rb_5}$, $M: \{(11)1\}$, $\Omega: \{1111\}$, and the double eigenvalue of $M$ is also an eigenvalue of $\Omega$, hence
\be
 C: \{(11111)(11)111\} .
\ee

\end{enumerate}

\item For $\wb_3^2=-(\Rb_3-\Rb_4)(\Rb_3-\Rb_5)$,  $M: \{(11)1\}$, $\Omega: \{1111\}$, and
\be
 C: \{(1111)(11)1111\} ,
\ee where the last pair becomes $\{X\bar X\}$ if
$\Rb_3^2+4(\Rb_3-\Rb_4)(\Rb_3-\Rb_5)<0$. Note also that $F_3^*=0$
identically here, therefore the spin type is
$\{111\}_\para[\Rb\neq0]$ (including all the following special
subcases unless stated otherwise). For $\Rb_3=0$ (so that
$\wb_3^2=-\Rb_4\Rb_5$) one gets $M: \{(00)1\}$, $\Omega: \{11X\bar
X\}$ and $C: \{(0000)(11)11X\bar X\}$, which describes the case
$\rank(M)=1$, $\rank(\Omega)=4$ of~(\ref{rankM=1}) (the more special
subcase $M: \{(000)\}$ has been already mentioned).

\begin{enumerate}

\item If $\Rb_3=\Rb_4-\Rb_5$, $M: \{(11)1\}$, $\Omega: \{(11)10\}$, hence
\be
 C: \{(1111)(11)(11)10\} .
\ee

\item If $4\Rb_5(\Rb_3-\Rb_4)=\Rb_3(5\Rb_3-4\Rb_4)$, $M: \{(11)1\}$, $\Omega: \{(11)11\}$, hence
\be
 C: \{(1111)(11)(11)11\} .
\ee

\begin{enumerate}

\item If, in addition, $\sqrt{3}\Rb_3=\pm2\Rb_4$, the double $M$-eigenvalue coincides with one of the single $\Omega$-eigenvalues, so that
\be
 C: \{(11111)(11)(11)1\} .
\ee

\item If $9\Rb_3=2(1\pm\sqrt{10})\Rb_4$, the double $M$-eigenvalue coincides with the double $\Omega$-eigenvalue, so that
\be
 C: \{(111111)(11)11\} .
\ee The spin type is $\{111\}_\para[\Rb=0]$.

\item If $\Rb_4=0$, the single $M$-eigenvalue coincides with one of the single $\Omega$-eigenvalues, therefore
\be
 C: \{(1111)(111)(11)1\} .
\ee

\item If $3\Rb_3=(-1\pm\sqrt{13})\Rb_4$, the single $M$-eigenvalue coincides with the double $\Omega$-eigenvalue, which implies
\be
 C: \{(1111)(1111)11\} .
\ee

\end{enumerate}

\item If $2\Rb_3=\Rb_5-\Rb_4$, $M: \{(11)1\}$, $\Omega: \{1111\}$, where the double $M$-eigenvalue is also an $\Omega$-eigenvalue, so that
\be
 C: \{(11111)(11)111\} ,
\ee where the last pair becomes $\{X\bar X\}$ if
$13\Rb_4+6\Rb_4\Rb_5-3\Rb_5<0$ (with $\wb_3^2>0$).

\item Also for $4\Rb_3=3(\Rb_4+\Rb_5)\pm\sqrt{13\Rb_4+10\Rb_4\Rb_5+13\Rb_5}$, $M: \{(11)1\}$, $\Omega: \{1111\}$, with the double $M$-eigenvalue being also an $\Omega$-eigenvalue, so that
\be
 C: \{(11111)(11)111\} .
\ee

\item If $\Rb_4=0$, $M: \{(11)1\}$, $\Omega: \{1111\}$, now sharing the single $M$-eigenvalue, i.e.
\be
 C: \{(1111)(111)111\} ,
\ee where the last pair becomes $\{X\bar X\}$ if
$\Rb_3(5\Rb_3-4\Rb_5)<0$ (with $\wb_3^2>0$).

\item If $\Rb_3=\pm\sqrt{\Rb_4^2+\Rb_4\Rb_5+\Rb_5^2}$, $M: \{(11)1\}$ and $\Omega: \{1111\}$ again share the single $M$-eigenvalue, thus
\be
 C: \{(1111)(111)111\} .
\ee

\end{enumerate}

\item For $4\wb_3^2=\Rb_3^2$, $M: \{111\}$, $\Omega: \{(11)11\}$. Further possible subcases can easily be studied as illustrated above, and listing all possibilities is not particularly illuminating.

\item For $\wb_3^2=\Rb_4(\Rb_3-\Rb_4)$, one has again $M: \{111\}$, $\Omega: \{(11)11\}$.

\item  There are four possible choices of parameters corresponding to $M$ and $\Omega$ having
a common eigenvalue (which is thus a multiple eigenvalue of $C$).
These are the cases (the common roots $x_0$ are indicated between
square brackets):
\begin{enumerate}
 \item $\wb_3^2=-(\Rb_3+\Rb_5)(\Rb_3-\Rb_4+2\Rb_5)$ \; \qquad [$2x_0=\Rb_3+\Rb_5-\Rb_4$];
 \item $4\wb_3^2=-(\Rb_4+\Rb_5)(2\Rb_3+\Rb_4+\Rb_5)$;
 \item  $3\wb_3^2=\Rb_3^2-4(\Rb_4^2+\Rb_5^2)-7\Rb_4\Rb_5\pm(\Rb_4+\Rb_5)\sqrt{D}$,\;
 \quad $D\equiv 3(4\Rb_4^2+4\Rb_5^2+4\Rb_4\Rb_5-\Rb_3^2)$;
 \item $2\Rb_3=\Rb_4-\Rb_5$.
\end{enumerate}
The common roots $x_1$ are given by: (a) $2x_1=\Rb_3+\Rb_5-\Rb_4$,
(b) and (d) $2x_1=\Rb_3$, (c) $x_1=-(\Rb_4+\Rb_5)/2+\sqrt{D}/6$.
Further specializations may give more common roots; for instance, if
$\Rb_4=-\Rb_5$ additionally holds in case (c), such that the type
becomes $\{110\}_\para$, then $x_2=-\sqrt{D}/6$ is another common
root. In general $M: \{111\}$, $\Omega: \{1111\}$ in all the above
cases, but some degeneracy is possible in special subcases.

\end{enumerate}

\paragraph{\underline{$\{111/0\}_\para[\Rb=0]$}}  This case is analogous to the case $\{111/0\}_\para[\Rb\neq0]$ discussed above, and can be analyzed similarly with the additional condition $\Rb=0$. Some special subcases have been already mentioned above.

\paragraph{\underline{$\{(000)\}_\para[\Rb\neq0]$}} Here we have $\Rb_3=\Rb_4=\Rb_5=\Rb/3\neq0$.
It follows from~(\ref{w3_eigenvalues}) that  $M: \{1X\bar X\}$ and
$\Omega: \{(11)11\}$, so that
\be
 C: \{(11)(11)(XX)(\bar X\bar X)11\} ,
\ee
where the last pair becomes $\{Z\bar Z\}$ if $4\wb_3^2>\Rb_3^2$.

$M$ can not have any degenerate eigenvalues. For $\Rb_3^2=4\wb_3^2$,
however, $\Omega$ has two double eigenvalues, namely $\Rb_3$ and
$\Rb_3/2$, therefore \be
 C: \{(11)(11)(11)(XX)(\bar X\bar X)\} .
\ee

\paragraph{\underline{$\{(000)\}_\para[\Rb=0]$}} If all $\Rb_i=0$ then $M: \{0[1],i\wb_3/2[1],-i\wb_3/2[1]\}$ and $\Omega: \{0[2],i\wb_3[1],-i\wb_3[1]\}$, so that
\be
 C: \{(0000)(XX)(\bar X\bar X)Z\bar Z\} .
\label{AppPM}\ee

Observe that, in the case of type {\bf D}, this is the spin type
that defines a {\em purely magnetic} type {\bf D} Weyl tensor. It
has recently been shown that (in any higher dimensions) such
spacetimes cannot occur, if one adds also the {\em
Ricci-flat/Einstein} condition~\cite{HOW}. So far no type {\bf D}
purely magnetic spacetimes are known in 5D (or in any dimension
higher than 4); in 4D the only known such spacetimes are locally
rotationally symmetric, see \cite{LozCarm03}.

\paragraph{\underline{$\{(11)1)\}_\para[\Rb\neq0]$}} This spin type occurs for $\Rb_5=\Rb_4$.   We observe, in particular, that in the type~{\bf D} case this is the spin type of both five-dimensional Myers-Perry black holes and Kerr black strings, and in fact of a ``generic'' five-dimensional Kerr-Schild spacetime (cf. subsection~6.4 of \cite{PraPraOrt07} and subsection~5.5 of \cite{OrtPraPra09}). From~(\ref{w3_eigenvalues}) one gets $M: \{1X\bar X\}$ and $\Omega: \{(11)11\}$, so that in general
\be
 C: \{(11)(11)(XX)(\bar X\bar X)11\} ,
\ee where the last pair becomes $\{Z\bar Z\}$ if $4\wb_3^2>\Rb_3^2$.
The Myers-Perry black hole has this general eigenvalue spectrum. One
of the non-degenerate eigenvalues of $\Omega$ becomes zero if
$4\wb_3^2=\Rb_3^2-4\Rb_4^2$, in which case then \be
 C: \{(11)(11)(XX)(\bar X\bar X)10\} .
\ee

The eigenvalues of $M$ can never be degenerate for this spin type.
Nevertheless, further degeneracies are possible in the following
cases.

\begin{enumerate}

\item For $4\wb_3^2=\Rb_3^2$,  $M: \{1X\bar X\}$, $\Omega: \{(11)(11)\}$, and
\be
 C: \{(11)(11)(XX)(\bar X\bar X)(11)\} .
\ee

This becomes even more special if one of the followings additionally
holds.

\begin{enumerate}

\item If $\Rb_3=2\Rb_4$ then $M: \{1X\bar X\}$, $\Omega: \{(1111)\}$ (or $\Omega: \{\Rb_4[4]\}$), and
\be
 C: \{(1111)(11)(XX)(\bar X\bar X)\} .
\ee

\item If $\Rb_4=0$ then $M: \{0X\bar X\}$, $\Omega: \{(11)(00)\}$, and
\be
 C: \{(0000)(11)(11)(XX)(\bar X\bar X)\} .
\ee

\end{enumerate}

\item For $\wb_3^2=\Rb_4(\Rb_3-\Rb_4)$,  $M: \{1X\bar X\}$, $\Omega: \{(111)1\}$, and
\be
 C: \{(111)(11)(XX)(\bar X\bar X)1\} .
\ee The non-degenerate eigenvalue of $\Omega$ vanishes for
$\Rb_3=4\Rb_4$, in which case \be
 C: \{(111)(11)(XX)(\bar X\bar X)0\} .
\ee

\item For $\Rb_3=0$ we get  $M: \{X\bar X0\}$, $\Omega: \{(00)Z\bar Z\}$, thus
\be
 C: \{(0000)(XX)(\bar X\bar X)Z\bar Z\} .
\label{AppIm}\ee This is the eigenvalue spectrum for the Kerr black
string (away from the equatorial plane, where it is
(\ref{blackstring_eqplane})).

\item For $\wb_3^2=-\Rb_4(\Rb_3+\Rb_4)$, we get  $M: \{1X\bar X\}$ and $\Omega: \{(11)11\}$, but $M$ and $\Omega$ share one non-degenerate eigenvalue (i.e., $-\Rb_3/2$), so that
\be
 C: \{(111)(11)(XX)(\bar X\bar X)1\} .
\ee The only non-degenerate eigenvalue of $C$ vanishes for
$\Rb_3=-4\Rb_4$, in which case \be
 C: \{(111)(11)(XX)(\bar X\bar X)0\} .
\ee

\item For $\Rb_4=0=\Rb_3^2-3\wb_3^2$, we have  $M: \{1X\bar X\}$ and $\Omega: \{(11)X\bar X\}$, so that
\be
 C: \{(XXX)(\bar X\bar X\bar X)(11)(11)\} .
\ee

\end{enumerate}

\paragraph{\underline{$\{(11)1)\}_\para[\Rb=0]$}} This can also be understood as a subcase of the previous spin type, with again $\Rb_5=\Rb_4$, and the additional condition $\Rb_3=-2\Rb_4$. Similarly, we have in general $M: \{1X\bar X\}$ and $\Omega: \{(11)11\}$, which gives
\be
 C: \{(11)(11)(XX)(\bar X\bar X)11\} ,
\ee where the last pair becomes $\{Z\bar Z\}$ if $\wb_3^2>\Rb_4^2$.
No eigenvalues can vanish here.

The only case with further degeneracy now arises for
$\wb_3^2=\Rb_4^2$, giving $M: \{1X\bar X\}$ and $\Omega:
\{(11)(11)\}$ with a common eigenvalue (namely, $\Rb_4$), so that
\be
 C: \{(1111)(11)(XX)(\bar X\bar X)\} .
\ee

\paragraph{\underline{$\{(11)1)\}_\p[\Rb\neq0]$}} This is defined by $\Rb_4=\Rb_3$. Hence $M: \{111\}$ (or $M: \{1X\bar X\}$ if $(\Rb_3-\Rb_5)^2<4\wb_3^2$) and $\Omega: \{1111\}$ (or $\Omega: \{11Z\bar Z\}$ if $4\wb_3^2>\Rb_3^2$), so that
\be
 C: \{(11)(11)(11)1111\} ,
\ee or $C: \{(11)(XX)(\bar X\bar X)1111\}$ or $C: \{(11)(XX)(\bar
X\bar X)11Z\bar Z\}$ or $C: \{(11)(11)(11)11Z\bar Z\}$ according to
the specific values of the parameters. One of the $M$-eigenvalues
vanishes for either $\Rb_3=0$ or $\wb_3^2=-\Rb_3\Rb_5$, while
$\Omega$ has a zero eigenvalue for $\Rb_5=0$ or $\Rb_5=2\Rb_3$ or
$4\wb_3^2=-\Rb_5(2\Rb_3+\Rb_5)$.

Additional degeneracies of the eigenvalues are possible in the
following cases.

\begin{enumerate}

\item For $4\wb_3^2=(\Rb_3-\Rb_5)^2$ $M$ has a double eigenvalue, i.e.  $M: \{(11)1\}$, $\Omega: \{1111\}$, and
\be
 C: \{(1111)(11)1111\} ,
\ee where the last pair becomes $\{X\bar X\}$ if
$\Rb_5(2\Rb_3-\Rb_5)<0$. The single $M$-eigenvalue vanishes for
$\Rb_3=0$, in which case then $M: \{(11)0\}$, $\Omega: \{1111\}$,
and $C: \{(1111)(00)11X\bar X\}$. Other $C$-eigenvalues can possibly
vanish only in more degenerate cases, which are all now listed.


\begin{enumerate}

\item If $\Rb_3=5\Rb_5$ then $M: \{(11)1\}$, $\Omega: \{(11)11\}$, so that
\be
 C: \{(1111)(11)(11)11\} .
\ee

\item If $\Rb_5=0$ then $M: \{(11)1\}$, $\Omega: \{(11)10\}$, and
\be
 C: \{(1111)(11)(11)10\} .
\ee

\item If $\Rb_5=2\Rb_3$ then $M: \{(11)1\}$, $\Omega: \{(11)10\}$, and
\be
 C: \{(1111)(11)(11)10\} .
\ee

\item If $\Rb_5=5\Rb_3$ then $M: \{(11)1\}$, $\Omega: \{11X\bar X\}$, the repeated eigenvalue of $M$ (i.e., $-3\Rb_3/2$) is also an eigenvalue of $\Omega$, so that
\be
 C: \{(11111)(11)1X\bar X\} .
\ee

\item If $\Rb_3=-3\Rb_5$ then $M: \{(11)1\}$, $\Omega: \{11X\bar X\}$, the repeated eigenvalue of $M$ (i.e., $\Rb_5/2$) is also an eigenvalue of $\Omega$, so that
\be
 C: \{(11111)(11)1X\bar X\} .
\ee

\item If $\Rb_5=-\Rb_3$ then $M: \{(00)1\}$, $\Omega: \{11X\bar X\}$, the single eigenvalue of $M$ (i.e., $-\Rb_3/2$) is also an eigenvalue of $\Omega$, so that
\be
 C: \{(0000)(111)1X\bar X\} .
\ee

\end{enumerate}

\item For $4\wb_3^2=\Rb_3^2$ it is $\Omega$ that has a double (real) eigenvalue, i.e.  $M: \{111\}$, $\Omega: \{(11)11\}$, and
\be
 C: \{(11)(11)(11)(11)11\} ,
\ee where $\{(11)(11)\}$ becomes $\{(XX)(\bar X\bar X)\}$ if
$\Rb_5(\Rb_5-\Rb_3)<0$. One eigenvalue of $M$ vanishes if
$\Rb_3=-4\Rb_5$, so that $ C: \{(11)(11)(00)(11)11\}$. The vanishing
of a single $\Omega$-eigenvalue leads to $M$ having a double
eigenvalue and it has been thus already discussed above, while the
vanishing of the double eigenvalue of $\Omega$ leads to $M$ and
$\Omega$ sharing a single eigenvalue and is discussed below.

Further degeneracy (omitting the already discussed cases $M:
\{(11)1\}$) is possible when other $\Omega$-eigenvalues coincide, or
when $\Omega$ and $M$ have some common eigenvalues, as we now
analyze.

\begin{enumerate}

\item If $\Rb_3=2\Rb_5$ then $M: \{1X\bar X\}$, $\Omega: \{(111)1\}$, so that
\be
 C: \{(111)(11)(XX)(\bar X\bar X)1\} .
\ee

\item If $\Rb_5=3\Rb_3$ or $8\Rb_5=25\Rb_3$ or $4\Rb_5=(-2\pm\sqrt{2})\Rb_3$, a single eigenvalue of $\Omega$ is also an eigenvalue of $M$, so that $M: \{111\}$, $\Omega: \{(11)11\}$, with
\be
 C: \{(111)(11)(11)(11)(11)1\} .
\ee

\item Also for $\Rb_5=-\Rb_3$ a single eigenvalue of $\Omega$ is also an eigenvalue of $M$, but additionally the double eigenvalue of $\Omega$ vanishes, so that $M: \{111\}$, $\Omega: \{(00)11\}$ and
\be
 C: \{(111)(11)(11)(00)1\} .
\ee

\item If $\Rb_5=-2\Rb_3$ or $4\Rb_5=(-5\pm\sqrt{7})\Rb_3$, the double eigenvalue of $\Omega$ is also an eigenvalue of $M$, so that $M: \{111\}$, $\Omega: \{(11)11\}$, with
\be
 C: \{(1111)(11)(11)11\} .
\ee

\end{enumerate}

\item Also for $\wb_3^2=\Rb_5(\Rb_3-\Rb_5)$ we have  $M: \{111\}$, $\Omega: \{(11)11\}$ with
\be
 C: \{(11)(11)(11)(11)11\} ,
\ee where $\{(11)(11)\}$ becomes $\{(XX)(\bar X\bar X)\}$ if
$(\Rb_3-\Rb_5)(\Rb_3-5\Rb_5)<0$. No eigenvalue can vanish here.

For $\Rb_3=2\Rb_5$ this further specializes to $M: \{1X\bar X\}$,
$\Omega: \{(111)1\}$ with \be
 C: \{(111)(11)(XX)(\bar X\bar X)1\} .
\ee

\item For $\Rb_5=3\Rb_3$ or $\Rb_5=-\Rb_3$ or $\wb_3^2=2\Rb_3(\Rb_5-3\Rb_3)$, $M$ and $\Omega$ are of general Segre type, however they share one eigenvalue, so that
\be
 C: \{(111)(11)(11)111\} .
\ee No eigenvalue can vanish here (unless one allows for more
special cases, see the appropriate paragraphs). Some eigenvalues can
be complex, in which case one should use the more precise notation
$\{X\bar X\}$ etc., but we omit these details here.

\item Also for $\wb_3^2=-2\Rb_5(\Rb_3+\Rb_5)$, $M$ and $\Omega$ are of general Segre type with a common eigenvalue, and
\be
 C: \{(111)(11)(11)111\} ,
\ee where the last pair becomes $\{X\bar X\}$ if
$\Rb_3^2+8\Rb_5(\Rb_3+\Rb_5)<0$ (along with $\wb_3^2>0$).

An $M$-eigenvalue vanishes for $\Rb_3=-2\Rb_5$, hence $C:
\{(111)(11)(00)X\bar X\}$. For $7\Rb_5=-6\Rb_3$ an
$\Omega$-eigenvalue is zero, so that $C: \{(111)(11)(11)110\}$.

\item Also for $4\wb_3^2=-(\Rb_3+\Rb_5)(3\Rb_3+\Rb_5)$, $M$ and $\Omega$ are of general Segre type with a common eigenvalue, and
\be
 C: \{(111)(11)(11)111\} .
\ee

No $M$-eigenvalue can vanish here, whereas an $\Omega$-eigenvalue is
zero when $2\Rb_5=-3\Rb_3$, in which case $C: \{(111)(11)(11)110\}$.

For $23\Rb_5=-45\Rb_3$, $M$ and $\Omega$ are still of generic type
but they share {\em two} eigenvalues, so that \be
 C: \{(111)(111)(11)11\} .
\ee

\item The last case where $M$ and $\Omega$ are of general Segre type with a common eigenvalue arises for $3\wb_3^2=-(\Rb_3+\Rb_5)(3\Rb_3+4\Rb_5\pm\sqrt{9\Rb_3^2+12\Rb_3\Rb_5+12\Rb_5^2}$, with
\be
 C: \{(111)(11)(11)111\} .
\ee No $M$-eigenvalue can vanish here, whereas an
$\Omega$-eigenvalue is zero when $23\Rb_5=(-25\pm\sqrt{73})\Rb_3$,
thus giving $C: \{(111)(11)(11)110\}$.

For $23\Rb_5=-45\Rb_3$, $M$ and $\Omega$ share two eigenvalues, but
this case has been just discussed above.

\end{enumerate}

\paragraph{\underline{$\{(11)1)\}_\p[\Rb=0]$}} This is defined by $\Rb_4=\Rb_3$, $\Rb_5=-2\Rb_3$. Hence $M: \{111\}$ (or $M: \{1X\bar X\}$ if $4\wb_3^2>9\Rb_3^2$) and $\Omega: \{1111\}$ (or $\Omega: \{11Z\bar Z\}$ if $4\wb_3^2>\Rb_3^2$), so that
\be
 C: \{(11)(11)(11)1111\} ,
\ee or $C: \{(11)(XX)(\bar X\bar X)11Z\bar Z\}$ or $C:
\{(11)(11)(11)11Z\bar Z\}$ according to the specific values of the
parameters. For $\wb_3^2=2\Rb_3^2$ we have $M: \{110\}$ and $\Omega:
\{11Z\bar Z\}$, hence $C: \{(11)(11)(00)11Z\bar Z\}$. More special
cases may arise as follows.

\begin{enumerate}

\item For $4\wb_3^2=9\Rb_3^2$, $M: \{(11)1\}$, $\Omega: \{11Z\bar Z\}$ so that
\be
 C: \{(1111)(11)11Z\bar Z\} .
\ee

\item For $4\wb_3^2=\Rb_3^2$, $M: \{111\}$, $\Omega: \{(11)11\}$, and the double $\Omega$-eigenvalue is also an eigenvalue of $M$, hence
\be
 C: \{(1111)(11)(11)11\} .
\label{AppLast}\ee

\end{enumerate}

\subsection{\bf Case $\wb_4=0\neq\wb_3\wb_5$.} The possible spin types are $\{111/0\}_\p$ and $\{(11)1\}_{\mbox{g}}$.
Since $\Rb_i-\Rb_j=\Sb_i-\Sb_j$, at most two of the $\Rb_i$ can take the same value.
In particular, we can assume that at least one $\Rb_i$ is non-zero.
Note also that the combination $\rank(M)=1$, $\rank(\Omega)=4$ is not permitted here
(cf.~(\ref{rankM=1})), so that, in particular, $M^2\neq0$. Both $\sf M$ and $\sf\Omega$ can be nilpotent (with $M^2\neq0$, $\Omega^3=0$), which {\em includes the special situation~(\ref{C nilpotent}) when $\sf C$ is nilpotent} (in which case the spin type specializes to $\{110\}_{\bot 0}[\Rb=0]$, cf.\ supra)}.

\subsection{\bf General case $\wb_3\wb_4\wb_5\neq0$.} Here all of the $\Rb_i$ take different values,
otherwise (after a suitable spin) this would reduce to the previous case. Accordingly,
the only possible spin type is the most general one; i.e., $\{111/0\}_{\mbox{g}}$. In particular, we can assume that at least two $\Rb_i$ are non-zero. The cases $\rank(M)=1$ (which implies $\rank(\Omega)=4$) and $\rank(\Omega)=2$ (which implies $\rank(M)=2$) are not permitted here (cf. table~\ref{Table: type II M/Omega}). Both $\sf M$ and $\sf\Omega$ can be nilpotent (with $M^2\neq0$, $\Omega^3\neq0$), but not simultaneously.





\end{document}